\documentstyle[12pt,a4,graphicx,subfigure]{article}
\begin{document}

%========================================================================
%          MACROS FOR REFERENCES
%========================================================================

\def\NPB#1#2#3{{\it Nucl.\ Phys.}\/ {\bf B#1} (#2) #3}
\def\NPA#1#2#3{{\it Nucl.\ Phys.}\/ {\bf A#1} (#2) #3}
\def\PLB#1#2#3{{\it Phys.\ Lett.}\/ {\bf B#1} (#2) #3}
\def\PRD#1#2#3{{\it Phys.\ Rev.}\/ {\bf D#1} (#2) #3}
\def\PRL#1#2#3{{\it Phys.\ Rev.\ Lett.}\/ {\bf #1} (#2) #3}
\def\PRT#1#2#3{{\it Phys.\ Rep.}\/ {\bf#1} (#2) #3}
\def\MODA#1#2#3{{\it Mod.\ Phys.\ Lett.}\/ {\bf A#1} (#2) #3}
\def\IJMP#1#2#3{{\it Int.\ J.\ Mod.\ Phys.}\/ {\bf A#1} (#2) #3}
\def\nuvc#1#2#3{{\it Nuovo Cimento}\/ {\bf #1A} (#2) #3}
\def\RPP#1#2#3{{\it Rept.\ Prog.\ Phys.}\/ {\bf #1} (#2) #3}
\def\APJ#1#2#3{{\it Astrophys.\ J.}\/ {\bf #1} (#2) #3}
\def\APP#1#2#3{{\it Astropart.\ Phys.}\/ {\bf #1} (#2) #3}
\def\etal{{\it et al\/}}
\def\AEF{A.E. Faraggi}

\newcommand{\n}{\hspace*{-2.5mm}}
\newcommand{\bev}{\begin{verbatim}}
\newcommand{\beq}{\begin{equation}}
\newcommand{\beqa}{\begin{eqnarray}}
\newcommand{\beqn}{\begin{eqnarray}}
\newcommand{\eeqn}{\end{eqnarray}}
\newcommand{\eeqa}{\end{eqnarray}}
\newcommand{\eeq}{\end{equation}}
\newcommand{\Eev}{\end{verbatim}}\newcommand{\bec}{\begin{center}}
\newcommand{\eec}{\end{center}}
\newcommand\N{{\rm I\kern-.18em N}}
\newcommand\R{{\rm I\kern-.21em R}}
\newcommand{\ve}{\varepsilon}
\newcommand{\COMMA}{\hspace{0.1cm},}
\newcommand{\STOP}{\hspace{0.1cm}.}
\newcommand{\SPACE}{\hspace{0.4cm}}

\def\ie{{\it i.e.}}
\def\eg{{\it e.g.}}
\def\half{{\textstyle{1\over 2}}}
\def\nicefrac#1#2{\hbox{${#1\over #2}$}}
\def\third{{\textstyle {1\over3}}}
\def\quarter{{\textstyle {1\over4}}}
\def\m{{\tt -}}
\def\mass{M_{l^+ l^-}}
\def\p{{\tt +}}
\def\slash#1{#1\hskip-6pt/\hskip6pt}
\def\slk{\slash{k}}
\def\GeV{\;{\rm GeV}}
\def\TeV{\;{\rm TeV}}
\def\y{\;{\rm y}}

\def\l{\langle}
\def\r{\rangle}
\newcommand{\lsim}   {\mathrel{\mathop{\kern 0pt \rlap
  {\raise.2ex\hbox{$<$}}}
  \lower.9ex\hbox{\kern-.190em $\sim$}}}
\newcommand{\gsim}   {\mathrel{\mathop{\kern 0pt \rlap
  {\raise.2ex\hbox{$>$}}}
  \lower.9ex\hbox{\kern-.190em $\sim$}}}
\renewcommand{\thefootnote}{\fnsymbol{footnote}}
\setcounter{footnote}{0}
\begin{titlepage}
\samepage{
\setcounter{page}{1}

%\rightline{OUTP-03-22P}
%\rightline{\tt hep-ph/0308169}

\vspace{3.5cm}
\begin{center}
{\Large \bf Cosmic Ray Signals
from Mini Black Holes in Models with Extra Dimensions:\\
\vspace{0.3cm}
An Analytical/Monte Carlo Study}

\vspace{1.5cm}
 {\large
Alessandro Cafarella$^1$\footnote{Alessandro.Cafarella@le.infn.it},
Claudio Corian\`{o}$^1$\footnote{Claudio.Coriano@le.infn.it} and
T. N. Tomaras$^{2}$\footnote{tomaras@physics.uoc.gr}}

\vspace{.25cm}
{\it $^1$Dipartimento di Fisica,
 Universita' di Lecce,\\
 I.N.F.N. Sezione di Lecce,
Via Arnesano, 73100 Lecce, Italy\\}
\vspace{.25cm}
{\it $^2$Department of Physics and Institute for Plasma Physics,
University of Crete and FORTH\\
Heraklion, Crete, Hellas\\}
\vspace{.25cm}
\end{center}
\begin{abstract}
We present a study of the multiplicities, of the lateral
distributions and of the ratio of the electromagnetic to the hadronic
components in the air showers, generated by the collision in the atmosphere
of an incoming high energy cosmic ray and mediated by the formation of a
mini black hole, predicted in TeV scale gravity
models with large extra dimensions.
The analysis is performed via a large scale simulation of the resulting
cascades over the entire range ($10^{15}-10^{19}$eV)
of ultra high initial energies, for several values of the number of large extra 
dimensions, for a variety of altitudes of the initial interaction and 
with the energy losses in the bulk taken into account.
The results are compared with a representative of the standard 
events, namely the shower due to the collision of a primary proton with 
a nucleon in the atmosphere.
Both the multiplicities and the lateral distribution of the showers
show important differences between the two cases and, consequently, may be
useful for the observational characterization of the events. The
electromagnetic/hadronic ratio is strongly fluctuating and, thus,  
less decisive for the altitudes considered.
\end{abstract}

\smallskip}
\end{titlepage}

\section{Introduction}
In the almost structureless fast falling with energy inclusive 
cosmic ray spectrum, two kinematic regions have drawn considerable 
attention 
for a long time \cite{explanations}. These regions are the only ones 
in which the spectral index of the cosmic ray flux shows
a sharper variation as a function of energy, probably signaling 
some ``new physics'', according to many. These two regions, 
termed the {\em knee} and the {\em ankle} \cite{uhecr} have been 
puzzling theorists and experimentalists alike and
no clear and widely accepted explanation of this 
unusual behaviour in the propagation of the primaries
- prior to their impact with the earth atmosphere - exists yet. 
A large experimental effort \cite{auger,euso} in the next several 
years will hopefully clarify several of the issues related 
to this behaviour. 

While the {\em ankle} is mentioned in the debate regarding the 
possible existence of the so called Greisen, Zatsepin and Kuzmin 
(GZK) cutoff \cite{gzk}, due to the interaction of the
primaries with the cosmic background radiation, the proposed 
resolutions of this puzzle are several, ranging from a resonant 
Z-burst mechanism \cite{zburst} to string relics and other exotic 
particle decays \cite{berez, ben, sb, cfp}. 
The existence of data beyond
the cutoff has also been critically discussed \cite{BW}.

Given the large energy involved in the first stage of the 
formation of the air showers, the study of the properties of the 
cascade should
be sensitive to any new physics between the electroweak scale and the
original collision scale. Especially in the highest energy region 
of the spectrum, the energy available in the interaction of the 
primaries with the atmospheric nuclei is far above 
any conceivable energy scale attainable at future ground-based accelerators. 
Therefore, the possibility of detecting supersymmetry, for instance, 
in cosmic ray showers has also been contemplated \cite{fragfun}.
Thus, it is not surprising, that most of the attempts to explain these 
features of the cosmic ray spectrum typically 
assume some form of new physics at those energies.
 
With the advent of theories with a low fundamental scale of 
gravity \cite{ED} and large compact or non-compact extra dimensions,
the possibility of copiously producing mini black holes (based on 
Thorne's hoop conjecture \cite{Thorne}) in collisions involving hadronic 
factorization scales above 1 TeV has received considerable 
attention \cite{DL} \cite{Sarcevic} and these ideas, naturally, 
have found their way also in the literature of high energy cosmic 
rays \cite{Anchordoqui,pp4} and astrophysics \cite{DR}. For instance,
it was recently suggested that the long known {\em Centauro} events 
might be understood as evaporating mini black holes, produced 
by the collision of a very energetic primary (maybe a neutrino) with 
a nucleon (quark) in the atmosphere \cite{Theodore}. Other proposals
\cite{Ewa} also either involve new forms of matter (for example 
strangelets) or 
speculate about major changes in the strong interaction dynamics 
\cite{White}.  

While estimates for the frequencies of these types of processes 
both in cosmic rays \cite{Cavaglia,Anchordoqui,Theodore} and at colliders 
\cite{Sarcevic} have been presented, detailed studies of the multiplicities of
the particles collected at the detectors, generated by the extensive 
atmospheric air showers following the first impact of the primary rays,
are far from covering all the main features of the cascade \cite{refhtml}. 
These studies 
will be useful in order to eventually disentangle new physics 
starting from an analysis of the geometry of the shower, of the multiplicity 
distributions of its main sub-components \cite{CCF}
and of its directionality from deep space. 
For instance, the study of the location of the 
maxima of the showers at positions which can be
detected by fluorescence mirrors \cite{review_Anchordoqui},
generated as they go across the atmosphere,
and their variations 
as a function of the parameters of the underlying physical theory, 
may help in this effort \cite{Cavaglia}; other observables which also 
contain potential new information are the multiplicities of the 
various particle sub-components and the opening of the showers as 
they are detected on the ground \cite{CCF}. 
We will focus on this last type of observables.

To summarize: in the context of the TeV scale gravity with large extra dimensions
it is reasonable to assume that mini black holes, black holes with mass
of a few TeV, can form at the first impact of ultra high energy
primary cosmic rays with nucleons in the atmosphere. The black hole 
will evaporate into all types of particles of 
the Standard Model and gravity. The initial partons will hadronize and
all resulting particles as they propagate in the atmosphere will develop
into a shower(s), which eventually will reach the detectors. 
The nature and basic characteristics of these showers is the 
question that is the main subject of the present work. What is the signature 
on the detector of the showers arising from the decay of such mini
black holes and how it compares with a normal (not black hole mediated)
cosmic ray event, due, for instance, to a primary proton with the 
same energy colliding with an
atmospheric nucleon (the "benchmark" event used here). The comparison will be 
based on appropriate observables of the type mentioned above.

Our incomplete control of the quantum gravity/string theory effects, of the physics
of low energy non-perturbative QCD and of the nature of the quark-gluon
plasma phase in QCD, makes a fully general analysis of the above phenomena 
impossible at this stage. To proceed, we made the following 
simplifying assumptions and approximations. (1) The brane 
tension was assumed much smaller than the fundamental gravity scale, so it does
not modify the flat background metric. It is not clear at this point
how severe this assumption is, since it is related to the 
``cosmological constant problem'' and to the concrete realization 
of the Brane-World scenario. (2) The black hole 
was assumed to evaporate instantly, leading to initial ``partons'', whose 
number and distributions are obtained semiclassically. No virtual holes
were discussed and no back reaction was taken into account. 
(3) The initial decay products were assumed
to fly away and hadronize, with no intermediate formation of a quark-gluon plasma 
or of a disoriented chiral condensate (DCC). (4) We used standard
simulation programs for the investigation of the extensive air showers produced
in the cases of interest. To this purpose, we have decided
to use the Monte Carlo program CORSIKA \cite{CORSIKA} with the 
hadronic interaction implemented in SIBYLL \cite{SIBYLL} 
in order to perform this comparison, selecting a benchmark process 
which can be realistically simulated by this Monte Carlo, though other 
hadronization models are also available \cite{QGSJET}.
Finally, (5) a comment is in order about our selection of benchmark process
and choice of interesting events. 
In contrast to the case of a hadronic primary, the mini black hole 
production cross section due to the collision of a $\geq 10^3$ TeV neutrino
with a parton is of the order of the weak interaction neutrino-parton
cross section \cite{Theodore}. It would, thus, be interesting to compare the atmospheric 
showers of a normal neutrino-induced cosmic ray event to one with a 
black hole intermediate state.  
Unfortunately, at present neutrinos are not available as primaries 
in CORSIKA, a fact which sets a limitation on our benchmark study. 
However, it has to be mentioned that neutrino scattering off 
protons is not treated coherently at very high energy, since effects of 
parton saturation have not yet been implemented in the existing
codes \cite{CCF}. As shown in \cite{KK} these effects tend
to lower the cross section in the neutrino case.  For 
a proton-proton impact, the distribution of momenta among the partons 
and the presence of a lower factorization scale should render this 
effect less pronounced. For these reasons we have selected 
as benchmark process a proton-to-air collision at the same depth ($X_0$) 
and with the same energy as the corresponding ``signal event''. 
In order to reduce the large statistical fluctuations in the
formation of the extensive air showers after the collisions,
we have chosen at a first stage, in the bulk of our work, to simulate collisions taking place in the lower part of the atmosphere,
up to 1 km above the detector, in order to see whether 
any deviation from a standard scattering scenario can be identified. 
Another motivation for the analysis of such deeply penetrating 
events is their relevance in the study of the possibility to interpret the 
Centauro events as evaporating mini black holes \cite{Theodore}.
A second group of simulations have been performed at a higher altitude, for comparison. 

The present paper consists of seven sections, of which this Introduction
is the first. In Section 2 we briefly describe the D-brane world scenario,
in order to make clear the fundamental theoretical assumptions in our
study. A brief review of the properties of black holes
and black hole evaporation is offered here, together with all basic 
semiclassical formulas used in the analysis, with the dependence on the 
large extra dimensions shown explicitly.
In Section 3 a detailed
phenomenological description of the modeling of the decay of the 
black hole is presented, which is 
complementary to the previous literature and provides an independent 
characterization of the structure of the decay. Incidentally, a Monte 
Carlo code for black hole decay 
has also been presented recently \cite{Webber}.
We recall that this description 
-as done in all the previous works on the subject-
is limited to the {\em Schwarzschild phase} of the lifetime of the mini 
black hole. 
The modeling of the radiation emission from the black hole - as obtained in
the semiclassical picture - (see \cite{kanti2} for an overview) is
performed here independently, using semi-analytical methods, and has been
included in the computer code that we have written and used, and which is 
interfaced with CORSIKA. 
Recent computations of the greybody factors
for bulk/brane emissions \cite{kanti2}, which match
well with the analytical approach of \cite{kanti1} valid in the low energy
limit of particle emission by the black hole, have also 
been taken into account.
Section 4 contains our modeling of the 
hadronization process.
The hadronization of the partons emitted by the black hole
is treated analytically in the black hole rest frame, by solving
the evolution equations for the parton fragmentation functions, making use of a
special algorithm \cite{CC} and of a specific set of initial conditions
for these functions \cite{kkp}.
After a brief discussion in Section 5 of the transformation of the 
kinematics of the black hole decay event from the black hole frame 
to the laboratory frame, we proceed in Section 6 with a Monte Carlo  
simulation of the extensive air showers of the particles produced by taking 
these particles as primaries. 
The simulations are quite intensive and have been 
performed on a small computer cluster. 
As we have already mentioned, in this work we focus on the multiplicities, on the 
lateral distributions of the events and on the ratio of electromagnetic 
to hadronic energies and multiplicities and scan the entire ultra high energy 
part of the cosmic ray spectrum. 
Our results are summarized in a series of plots and are commented upon in
the final discussion Section 7.

\section{TeV Scale Gravity, Large Extra Dimensions and Mini Black Holes}

The theoretical framework of the present study is the D-brane 
world scenario \cite{ED}. 
The World, in this scenario, 
is 10 dimensional, but all the Standard Model matter and forces
are confined on a $4+n_L$ dimensional hypersurface (the $D_{3+n_L}$-brane).
Only gravity with a characteristic scale $M_*$ can propagate in the bulk.
The $n_L$ longitudinal dimensions are constrained experimentally to be
smaller than ${\cal O}(TeV^{-1})$. However, for our purposes these dimensions 
may be neglected, since the Kaluza-Klein excitations related to these have 
masses at least of ${\cal O}(TeV^{-1})$, too large to affect our discussion 
below.
Consistency with the observed Newton's law, on the other hand, leads to the 
relation $M_{Pl}^2=M_*^{n+2}V_n$, between $M_{Pl}\simeq 10^{19}$ GeV, the 
fundamental gravity scale $M_*$ and the volume $V_n$ of the $n=6-n_L$ 
dimensional compact or non-compact transverse space.  
A natural choice for $M_*$, dictated a priori by the ``gauge hierarchy'' 
puzzle,
is $M_*={\cal O}(M_W)={\cal O}$(1 TeV), while the simplest choice for
the transverse space is an $n$-dimensional torus with all radii 
equal to $R$. Thus, one obtains a condition
between the number $n$ and the size $R(n)$ of the transverse dimensions.
Notice that under the above assumptions and for all values of $n$, $R$ 
is much larger than $10^{-33}$cm,
the length scale at which one traditionally expects possible deviations
from the 3-dimensional gravity force, and the corresponding dimensions 
are termed ``large extra dimensions'' (LED). 
For $n=2$ one obtains $R(n=2)$ of the order of a fraction of a mm. 
At distances much smaller than $R$ one should observe $3+n$-dimensional
Newton's law, for instance, as in torsion balance experiments \cite{Hoyle}. 
Current bounds on the size of these large extra dimensions and on $M_*$
come from various arguments, mostly of astrophysical
(for instance $M_* > 1500$ TeV for $n=2$) or cosmological ($M_* > 1.5$ TeV 
for $n=4$) origin \cite{kanti2}. A larger number of LED ($n$) translates into 
a reduced lower bound on $M_*$. It should be pointed out, that in general 
it is possible, even if ``unatural'', that the transverse space has a few 
dimensions large and the others small. 
Here we shall assume a value of $M_*$ of
order 1TeV, neglect the small extra dimensions and treat the number of 
LED ($n$) as a free parameter. 

The implications of the existence of LED are quite direct
in the case of black hole physics. The black hole is effectively
4-dimensional if its horizon ($r_H$) is larger than the size of the extra 
dimensions. In the opposite case ($r_H\ll R$, or equivalently for black hole masses
$M_{BH}\ll 10^{13}$kg for $n=6$ \cite{Theodore}) it is $4+n$ dimensional, it 
spreads over the full space and its properties are those of a genuine 
higher dimensional hole. 
According to some estimates, over which however there is no universal 
consensus \cite{suppress}, black holes should be produced copiously \cite{DL}
\cite{Giddings} in particle collisions, whenever the center of mass energy 
available in the 
collision is considerably larger than the effective scale $M_*$ ($\sqrt{s} >> M_*$). 
With $M_*\sim$1 TeV, one may contemplate the possibility of producing 
black holes with masses of order a few TeV. 

Their characteristic temperature $T_H$ is 
inversely proportional to the radius $r_H$ 
of the horizon, or roughly of order $M_*$ and evaporate by emitting particles, 
whose mass is smaller than $T_H$. The radiation emitted depends both on the 
spin of the emitted particle, on the dimension of the ambient space 
and on the amount of back-scattering outside the horizon,
contributions which are commonly included in the so called 
``greybody factor'', which are particularly relevant 
in the characterization of the spectrum at lower and at intermediate energy. 
A main feature of the decaying mini black hole 
is its large partonic multiplicity, 
with a structure of the event which is approximately spheroidal in the 
black hole rest frame. 

Once produced, these mini black holes evaporate almost instantly.
The phenomenological study of 4-dimensional black holes of large mass 
and, in particular, 
of their Hawking radiation \cite{Hawking} 
\cite{Page} \cite{MacGibbon}, as well as the study 
of the scattering of states of various spins ($s=0,1/2,1$) on a 
black hole background, all performed in the semiclassical 
approximation, have a long history.
For rotating black holes one identifies four phases characterizing its decay, 
which are
(1) the balding phase (during which the hole gets rid of its hair); 
(2) the spin-down phase (during which the hole slows down its rotational
motion);
(3) the Schwarzschild phase (the usual semiclassically approximated 
evaporation phase) and, finally, 
(4) the Planck phase (the final explosive part of the evaporation 
process, with important quantum gravitational contributions). 
Undoubtedly, the best understood among these phases 
is the Schwarzschild phase, which is characterized by the emission of 
a (black body) energy spectrum which is approximately thermal, 
with a superimposed energy-dependent modulation, especially at larger values 
of the energy. The modulation is a function of the spin and is calculable 
analytically only at small energies. 
Extensions of these results to 4+n dimensions are now available,
especially in the Schwarzschild phase, where no rotation and no charge 
parameter characterize the background black hole solutions. 
Partial results exist for the spin down phase, where the behaviour 
of the greybody factors have been studied (at least for 1 additional 
extra dimension) both analytically and numerically.
The Planck phase, not so relevant for a hole of large mass (say of the mass 
of the sun ($M\sim 2\times 10^{33}$ gr) which emits in the nano-Kelvin region,
is instead very relevant for the case of mini-black holes, for which the 
separation between the mass of the hole and the corresponding (effective) 
Planck mass $M_*$ gets drastically reduced as the temperature 
of the hole raises and the back-reaction of the metric has to be taken into account.

In the discussion below we shall use the semiclassical formulas derived
for large black holes in the Schwarzschild phase and naively extrapolate them 
to the mini black holes as well. 
This is not, we believe, a severe approximation for the phenomena we shall discuss.
As the hole evaporates, it looses energy, its mass decreases, its temperature
increases and the rate of evaporation becomes faster. Thus, the lifetime of
the hole is actually shorter than the one derived ignoring the back reaction.
As we shall see below, the naive lifetime is already many orders of magnitude
smaller than the hadronization time. This justifies the use
of the ``sudden approximation'' we are making of the decay process
and explains why the neglect of the back reaction is not severe.

We recall that the metric of the $4+n$
dimensional hole in the Schwarzschild phase is given by 
\cite{Myers}
%%%%%%%%
\begin{equation}
ds^2 = \left[1-\left(\frac{r_H}{r}\right)^{n+1}\right]\,dt^2 -
\left[1-\left(\frac{r_H}{r}\right)^{n+1}\right]^{-1}\,dr^2 - 
r^2 d\Omega_{2+n}^2\,,
\label{metric-n}
\end{equation}
%%%%%%%%%
where $n$ denotes the number of extra spacelike dimensions, 
and $d\Omega_{2+n}^2$ is the area of the 
($2+n$)-dimensional unit sphere which, using coordinates
 $0 <\varphi < 2 \pi$ and $0< \theta_i < \pi$, with $i=1, ..., n+1$
takes the form 
%%%%%%%%
\begin{equation}
d\Omega_{2+n}^2=d\theta^2_{n+1} + \sin^2\theta_{n+1} \,\biggl(d\theta_n^2 +
\sin^2\theta_n\,\Bigl(\,... + \sin^2\theta_2\,(d\theta_1^2 + \sin^2 \theta_1
\,d\varphi^2)\,\Bigr)\biggr)\,.
\label{unit}
\end{equation}
The temperature $T_H$ of the black hole is related to the size 
of its horizon by \cite{Myers}
%%%%%%%%%%
\begin{equation}
T_H=\frac{n+1}{4\pi\,r_H}\,
\label{temp}
\end{equation}
and the formula for the horizon $r_H$ can be expressed in general in terms 
of the mass of the black hole $M_{BH}$ and the gravity scale $M_*$ \cite{Myers}
%%%%%%%%%%%%
\begin{equation}
r_H= \frac{1}{\sqrt{\pi}M_*}\left(\frac{M_{BH}}{M_*}\right)^
{\frac{1}{n+1}}\left(\frac{8\Gamma\left(\frac{n+3}{2}\right)}{n+2}\right)
^{\frac{1}{n+1}}\,.
\label{horizon}
\end{equation}
%%%%%%%%%%%%
For $n=0$ and $M_*=M_{Pl}\simeq 10^{19}$ GeV it reproduces the usual formula 
for the horizon ($r_H=2 G M_{BH}$) of a 4 dimensional black hole. 
For $n>0$ the relation 
between $r_H$ and $M_{BH}$ becomes nonlinear and the presence of $M_*$ in 
the denominator of 
Eq.~(\ref{horizon}) in place of $M_{Pl}$ increases the horizon size 
for a given $M_{BH}$. For $M_{BH}/M_*\sim 5$ and $M_*=1$ TeV the size of the 
horizon is around $10^{-4}$ fm and decreases with increasing $n$.

In the Schwarzschild/spin-down phase, the number of particles emitted per unit 
time by the black hole as a function of energy 
is expressed in terms of the absorption/emission cross sections 
$\sigma^{(s)}_{j,n}(\omega)$ (or equivalently of the greybody factors 
$\Gamma(\omega)$), which, apart from $n$, depend on the spin $(s)$ of the 
emitted particle, the angular momentum $(j)$ of the partial wave and the 
corresponding energy $(\omega)$, 
\begin{equation}
\label{flux}
\frac{dN^{(s)}(\omega)}{dt d\omega} = 
\sum_{j} \frac{\sigma^{(s)}_{j,n}(\omega)}{2 \pi^2}
{\omega^2 \over \exp\left(\omega/T_{H}\right) \pm 1}d\,\omega  \,.
\label{rate}
\end{equation}
Multiplying the rate of emitted particles per energy interval 
$dN^{(s)}(\omega)/dt d\omega$ by the particle energy $\omega$ one obtains for
the power emission density 
\begin{equation}
\frac{d E^{(s)}(\omega)}{dt d\omega} = \sum_{j} 
\frac{\sigma^{(s)}_{j,n}(\omega)}{2 \pi^2}\,
{\omega^3  \over \exp\left(\omega/T_{H}\right) \pm 1}d\,\omega\,
\label{power}
\end{equation}
where the sum is over all Standard Model particles and the +($-$) in the denominator 
correspond to fermions (bosons), respectively.
$\sigma^{(s)}_{j,n}$ are the cross sections for the various partial waves and 
depend on the spin $s$ of each particle. 
We recall, that in the geometric optics approximation a black hole acts as a perfect 
absorber of slightly larger radius $r_c$ than $r_H$ \cite{Sanchez}, 
which can be identified as the critical radius for null geodesics 
\beq
r_c=\left(\frac{n+3}{2}\right)^{1/(n+1)}\sqrt{{n+3}\over {n+1}}\, r_H \,.
\label{radius}
\eeq

The optical cross section is then defined in function of  
$r_c$ (or equivalently $r_H$ via Eq.~(\ref{radius})), 
such that $A_k$, the effective surface area of the black hole hole 
projected over a k-dimensional sub-manifold 
becomes \cite{ehm}:
\begin{equation}
 A_k = \Omega_{k-2}\left( {d-1 \over 2} \right)^{{2\over d-3}}
 \left( {d-1\over d-3} \right)^{{k-2\over 2}} r_{H}^{k-2}
\label{ak}
\end{equation}
and 
\begin{equation}
 \Omega_k= {2\pi^{k+1\over 2} \over \Gamma({k+1\over 2})}.
\end{equation}
is the volume of a k-sphere. 

It is convenient to rewrite the greybody factors as a 
dimensionless constant $\Gamma_s=\sigma_s /A_4$ normalized to the 
effective area of the horizon $A_4$, obtained from (\ref{ak}) 
setting $k=4$ and $d= 4 + n$
\beq
A_4=4 \pi \left(\frac{n+3}{2}\right)^{2/(n+1)} \frac{n+3}{n +1}\, r_H^2,
\eeq
and replacing the particle cross section $\sigma$ in terms of a thermal 
averaged graybody factor $\Gamma_i$($\Gamma_{1/2}=2/3,\Gamma_{1}=1/4, 
\Gamma_0=1$, $i$ denoting the spin or species \cite{page}). 
Eqs.~(\ref{rate}) integrated over the frequency give (for particle $i$)

\beq
\frac{d N_i}{d t}=\alpha(n,r_H) \, T_H^3
\eeq
with
\beq
\alpha(n,r_H)=\frac{f_i}{2\pi^2} \,\Gamma_i \,\Gamma(3)\,\zeta(3)\, 
c_i\, A_4\, T_H^3, 
\eeq
where $c_i$ is the number of degrees of freedom of particle $i$ and $f_i$ 
is defined by the integral ($s_i$ is the spin)
\beq
\int_0^\infty d\, \omega\frac{\omega^2}{e^{\omega/T_H} - (-1)^{2 s_i}}= 
f_i\, \Gamma(3)\,\zeta(3)\, T_H^3
\eeq
from which $f_i=1$ $(f_i=3/4)$ for bosons (fermions). 
These numbers depend on the dimension of the brane, which in our case is 3.
$\Gamma(x)$ and $\zeta(x)$ are the Gamma and the Riemann function respectively.
Since $A_4$ depends on the temperature (via $r_H$), after some manipulations 
one obtains 
\beq
A_4 T_H^3= \frac{1}{4\pi} \left(\frac{n+3}{2}\right)^{2/(n+1)}(n+3)(n+1)\, T_H
\eeq
and 
\beq
\frac{d N_i}{d t}=\frac{f_i}{8\pi^3}\frac{(n+3)^{(n+3)/(n+1)}}{2^{2/(n+1)}} 
(n+1)\Gamma(3)\zeta(3) \Gamma_i c_i T_H.
\eeq
Summing over all the particles $i$ we obtain the compact expression \
\beq
\frac{d N }{d t}=\frac{1}{2\pi}\left(\sum_i{f_i}\,\overline{\Gamma}_i\, 
c_i\right)\,\Gamma(3)\,\zeta(3)\,T_H
\label{enne}
\eeq
with 
\beq
\overline{\Gamma}_i= 
\frac{\Gamma_i (n+1)\,(n+3)^{(n+3)/(n+1)}}{4 \pi^2 \, 2^{2/(n+1)}}.
\eeq

The emission rates are given by
\begin{equation}
\dot{N}_i \approx 4\times 3.7 \times 10^{21}\,
\frac{(n+3)^{(n+3)/(n+1)}(n+1)}{2^{2/(n+1)} \,}\,
\left(\frac{T_H}{{\rm GeV}}\right)\,\, {\rm s}^{-1} \,\,,
\end{equation}
\begin{equation}
\dot{N}_i \approx 4\times 3.7 \times 10^{21}\,
\frac{(n+3)^{(n+3)/(n+1)}(n+1)}{2^{2/(n+1)}}\,
\left(\frac{T_H}{{\rm GeV}}\right)\,\, {\rm s}^{-1} \,\,,
\end{equation}
\begin{equation}
\dot{N}_i \approx 4\times 1.85 \times 10^{20}\,
\frac{(n+3)^{(n+3)/(n+1)}(n+1)}{2^{2/(n+1)}
\,}\,
\left(\frac{T_H}{{\rm GeV}}\right)\,\, {\rm s}^{-1} \,\,,
\end{equation}
for particles with $s = 0,\, 1/2, \,1,$ respectively. Furthermore,
integration of Eq.~(\ref{power}) gives for the black hole mass evolution
\beqa
\frac{d M}{d t} &\equiv& -\frac{dE}{dt} = \beta(n, r_H)\, T_H^4 \nonumber \\
&=& \frac{1}{2\pi}\left(\sum_i{f_i}\,\overline{\Gamma}_i\, c_i\right)\,
\Gamma(4)\,\zeta(4)\,T_H^2, \nonumber \\
\label{emme}
\eeqa
with
\beq
\beta=\frac{1}{2 \pi^2}\sum_i \left(c_i\, \Gamma_i\, {f'}_i \right)\, A_4\, \Gamma(4)\,\zeta(4)
\eeq
where now ${f'_i}=1$ $(7/8)$ for bosons (fermions). 
Taking the ratio of the two equations (\ref{emme}) and (\ref{enne}) 
we obtain 
\beqa
\frac{d N}{d M}&=&\left(\frac{\alpha}{\beta}\right)\,\frac{1}{T_H}\nonumber \\
 &=& \rho \frac{4 \pi\theta(n)}{n+1}\frac{1}{M_*}
\left(\frac{M}{M_*}\right)^{\frac{1}{(n+1)}},
\label{better}
\eeqa
where we have defined 

\beq
\theta(n)= \left(\frac{8\Gamma\left(\frac{n+3}{2}\right)}{n+2}\right)^{\frac{1}{n+1}}\frac{1}{\sqrt{\pi}},
\eeq
and 
\beq
\rho=\frac{ \sum_i c_i\, f_i\, {\Gamma}_i\, \Gamma(3)\, \zeta(3)}
{ \sum_i c_i \,{f'}_i\, {\Gamma}_i\, \Gamma(4)\, \zeta(4)}.
\eeq
This formula does not include corrections from emission in the bulk. 

\begin{figure}
{\centering \resizebox*{11cm}{!}{\rotatebox{0}
{\includegraphics{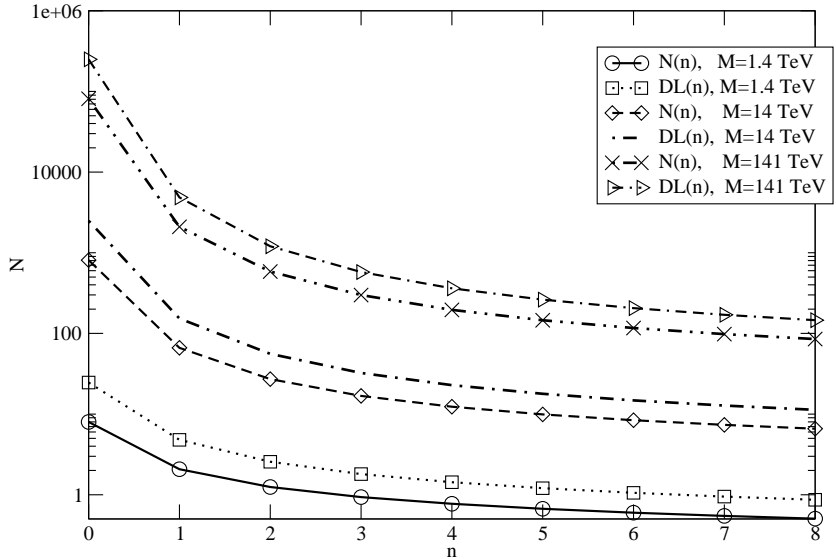}}} \par}

\caption{Multiplicities computed with Eq.~(\ref{improved}) (N(n)) 
and Eq.~(\ref{nav}) (DL(n)) for a varying number of extra dimensions $n$. 
DL(n) is the expression given in \cite{DL}.}
\label{DLL}
\end{figure}

In the ``sudden approximation'' in which the black hole decays at its original 
formation temperature one easily finds $ N = \left\langle
\frac{M_{BH}}{E} \right\rangle$, where $E$ is the energy spectrum
of the decay products, and using Boltzmann statistics 
$ N  \approx \frac{M_{BH}}{2T_H}$ one obtains the expression \cite{DL}

\begin{equation}
    N  = \frac{2\pi}{n+1}
    \left(\frac{M_{BH}}{M_*}\right)^\frac{n+2}{n+1}\,\theta(n).
\label{nav}
\end{equation}
This formula is approximate as are all the formulas for the multiplicities. 
A more accurate expression is obtained integrating Eq.~(\ref{better}) to obtain 

\beq
N=\rho \frac{4 \pi}{n+2}\left(\frac{M_{BH}}{M_*}\right)^\frac{n+2}{n+1}\theta(n),
\eeq
and noticing that the entropy of a black hole is given semiclassically by 
the expression 
\beqa
S_0 &=&\frac{n+1}{n+2} \frac{M}{T} \nonumber \\
&=& \frac{4 \pi}{n+2}\left(\frac{M_{BH}}{M_*}\right)^\frac{n+2}{n+1}\theta(n),
\nonumber 
\eeqa
one finds that 
\beq
N= \rho\, S_0, \nonumber \\
\label{improved}
\eeq
which can be computed numerically for a varying $n$. As one can see from Fig.~\ref{DLL}, 
the two formulas for the multiplicities are quite close, as expected, 
but Eq.~(\ref{nav}) gives larger values for the multiplicities 
compared to (\ref{improved}) as noted by \cite{Cavaglia1}. 
Other expressions for the multiplicities can be found in \cite{Cavaglia-Das-Casadio}.
Since the number of elementary states becomes quite large as we raise 
the black hole mass compared to the (fixed) gravity scale, and given 
the (large) statistical fluctuations induced by the formation of the airshower, which reduce the dependence on the multiplicity formula used, we will adopt 
Eq.~(\ref{nav}) in our simulations. Overall, in the massless approximation,
the emission of the various species for a 3-brane is characterized by 
approximately $2\%$ into spin zero, $85\%$ into spin half and $13\%$ 
into spin one particles, with similar contributions 
also for the power emissivities. These numbers change as we vary the 
dimension of the brane (d) and so does the formula for the emissivities, 
since the number of brane degrees of freedom $(c_i(d))$ 
has to be recomputed, together with the integrals on the emission spectra 
$(f_i(d))$ \cite{Cavaglia}.     

The integration of the equation for the power spectrum, 
in the massless approximation, can be used to compute the total time of decay 
(assuming no mass evolution during the decay)
%%%%%%%%%%
\begin{equation}
\tau \sim \frac{1}{M_*}\,\biggl(\frac{M_{BH}}{M_*}\biggr)^{(n+3)/(n+1)}\,
\label{lifetime}
\end{equation}
which implies that at an energy of approximately 1 TeV the decay time 
is of the order of $10^{-27}$ seconds. Therefore 
strong interaction effects and gravity effects appear to be widely separated 
and hadronization of the partons takes place after their crossing of the horizon. 
The black hole is assumed to decay isotropically (s-wave) to a set of 
$N$ elementary states, selected with equal probability from 
all the possible states available in the Standard Model. 
We mention that in most of the analysis presented so far \cite{pp4,Anchordoqui, Cavaglia} 
the (semiclassical) energy loss due to bulk emission has not been thoroughly analyzed. 
We will therefore correct our numerical studies by keeping into account some estimates 
of the bulk emission.

%\clearpage
\section{Modeling of the Black Hole Decay}

The amount of radiation emitted by the black hole in the ED is viewed, 
by an observer living on the brane, as missing energy compared to 
the energy available at the time when the black hole forms. 
From the point of view of cosmic ray physics 
missing energy channels imply 
reduced multiplicities in the final air shower and modified lateral distributions, 
these two features being among the main observables of the cosmic ray event. 
However, since the initial energy of the original cosmic ray 
is reconstructed by a measurement of the multiplicities, 
an event of reduced multiplicity will simply 
be recorded as an event of lower energy. It is then 
obvious that an additional and independent reconstruction of 
the energy of the primary cosmic ray is needed in order to correctly identify the energy of 
these events. 

In our study we will compute all the observables of the induced 
air shower using both the lab frame (LF) 
and the black hole frame (BHF) to describe the impact and the formation of the 
intermediate black hole resonance. 
Also, in the simulations that we will perform, the observation level at which  
we measure the properties of the air showers 
will be selected to take properly into account the actual position of a 
hypothetical experimental detector. The target of the first impact of mass $M$ 
is assumed to be a nucleon (or a quark) at rest in the atmosphere and the 
center of mass energy, corrected by emission loss in the bulk, is made 
promptly available for an instantaneous black hole formation and decay.
We will also assume that the energy $E_1$ of the incoming primary 
varies over all the highest part of the cosmic ray spectrum, from $10^{15}$ eV up to
$10^{20}$ eV. 

We denote by 
$\beta$ the speed of the black hole in the lab frame. In our notations, $E^*$
is the typical energy of 
each elementary state in the decay products (parton, lepton) 
in the BH frame and $P^*$ is its corresponding momentum. 

We will assume that a black hole decays ``democratically'' into all the 
possible partonic states, proportionally to the number of Standard Model 
states which are available to it at a given energy. 

The energy per partonic channel will be appropriately weighted   
and we will assume that each parton $(f)$ will decay into a final state hadron 
$h$ (carrying a fraction $x$ of the original momentum), 
with a probability distribution given by the corresponding fragmentation 
function $D_f^h(x,Q^2)$, which is evolved from some low energy 
input scale $Q_0$ up to the relevant scale characterizing the decay. This is given by 
the available energy per fundamental state, equally distributed among all the 
states.

The quantification of the injection spectrum 
involves a computation of the relevant probabilities 
for the formation of all the possible hadronic/leptonic states prior to the simulation 
of the air shower. Let's briefly elaborate on this.

To move from the parton level to hadron level, 
we let $D_q^h(x,Q^2)$, $D_{\overline{q}}^h(x,Q^2)$, 
and $D_g^h(x,Q^2)$ be the fragmentation functions 
of $N_F$ quarks $q$, antiquarks $\overline{q}$, and of the gluon $g$, 
respectively, into
some hadron $h$ with momentum fraction $x$ at the scale $Q$. 
From the fragmentation 
functions we obtain, for each hadron $h$, the mean multiplicity of the corresponding s-wave  and the corresponding average energy and momentum. Specifically 
we obtain
\beqa
  <D_h> &=&\sum_f\int_{z_{min}}^1 dz\, D_f^h(z,Q^2)
\eeqa
for the probability of producing a hadron $h$, and 
\beqa
E^*_h &=& \sum_f \int_{z_{min}}^1 z\, dz\, D_f^h(z,Q^2)
\eeqa
for the average energy of the same hadron. We recall that $z_{min}$
is the minimal fraction of energy a hadron $(h)$, of mass $m_h$, can take at a scale $Q$, 
and can be defined as $z_{min}=m_h/(Q/2)$. In practical applications 
one can take the nominal value $z_{min}=0.05$ for every hadron, without affecting
much the mean multiplicities and the related probabilities.  
This implies that 

\beq
<D_r^h> + \sum_f <D_f^h> + <D_g^h> + <D_\gamma^h>\equiv \textrm{Pr}_h \nonumber \\
\eeq
together with the condition $\sum_h \textrm{Pr}_h =1$, where the sum runs over all the 
types of hadrons allowed by the fragmentation. In all the equations above, 
the fragmentation takes place at the typical scale $Q=E/N$, scale at 
which the moments are computed numerically. 
For the identification of the probabilities it is convenient 
to organize the 123 fundamental states of the Standard Model 
into a set of flavour states 
$(q_f)$, with $f$ running over all the flavours except for the top quark, 
where in $(q_f)$ we lump antiquark states and color states, 
plus some additional states. The weight of the $(q_f)$ set is 
$p_f=2\times 2\times 3/123$, where the factors 2 and 3 refer to spin, quark-antiquark 
degeneracy and color. It is worth to
recall that quark and antiquark states of the same flavour 
have equal fragmentation functions in all the hadrons, and this 
justifies the $q/\bar{q}$ degeneracy of the set.
The additional states are the gluon (g) with a weight $p_g=2\times 8/123$, the photon 
$(\gamma)$, with a weight $p_\gamma=2/123$ and the remaining states $(r)$ 
in which we lump all the states which have been unaccounted for, 
whose probabilities 
$p_r$ are computed by difference. These include the top and the antitop 
$(12/123)$, the W's and Z $(9/123)$ and the leptons $(24/123)$.
The fragmentation functions into hadrons, corresponding to these states, 
$<D_r^h>$ are computed by difference from the remaining ones
$<D_g^h>$, $<D_f^h>$ and $<D_\gamma^h>$, which are known at any scale 
$Q$ from the literature. 
Beside the favour index $f=u,d,c,s,b$, introduced above, we introduce a second index 
$i$ running over the $(r)$ states, the photon and the gluons $(i=g,\gamma,r)$. 

The probability of generating a specific sequence of $N$ states in the course 
of the evaporation of the black hole is then given by a multinomial distribution of the form
\beq
f(n_f,n_i,p_f,p_i)= {N!\over \prod_f n_f! \prod_i n_i!}\prod_f 
p_f^{n_f}\prod_i {p_i}^{n_i}
\eeq
which describes a typical multi-poissonian process with $N$ trials. 
Notice that, to ensure proper normalization, we need to require that 
\beqa
\prod_i n_i! &=& n_g! n_\gamma! n_r! \nonumber \\
&=& n_g! n_\gamma! (N - \sum_f n_f - n_g - n_\gamma)! 
\eeqa

The computation of the cumulative probabilities to produce any number of hadrons of type 
$h$ by the decay of the black hole are obtained from the multinomial 
distribution multiplied by the fragmentation probabilities of each elementary 
state into $(h)$ and summing over all the possible sequences
\beq
\textrm{Pr}_{\textrm{cum h}}\equiv \sum_{n_f, n_i} {N!\over \prod_f n_f! \prod_i n_i!}\prod_f 
\left(p_f <D_f^h>\right)^{n_f}\prod_i \left(p_i <D_i^h>\right)^{n_i}.
\label{probab}
\eeq
A possible way to compute 
$\textrm{Pr}_{\textrm{cum h}}$ when
$N$ is large is to multiply the multinomial distribution by a suppression 
factor  $\textrm{Exp}[{-\Lambda( \sum_i n_i + \sum_f n_f -N)}]$, with $\Lambda$
a very large number, and interpret this factor as a Boltzmann weight, as in 
standard Monte Carlo computations of the partition function 
for a statistical system. Simulations can be easily done by a 
Metropolis algorithm and the configurations of integers selected are those for 
which the normalization condition $N= \sum_i n_i + \sum_f n_f$ is satisfied. 
In our case, since we are interested only in the mean number of hadrons produced 
in the decay and in their thermal spectrum, 
the computation simplifies if we average over all the relevant configurations.

%\clearpage
\section{Fragmentation and the Photon Component}

\begin{figure}
{\centering \resizebox*{9cm}{!}{\rotatebox{-90}{\includegraphics{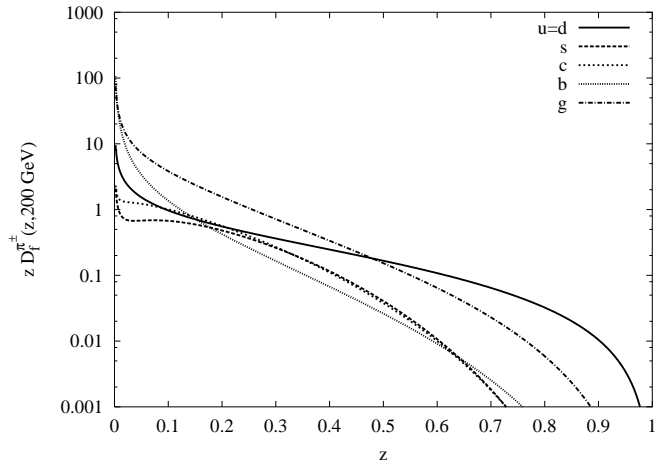}}} \par}

\caption{Fragmentation functions into \protect\( \pi ^{\pm }\protect \) at
\protect\( 200\, \textrm{GeV}\protect \).}
\label{frag1}
\end{figure}

\begin{figure}
{\centering \resizebox*{9cm}{!}{\rotatebox{-90}{\includegraphics{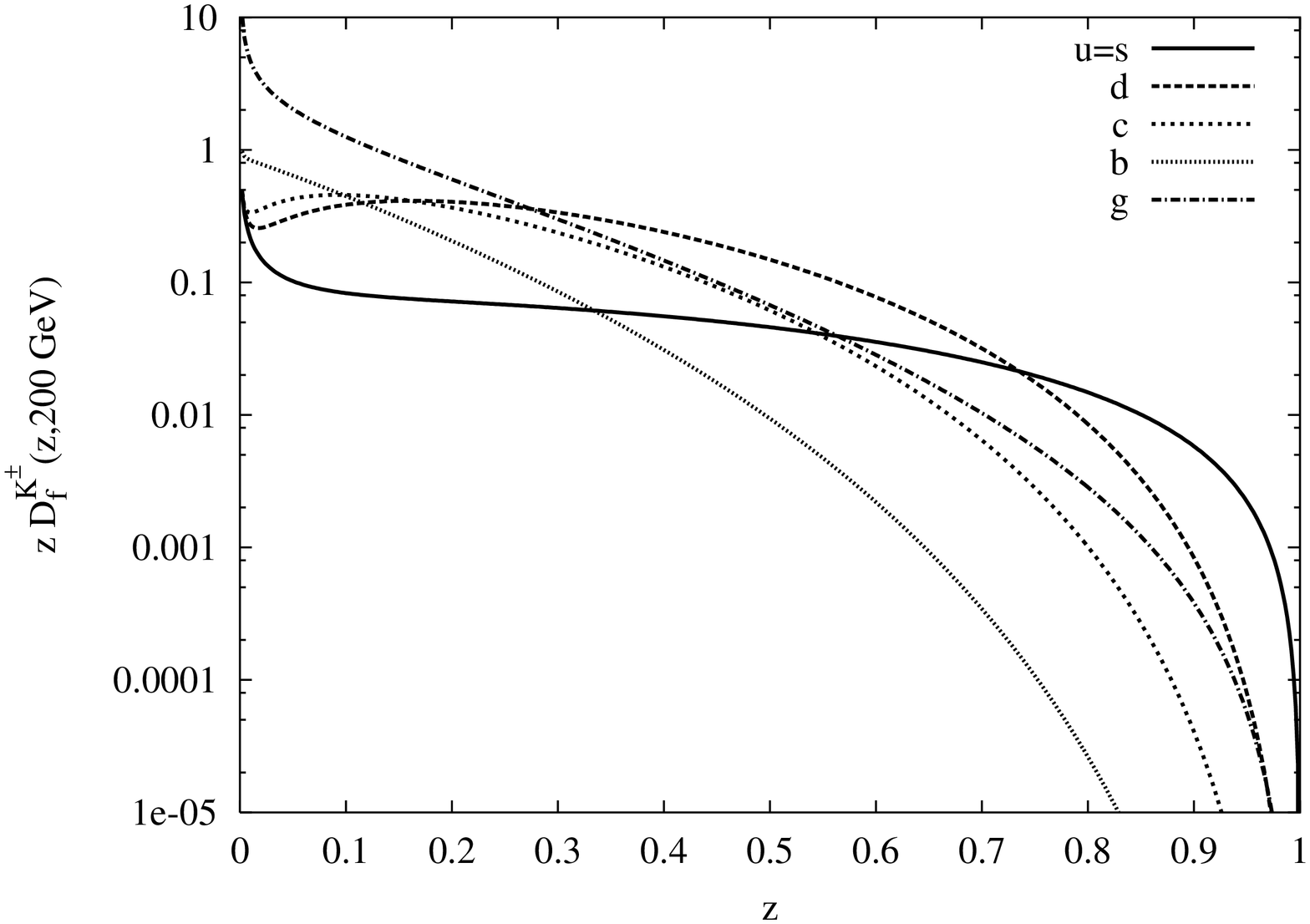}}} \par}

\caption{Fragmentation functions into \protect\( K^{\pm }\protect \) 
at \protect\( 200\, \textrm{GeV}\protect \).}
\label{frag2}
\end{figure}

\begin{figure}
{\centering \resizebox*{9cm}{!}{\rotatebox{-90}{\includegraphics{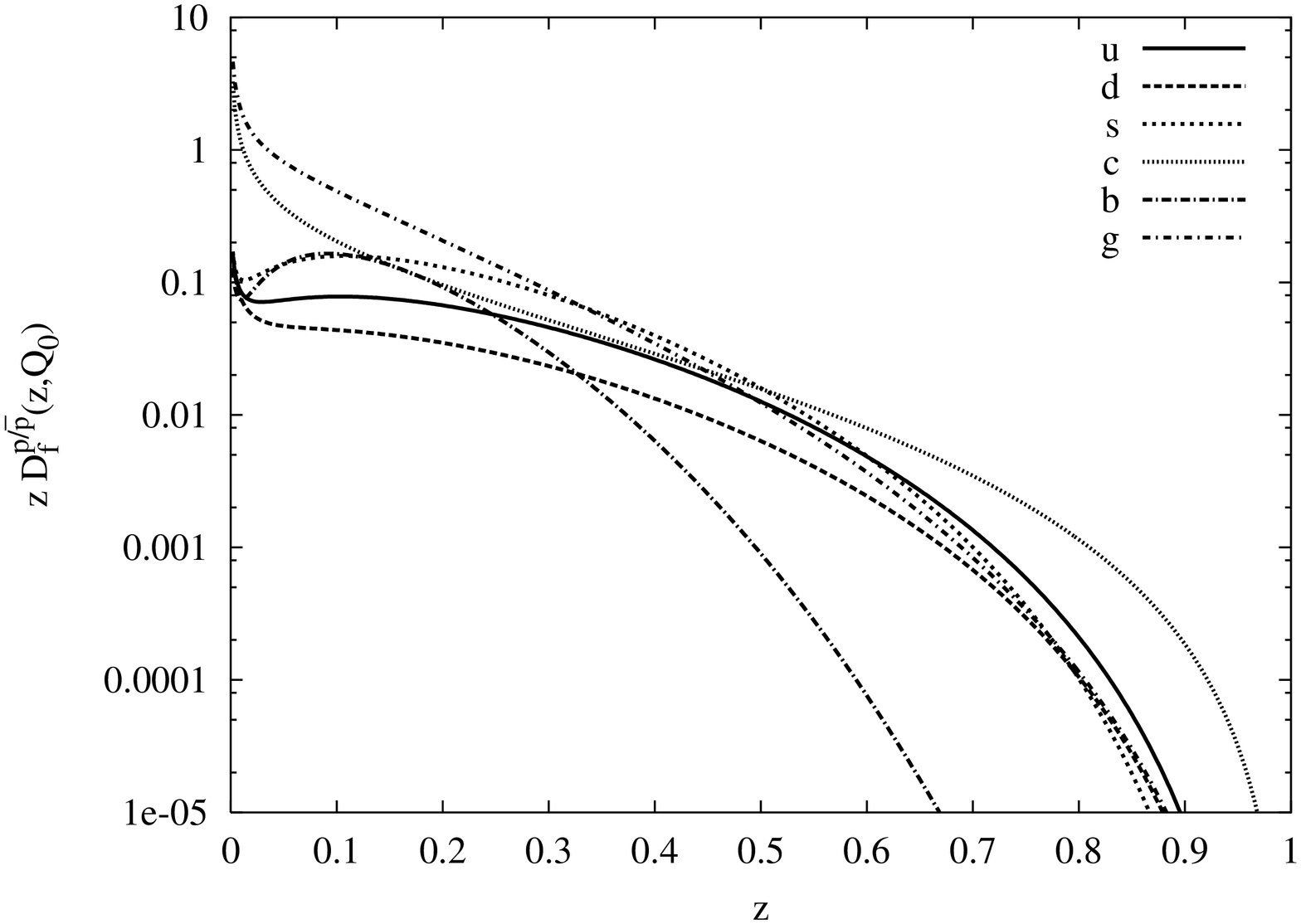}}} \par}

\caption{Fragmentation functions into \protect\( p/\overline{p}\protect \)
at \protect\( 200\, \textrm{GeV}\protect \).}
\label{frag3}
\end{figure}

The evolution with $Q^2$ of the fragmentation functions is 
conveniently formulated in terms of the linear combinations
\begin{eqnarray}
D_{\Sigma}^h(x,Q^2)&\n=\n&\sum_{i=1}^{N_F}\left( 
      D_{q_i}^h(x,Q^2)+D_{\overline{q}_i}^h(x,Q^2) \right)
\COMMA
\\
\nonumber\\
D_{(+),i}^h(x,Q^2)&\n=\n&D_{q_i}^h(x,Q^2) 
     + D_{{\overline{q}},i}^h(x,Q^2)
   -\frac{1}{N_F} D_{\Sigma}^h(x, Q^2)
\COMMA
\\
\nonumber\\
D_{(-),i}^h(x,Q^2)&\n=\n&D_{q_i}^h(x,Q^2)
     - D_{{\overline{q}},i}^h(x,Q^2)
\COMMA
\end{eqnarray}
as for these the gluon decouples from the 
non--singlet $\scriptstyle (+)$ and the
asymmetric $\scriptstyle (-)$
combinations, leaving only the singlet and the gluon 
fragmentation functions coupled;
\begin{eqnarray}
Q^2\frac{d}{d Q^2}
D_{(+),i}^h(x,Q^2)
&\n=\n&\left[P_{(+)}\left(\alpha_s(Q^2)\right)\otimes
D_{(+),i}^h(Q^2)\right](x) 
\COMMA \label{APplus}
\\
\nonumber\\
Q^2\frac{d}{d Q^2}D_{(-),i}^h(x,Q^2)
&\n=\n&\left[P_{(-)}\left(\alpha_s(Q^2)\right)\otimes
D_{(-),i}^h(Q^2)\right](x)
\COMMA \label{APminus}
\\
\nonumber\\
Q^2\frac{d}{d Q^2}D_{\Sigma}^h(x,Q^2)
&\n=\n&\left[
P_{\Sigma}\left(\alpha_s(Q^2)\right)\otimes D_{\Sigma}^h(Q^2)
\right](x)
\nonumber\\
& & \hspace{1cm}
+2N_F\left[P_{q\to G}\left(\alpha_s(Q^2)\right)\otimes D_G^h(Q^2)
\right](x)
\COMMA 
\\
\nonumber\\
Q^2\frac{d}{d Q^2}D_G^h(x,Q^2)
&\n=\n&\frac{1}{2N_F}\left[
P_{G\to q}\left(\alpha_s(Q^2)\right)\otimes D_{\Sigma}^h(Q^2)
\right](x)
\nonumber\\
& & \hspace{1cm}
+\left[
P_{g\to g}\left(\alpha_s(Q^2)\right)D_g^h(Q^2)\right](x)
\STOP 
\end{eqnarray}
The kernels 
that appear in the equations above are defined by
\begin{eqnarray}
P_{(+)}\left(x,\alpha_s(Q^2)\right)&\n=\n&
P_{q\to q}^{V}\left(x,\alpha_s(Q^2)\right) +
P_{q\to \overline{q}}^{V}\left(x,\alpha_s(Q^2)\right)
\COMMA
\\
\nonumber\\
P_{\Sigma}\left(x,\alpha_s(Q^2)\right)&\n=\n&
P_{(+)}\left(x,\alpha_s(Q^2)\right)+
2N_FP_{q\to q}^{S}\left(x,\alpha_s(Q^2)\right)
\COMMA
\\
\nonumber\\
P_{(-)}\left(x,\alpha_s(Q^2)\right)&\n=\n&
P_{q\to q}^{V}\left(x,\alpha_s(Q^2)\right) -
P_{q\to \overline{q}}^{V}\left(x,\alpha_s(Q^2)\right), 
\end{eqnarray}
with $\alpha_s(Q^2)$ being the QCD coupling constant.
In the perturbative expansion of the splitting functions,
\begin{equation}
P(x,\alpha_s(Q^2))=\frac{\alpha_s(Q^2)}{2\pi}P^{(0)}(x)
+\left(\frac{\alpha_s(Q^2)}{2\pi}\right)^2 P^{(1)}(x)
+{\cal O}\left(\left(\frac{\alpha_s(Q^2)}{2\pi}\right)^3\right).
\end{equation}
The timelike kernels that we use are given by
\begin{eqnarray}
P_{q\to q}^{V,(0)}(x) &\n=\n& C_F\left[\frac{3}{2}\delta(1-x)
   +2\left(\frac{1}{1-x}\right)_+ -1 -x\right] \COMMA
\label{P0Vqq}
\\
P_{q\to \overline{q}}^{V,(0)}(x) &\n=\n& P_{q\to q}^{S,(0)}(x) = 0
\COMMA
\label{P0Vqqbar}
\\
P_{q\to G}^{(0)}(x) &\n=\n& C_F\left[\frac{1+(1-x)^2}{x}\right]
\COMMA
\label{P0qG}
\\
P_{q\to q}^{(0)}(x) &\n=\n& 2N_F T_R\left[x^2+(1-x)^2\right]
\COMMA
\label{P0Gq}
\\
P_{G\to G}^{(0)}(x) &\n=\n& \left(\frac{11}{6}N_C-\frac{2}{3}N_F T_R
  \right)\delta(1-x) 
\nonumber\\
& & \hspace{1cm}
+2N_C\left[\left(\frac{1}{1-x}\right)_+ 
  +\frac{1}{x}-2+x-x^2\right]
\STOP
\label{P0GG}
\end{eqnarray}
\smallskip
The formal solution of the equations is given by
\begin{equation}
D_a^h(x,Q^2)=D_a^h(x,Q_0^2)+\int_0^{\log(Q^2/Q_0^2)} d \log(Q^2/Q_0^2)
\frac{\alpha_s(Q^2)}{2\pi}
   \sum_b \left[P_{a\to b}(\alpha_s(Q^2)) 
    \otimes D_b^h(Q^2)\right]
\label{ap}
\end{equation}
 where $Q_0$ is the starting scale of the initial conditions, given by 
$D(x,Q_0^2)$.
At leadiong order in $\alpha_s$, we solve this equation using a special ansatz
\beq
D_f^{h}(x,Q^2)=\sum_n {A_n(x)\over n!} \log\left( {\alpha_s(Q^2)\over 
\alpha_s (Q_0^2)}\right)^n
\eeq
and generating recurrence relations at the $n+1-th$ order for the 
$A_{n+1}$ coefficients in terms of the $A_n$ \cite{CC}. It is easy to see 
that this corresponds to the numerical implementation 
of the formal solution 
\beq
D_f^{h}(x,Q^2)={\it{Exp}}\left(t P\otimes\right) D_f^{h}(x,Q_0^2)
\eeq
with $t=\left(\alpha_s(Q^2)/ \alpha_s (Q_0^2)\right)$, where the 
exponential is a formal expression for an infinite iteration of convolution products. 
We show in Figures~\ref{frag1}-\ref{frag3} results for some of the 
fragmentation functions into pions, kaons and protons 
computed for a typical parton scale of 200 GeV.  

The photon contributions to the decay of the black hole is treated separately. 
The evolution equation for the fragmentation functions of photons and parton 
fragmentation into photon 
$D_\gamma^\gamma(x,Q^2), \,\, D_q^\gamma(x,Q^2)$ satisy at leading order in 
$\alpha_{\rm{em}}$ (the QED fine structure constant) and $\alpha_s$ 
(the QCD coupling), the evolution 
equations \cite{Klasen}
\beq
\frac{\textrm{d} D_{\gamma}^\gamma(x,Q^2)}{\textrm{d}\ln Q^2} =
 \frac{\alpha}{2\pi} P_{\gamma\rightarrow \gamma}(x)\otimes D_{\gamma}^{\gamma}(x,Q^2)
\eeq
which can be integrated with the initial conditions $D_{\gamma}^\gamma(x,Q^2)=\delta (1-x)$,  
and 
\beq
 \frac{\textrm{d} D_q^{\gamma}(x,Q^2)}{\textrm{d}\ln Q^2} =
 \frac{\alpha}{2\pi} P_{\gamma\rightarrow \gamma}(x)\otimes D_{\gamma}^\gamma(x, Q^2)
\eeq
which can also be integrated with the result \cite{gehrmann}
\beq
D_{\gamma/q}(x,Q^2)=\frac{\alpha}{2\pi}P_{q\rightarrow \gamma}(x)\ln\frac{Q^2} 
{Q_0^2}+D_{q}^\gamma (x,Q_0^2).
\label{frag}
\eeq
In Eq.~(\ref{frag}) the second term is termed the hadronic boundary conditions, 
which come from an experimental fit, 
while the first term is the pointlike contributions, which can be obtained perturbatively.  
In \cite{ALEPH} a leading order fit was given 
\beq
D_q^\gamma(x,Q_0^2)=\frac{\alpha}{2\pi}\left[-\, P_{q\rightarrow \gamma }(x)
\ln(1-x)^2-13.26\right]
\eeq
at the starting scale $Q_0=0.14$ GeV. 
The kernels in these cases are given by simple modifications of the ordinary 
QCD kernels, for instance  
$P_{q\rightarrow \gamma}(x)= e_q^2/C_F P_{q\rightarrow g}(x)$ \cite{Klasen}.

For the fragmentation function of quarks
to photons with virtuality $M_\gamma$, the perturbative result is given in \cite{qiu}
\beq
D_{\gamma/q}(x,Q^2,M_{\gamma}^2) = e_q^2\frac{\alpha}{2\pi}\left[\frac{1+(1-x)^2}{x}
\ln\frac{xQ^2}{M_{\gamma}^2}- x\left( 1-\frac{P^2}{xQ^2}\right)\right],
\eeq
where $P^2$ is the virtuality of the photon. 
The gluon to photon transitions are neglected, since $P_{g\rightarrow \gamma}$ vanishes 
in leading order. 

We recall that each elementary
state emitted is characterized by an average energy 
given by \( \left\langle \varepsilon \right\rangle =M_{BH}/\left\langle N\right\rangle  \).

The leptonic component \( e^{\pm } \), \( \mu ^{\pm } \), produced by the decay
is left unaltered and provides an input for the air shower simulator as soon as 
these particles 
cross the horizon. The $\tau^\pm$s are left to decay into their main channels,
while the hadronization of the \( u,\, d,\, s,\, c \) quarks and the gluons 
is treated with our code, that evolves the
fragmentation functions to the energy scale \( \left\langle \varepsilon \right\rangle  \).
The top (\( t \)) quark is treated consistently 
with all its fundamental decays included; hadronization of the $b$ quark is treated 
with suitable fragmentation functions and also involves a suitable evolution. 
As we vary $M_{BH}$ and we scan over the spectrum of the incoming cosmic rays 
the procedure is repeated and rendered automatic by combining in a single algorithm 
all the intermediate steps. 
Tables 1 and 2 contain the results of a renormalization group analysis of the 
fragmentation functions for all the 
partons (except the top quark), where we show 
both the intial conditions at the input scale, whose lowest value is $Q=1.414$ GeV, 
and the results of the evolution, at a final scale of $Q=200$ GeV,  
the initial set being taken from ref.~\cite{kkp}.

\smallskip

\begin{table}
\begin{tabular}{|c||c|c|c|c|c|c|}
\hline
&
\( \pi ^{\pm } \)&
\( \pi ^{0} \)&
\( K^{\pm } \)&
\( K^{0}/\overline{K}^{0} \)&
\( p/\overline{p} \)&
\( n/\overline{n} \)\\
\hline
\hline
\emph{u}&
\( 0.451 \)&
\( 0.226 \)&
\( 0.048 \)&
\( 0.174 \)&
\( 0.067 \)&
\( 0.034 \)\\
&
\( 0.463 \)&
\( 0.231 \)&
\( 0.084 \)&
\( 0.252 \)&
\( 0.070 \)&
\( 0.035 \)\\
\hline
\emph{d}&
\( 0.451 \)&
\( 0.226 \)&
\( 0.174 \)&
\( 0.048 \)&
\( 0.034 \)&
\( 0.067 \)\\
&
\( 0.463 \)&
\( 0.231 \)&
\( 0.252 \)&
\( 0.084 \)&
\( 0.035 \)&
\( 0.070 \)\\
\hline
\emph{s}&
\( 0.391 \)&
\( 0.195 \)&
\( 0.068 \)&
\( 0.068 \)&
\( 0.139 \)&
\( 0.139 \)\\
&
\( 0.295 \)&
\( 0.147 \)&
\( 0.084 \)&
\( 0.084 \)&
\( 0.108 \)&
\( 0.108 \)\\
\hline
\emph{c}&
\( 0.329 \)&
\( 0.165 \)&
\( 0.167 \)&
\( 0.167 \)&
\( 0.085 \)&
\( 0.085 \)\\
&
\( 0.309 \)&
\( 0.155 \)&
\( 0.194 \)&
\( 0.194 \)&
\( 0.071 \)&
\( 0.071 \)\\
\hline
\emph{b}&
\( 0.438 \)&
\( 0.219 \)&
\( 0.129 \)&
\( 0.129 \)&
\( 0.042 \)&
\( 0.042 \)\\
&
\( 0.324 \)&
\( 0.162 \)&
\( 0.115 \)&
\( 0.115 \)&
\( 0.041 \)&
\( 0.041 \)\\
\hline
\emph{g}&
\( 0.303 \)&
\( 0.152 \)&
\( 0.253 \)&
\( 0.253 \)&
\( 0.020 \)&
\( 0.020 \)\\
&
\( 0.807 \)&
\( 0.404 \)&
\( 0.317 \)&
\( 0.317 \)&
\( 0.034 \)&
\( 0.034 \)\\
\hline
%\label{tabfrag1}
\end{tabular}

\caption{Initial conditions. For each couple of parton and hadron, 
the upper number in the box is the probability for 
the parton \protect\( f\protect \) to hadronize
into the hadron \protect\( h\protect \), \protect\( \left( \int _{z_{min}}^{1}D_{f}^{h}(z,Q)dz\right) /\sum _{h'}\left( \int _{z_{min}}^{1}D_{f}^{h'}(z,Q)dz\right) \protect \),
while the lower number is the average energy fraction of \protect\( h\protect \),
\protect\( \int _{z_{min}}^{1}zD_{f}^{h}(z,Q)dz\protect \). In this table 
the energy of the parton is \protect\( Q=1.414\, \textrm{GeV}\protect \) for
\protect\( u\protect \), \protect\( d\protect \), \protect\( s\protect \)
and \protect\( g\protect \), \protect\( Q=2m_{c}=2.9968\, \textrm{GeV}\protect \)
for \protect\( c\protect \) and \protect\( Q=2m_{b}=9.46036\, \textrm{GeV}\protect \)
for the \protect\( b\protect \) quark, generated via the set of ref.~\cite{kkp}}
\end{table}

\begin{table}
\begin{tabular}{|c||c|c|c|c|c|c|}
\hline
\( Q=200\, \textrm{GeV} \) &
\( \pi ^{\pm } \)&
\( \pi ^{0} \)&
\( K^{\pm } \)&
\( K^{0}/\overline{K}^{0} \)&
\( p/\overline{p} \)&
\( n/\overline{n} \)\\
\hline
\hline
\emph{u}&
\( 0.446 \)&
\( 0.223 \)&
\( 0.079 \)&
\( 0.166 \)&
\( 0.053 \)&
\( 0.033 \)\\
&
\( 0.385 \)&
\( 0.193 \)&
\( 0.077 \)&
\( 0.178 \)&
\( 0.047 \)&
\( 0.027 \)\\
\hline
\emph{d}&
\( 0.446 \)&
\( 0.223 \)&
\( 0.166 \)&
\( 0.079 \)&
\( 0.033 \)&
\( 0.053 \)\\
&
\( 0.385 \)&
\( 0.193 \)&
\( 0.178 \)&
\( 0.077 \)&
\( 0.027 \)&
\( 0.047 \)\\
\hline
\emph{s}&
\( 0.425 \)&
\( 0.213 \)&
\( 0.093 \)&
\( 0.093 \)&
\( 0.088 \)&
\( 0.088 \)\\
&
\( 0.295 \)&
\( 0.147 \)&
\( 0.077 \)&
\( 0.077 \)&
\( 0.070 \)&
\( 0.070 \)\\
\hline
\emph{c}&
\( 0.371 \)&
\( 0.185 \)&
\( 0.158 \)&
\( 0.158 \)&
\( 0.064 \)&
\( 0.064 \)\\
&
\( 0.295 \)&
\( 0.147 \)&
\( 0.150 \)&
\( 0.150 \)&
\( 0.051 \)&
\( 0.051 \)\\
\hline
\emph{b}&
\( 0.431 \)&
\( 0.216 \)&
\( 0.132 \)&
\( 0.132 \)&
\( 0.045 \)&
\( 0.045 \)\\
&
\( 0.292 \)&
\( 0.146 \)&
\( 0.101 \)&
\( 0.101 \)&
\( 0.036 \)&
\( 0.036 \)\\
\hline
\emph{g}&
\( 0.428 \)&
\( 0.214 \)&
\( 0.135 \)&
\( 0.135 \)&
\( 0.044 \)&
\( 0.044 \)\\
&
\( 0.577 \)&
\( 0.289 \)&
\( 0.175 \)&
\( 0.175 \)&
\( 0.057 \)&
\( 0.057 \)\\
\hline
\end{tabular}
\caption{For each couple of parton/hadron, the first number is the probability of fragmentation of the parton $f$ into the hadron
 $\left( \int _{z_{min}}^{1}D_{f}^{h}(z,Q)dz\right) /\sum _{h'}\left( \int _{z_{min}}^{1}D_{f}^{h'}(z,Q)dz\right)$,
while the second is the average energy fraction of $h$,
$ \int _{z_{min}}^{1}zD_{f}^{h}(z,Q)dz $. The
energy of the parton is $Q=200$ GeV. }
\label{incon}
\end{table}

This concludes the computation of the probabilities for each hadron/lepton 
present in the decay products of the mini black hole. It is 
reasonable to assume that these particles will be produced
spherically, since higher angular momenta are suppressed by the corresponding
centrifugal barrier. However, the analysis of the shower profile has to be 
performed in the lab frame. This requires the transformation of the 
initial configurations above to the laboratory frame, which is exactly what is 
discussed next.

%\clearpage
\section{Sphericity and Boost}

\begin{figure}
{\centering \resizebox*{12cm}{!}{\includegraphics{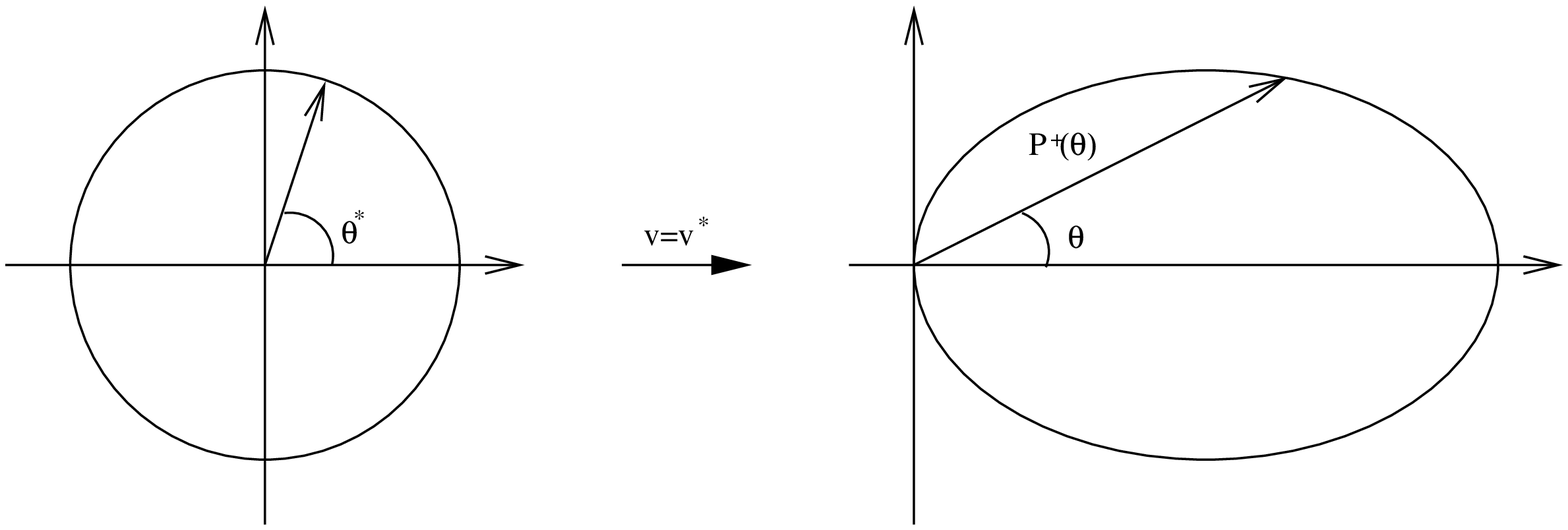}} 
\resizebox*{12cm}{!}{\includegraphics{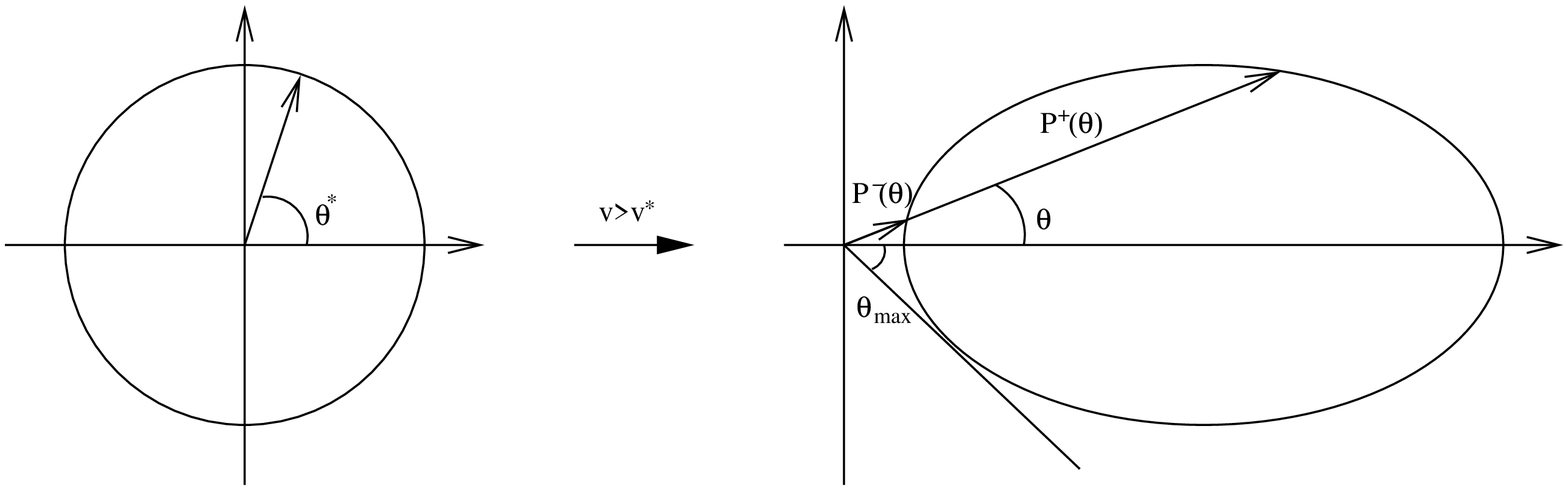}} \par}
\caption{An s-wave distribution in the BHF is transformed in the detector
frame to an elliptical one, whose detailed shape depends on the value 
of $g^*$}
\label{shapes}
\end{figure}

The transformation from the black hole frame (BHF) to the laboratory frame (LF)
is performed by a Lorentz boost with speed $\beta$, the speed of the black
hole in the LF. 
Assuming that the black hole is produced in the collision of a primary of 
energy $E_1$ in the LF and negligible mass compared to $E_1$, 
with a parton of mass $M$ in the atmosphere, one 
obtains $\beta=E_1/(E_1+M)$. 
A spherical distribution of a particular particle of mass $m$ 
among the decay products 
in the BHF is transformed to an elliptical one,
whose detailed form is conveniently parametrized by 
%\begin{figure}[t]
%{\centering
%\resizebox*{10cm}{!}{\rotatebox{-90}{\includegraphics{gstar.ps}}} \par}
%\caption{Plot of the boost parameter $g^*$ as a function of the number of
%elementary states in which the black hole decays.}
%\label{boostgstar}
%\end{figure}
\beq
g^*={\beta \over \beta^*}=
\frac{1-\frac{M}{E_1+M}}{\left(1 - {{m^2}\over {E^{*2}}}\right)^{1/2}}
\eeq
where  
$\beta^*=P^*/E^*$ is the speed of this particle in the BHF, 
the ratio of its BHF momentum and energy. Figure \ref{shapes}
depicts the relevant kinematics.
A particle emitted in the direction $\theta^*$ in the BHF, is seen in the 
direction $\theta$ in the LF, with
\beq
\tan\theta=\sqrt{1-\beta^2}\frac{\sin\theta^*}{g^*+\cos\theta^*}.
\label{theta}
\eeq
For values of $g^*\geq 1$ the shape of the 
1-particle distribution in the LF is characterized by a maximum 
angle of emission
\beq
|\tan\theta_{max}|=\sqrt{\frac{1-\beta^2}{g^{*2}-1}}\,, 
\label{thetamax}
\eeq
which for $g^*=1$ is equal to $90^o$. 
Only for $g^*<1$ there is backward emission in the LF. 
As a relevant example, let us consider the case of a hadron of mass
$m=1$ GeV and energy $E^*=100$ GeV emitted by
a black hole, formed by an initial primary 
of energy $E_1=1000$ TeV, which hit a quark of mass $M\sim 10$MeV to form 
a black hole. The gives $g^*=1.00005$ and the corresponding maximum angle in
the LF is
$\theta_{max}\simeq \tan\theta_{max}=1.4\times 10^{-2}$, giving an angular
opening of the decay products of about 2 degrees. 

As shown in Figure \ref{shapes}, for $g^*>1$, which is relevant for our 
purposes, the mapping from $\theta^*$ to $\theta$ is not one to one. In
a given direction $\theta$ in the LF, one receives particles emitted in
two different directions $\theta^{*\pm}$ in the BHF. They satisfy \cite{KB}
\beq
\frac{ d\cos\theta^{*\pm}}{d\cos\theta}=\left(\frac{P^\pm}{P^*}\right)^2 
\frac{1}{\left(\pm \cos\theta\sqrt{K}\right)}
\label{wh}
\eeq
where
\beq
K=1 + \gamma^2(1 - {g^*}^2)\tan^2 \theta,
\eeq
and with the momenta $P^{\pm}$ and energies $E^{\pm}$ 
of the two branches given by
\beq
P^{\pm}={P^* \cos\theta (g^* \pm \sqrt{K}) \over \gamma (1 - v^2 \cos^2\theta)}.
\eeq
and 
\beq
E^{\pm}=\frac{m\left(\gamma^* \pm v\cos\theta \left(v^{* 2}\gamma^{* 2} - 
v^2\gamma^2\sin^2\theta\right)^{1/2}\right)}{\gamma (1- v^2\cos^2\theta)}
\eeq
respectively. In the above formulas $\gamma^{-1}=\sqrt{1-\beta^2}$, $v(v^*)$
is the speed of the hadron in the LF (BHF), and $(\gamma^*)^{-1}=
m/E^*_h=\sqrt{1-v^{*2}}$.
For massless final state particles, in particular, these relations become 
\beq
P=E={P^*\over \gamma (1 - \cos\theta)}
\eeq
and reduce to the familiar Doppler formula when $\theta=0$.

The probability distribution $W_h^*(\cos\theta^*,\phi)$ of a hadron (h) 
as a function of the direction $\Omega=(\cos\theta^*,\phi)$ in the BHF, 
assumed spherically symmetric and normalized to the 
total probability $Pr_h$ of detecting this hadron among the decay products
with $N$ elementary states, is
\beq
W_h^*(\cos\theta^*)={ \textrm{Pr}_h \over 2}.
\eeq
The corresponding one in the LF is
\beq
W_h(\cos\theta)=\sum_{\pm}\frac{d\cos\theta^{*\pm}}{d\cos\theta}\,
W^*_h(\cos\theta^{*\pm})\,.
\eeq

In the special case $g^*=1$, the probability distribution simplifies to  
\beq
W_h(\cos\theta)=2 \textrm{Pr}_h{ \cos\theta \over 
\gamma^2(1- \beta^2\cos\theta^2)^2}\,,
\eeq
peaked in the forward direction, symmetric around the maximum value, 
obtained for $\theta=0$ and equal to $2 \gamma^2$, while the 
momentum distribution is
\beq
P(\theta)=m { \beta\gamma^*\cos\theta\over \gamma
(1- \beta^2\cos\theta^2)}\,.
\eeq   

\begin{figure}
%[t]
{\centering
\resizebox*{10cm}{!}{\rotatebox{-90}{\includegraphics{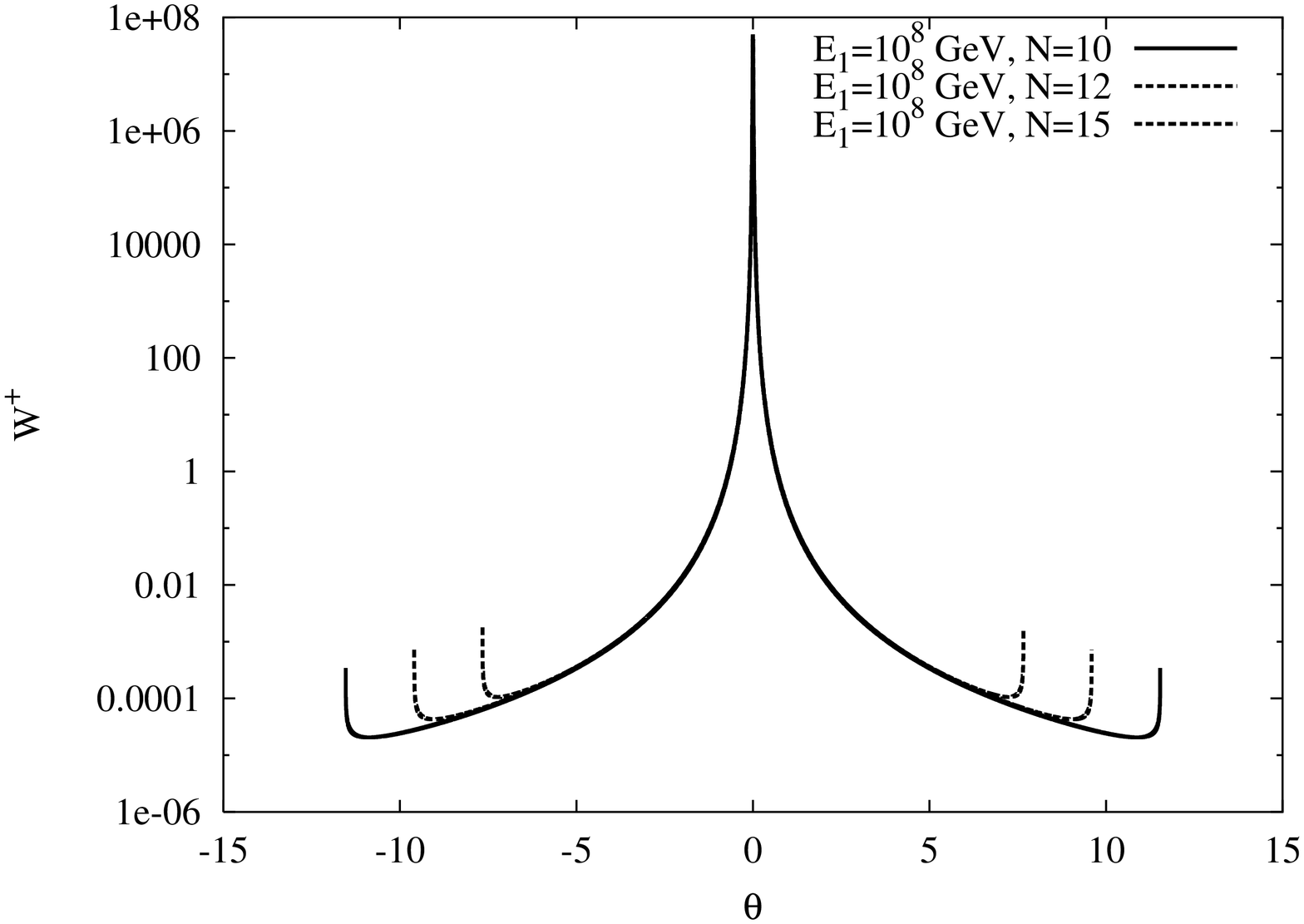}}} \par}
\caption{Plot of the $W^+$ branch of the probability distribution
for an incoming energy of the cosmic ray $E_1= 10^8$ GeV and for various 
values of $N$ of the elementary partonic states emitted during the decay
of the black hole}
\label{wplus}
\vspace{3cm}
{\centering
\resizebox*{10cm}{!}{\rotatebox{-90}{\includegraphics{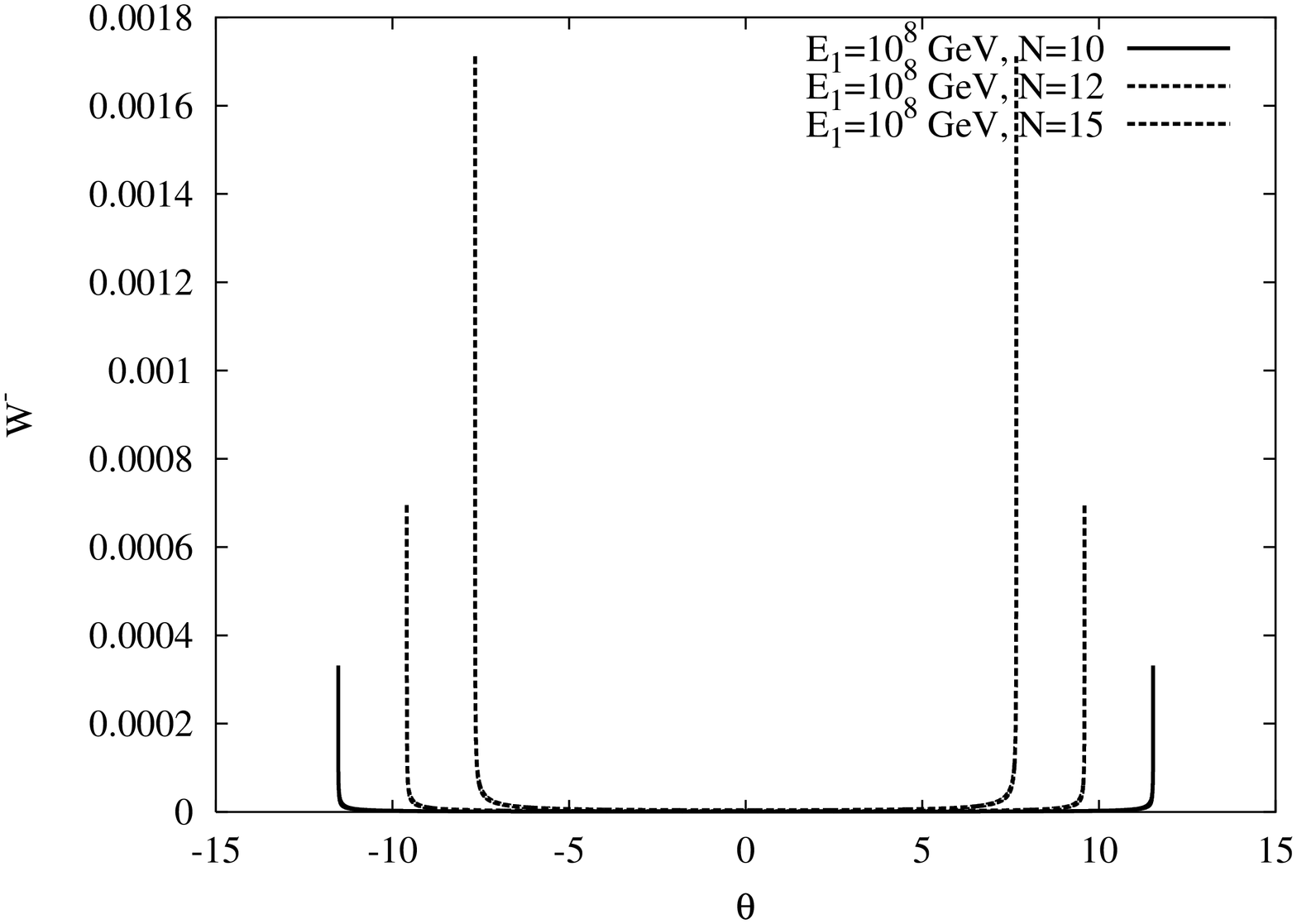}}} \par}
\caption{Same plot as above but for the $W^-$ branch of the distribution}
\label{wminus}
\end{figure}

\begin{figure}
%[tbh]
{\centering
\resizebox*{8cm}{!}{\rotatebox{-90}{\includegraphics{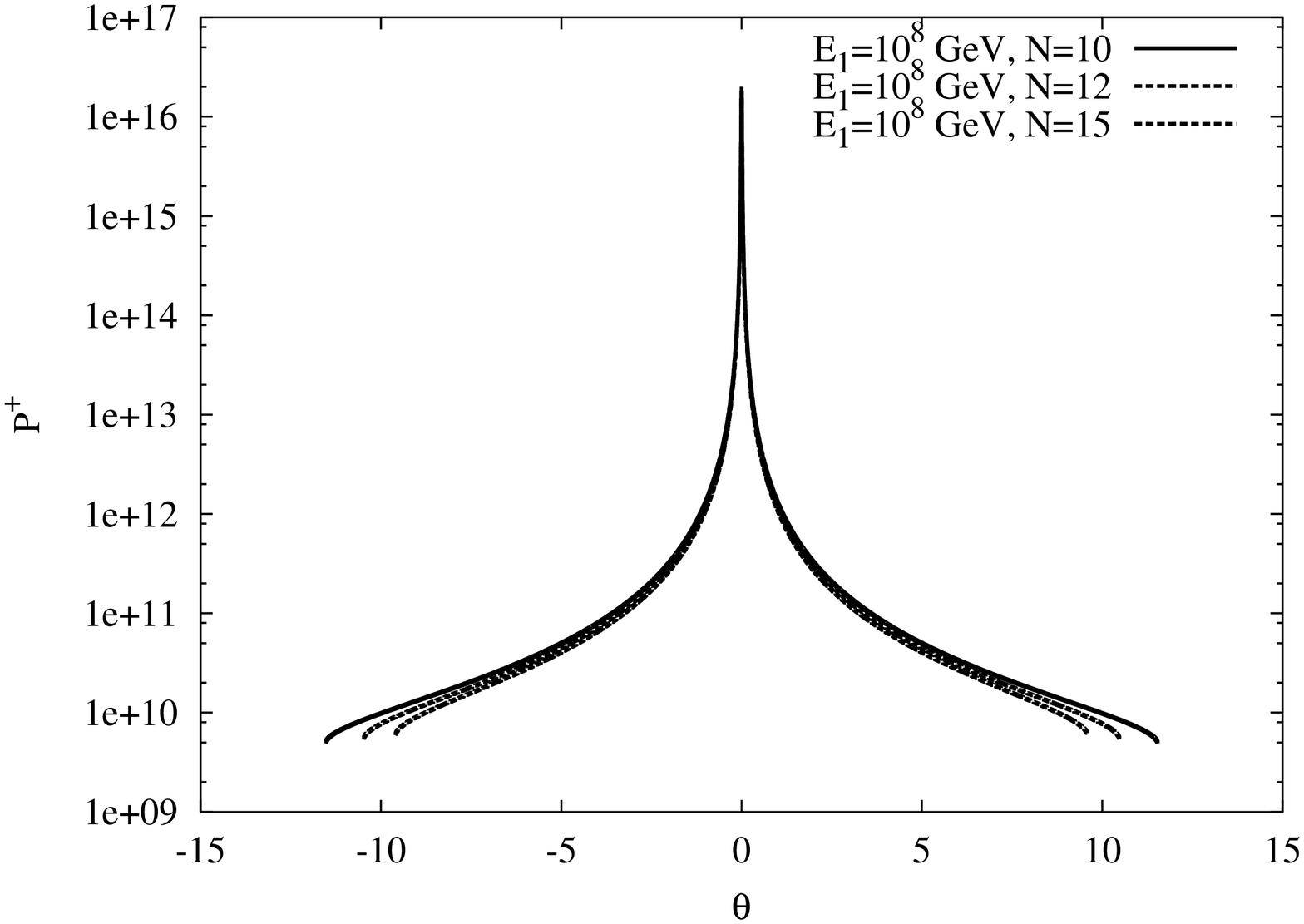}}} \par}
\caption{Plot of the $P^+$ branch of the momentum distribution in eV
for an incoming energy of the cosmic ray $E_1= 10^8$ GeV and for various 
values of $N$ of the elementary partonic states emitted during the decay
of the black hole}
\label{ppbranch}
\vspace{3cm}
{\centering
\resizebox*{8cm}{!}{\rotatebox{-90}{\includegraphics{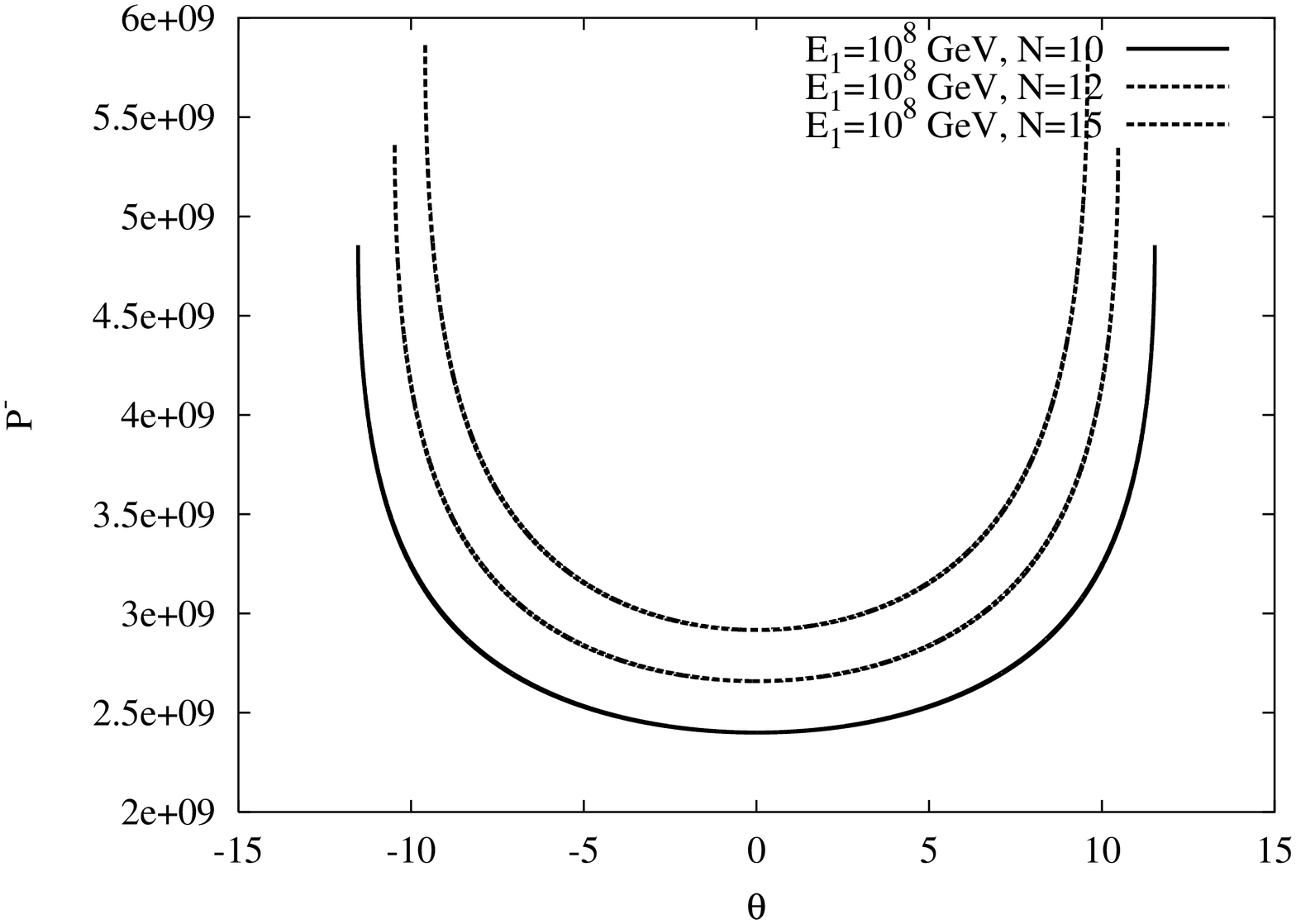}}} \par}
\caption{Same plot as above but for the $P^-$ branch of the momentum 
distribution}
\label{pmbranch}
\end{figure}

As we have already mentioned, 
the structure of the partonic event (and, similarly, of the hadronic 
event after fragmentation) is characterized by the formation of an 
elliptical distribution of partons, strongly boosted toward the detector 
along the vertical direction. Each uniform (s-wave) distribution 
is strongly elongated along the arrival direction (due to the large 
speed of the black hole along this direction) and is characterized by 
two sub-components ($W^\pm$), identified by a $\pm$ superscript. Their sum
is the total probability distribution given in (\ref{wh}).
The ``+'' momentum component is largely dominant and strongly peaked around 
the vertical direction with rather small opening angles and this 
behaviour can be analyzed numerically with its $N$ dependence. 
In the explicit identification of the two independent distributions 
$\pm$ in terms of the opening angle $\theta$, as measured in the LF, we 
use the relations 
\beq
W^\pm(\theta) = \frac{1}{2} W^\pm(\cos \theta^*) 
\mid {d\cos(\theta^*)\over d\cos \theta}^\pm\mid \sin\theta,
\eeq
where we have introduced a factor of 1/2 for a correct normalization of the 
new distribution in the $\theta$ variable. 
In Figures~\ref{wplus} and \ref{wminus} we show the structure of these 
distributions in the LF. Both are characterized by a very small 
opening angle ($\theta$) with respect to the azimuthal direction of the 
incoming cosmic ray, $W^+$ being the dominant one. Two similar 
plots (Figures~\ref{ppbranch} and \ref{pmbranch}) illustrate the two 
components $P^\pm$ as functions of the same angle.

One can easily check that we can 
now integrate symmetrically on both distributions to obtain 
the correct normalization (to $\textrm{Pr}_h$) for a given hadron (or parton)
\beq
\int_{-\theta_{max}}^{\theta_{max}}  W^\pm(\theta) d\theta = 
\int_{-{\theta_1}^*}^{{\theta_1}^*} W^+(\theta) d\theta +
\int_{{\theta_1}^*}^{\pi} W^-(\theta) d\theta +
\int_{-\pi}^{{-\theta_1}^*} W^-(\theta) d\theta 
={\textrm{Pr}_h}
\eeq
or, equivalently, using $\cos \theta$ as a distribution variable
\beq
W^\pm(cos \theta)=W^\pm(\cos \theta^*) \mid 
{d\cos \theta^*\over d\cos \theta}^\pm\mid
\sin\theta
\eeq
with
\beq
\int_{cos\theta_{max}}^{1}  W^\pm(\cos\theta) d \cos\theta =
\int_{\cos{\theta_1}^*}^{1} W^+(\cos\theta^*) d\ cos\theta^* +
\int_{-1}^{\pi} W^-(\cos\theta^*) d\cos\theta^*
={\textrm{Pr}_h}
\eeq
and ${\theta^*}_1$ obtained from (\ref{thetamax}).
%\beq
%{\sin\theta^*\over \cos\theta^* + g^*}= \tan\theta_{max}.
%\eeq

\clearpage
\section{Air Shower Simulations}

The simulation of the events is performed at the last stage, using 
an air shower simulator. We have used CORSIKA \cite{CORSIKA} with 
appropriate initial conditions on the spectrum of the incoming particles 
in order to generate the full event measured at detector level.
In most of the simulations we have assumed that the first impact takes place in the 
lower part of the atmosphere, not far from the level of the detector, 
at a varying altitude. The reason is that one of our interests is the investigation
of the possibility that the Centauro events may be related to evaporating
mini black holes, formed by the collision of weakly interacting particles (e.g.
neutrinos) which penetrate the atmosphere. Of course, we have simulated 
events happening at higher altitudes as well.  
We have performed two separate sets of simulations, the first set being 
benchmark events with an ``equivalent'' proton replacing the neutrino-nucleon 
event, colliding at the same height, the second being the signal event, 
i.e. the black hole resonance. The difference between the first and the 
second set is attributed to the different components of the final state 
prior to the development of the air shower.

We compute the average number of particles produced in the process
of BH evaporation using the formula
\beq
M_{BH}=E_{CM}f_{n}
\eeq
where \( E_{CM} \) is the energy in the center of mass frame in the
neutrino-nucleon collision and \( f_{n} \) is the fraction of \( E_{CM} \)
that is bound into the black hole as a function of the number \( n \)
of extra-dimensions. Numerical values for \( f_{n} \) in head-on
collisions are taken from Ref. \cite{EardleyGiddings} and 
reported in Table \ref{tab:eardley}.
\begin{table}
\begin{tabular}{|c|c|}
\hline
\( n \)&
\( f_{n} \)\\
\hline
\hline
0&
\( 0.70711 \)\\
\hline
1&
\( 0.66533 \)\\
\hline
2&
\( 0.63894 \)\\
\hline
3&
\( 0.62057 \)\\
\hline
4&
\( 0.60696 \)\\
\hline
\end{tabular}

\caption{Fraction \protect\( f_{n}\protect \) of \protect\( E_{CM}\protect \)
that is bound into the black hole as a function of the number \protect\( n\protect \)
of extra-dimension in head-on collisions.}
\label{tab:eardley}
\end{table}

The overall shower, defined as the superposition of the various
sub-components, develops according to an obvious cylindrical symmetry around 
the vertical z-axis near the center. We assume in all the studies that the 
incoming primary undergoes a collision with a nucleon in the atoms of the 
atmosphere at zero zenith with respect to the plane of the detector.

The model of the atmosphere that we have adopted consists 
of \( N_{2} \), \( O_{2} \) and
\( Ar \) with the volume fractions of \( 78.1\% \), \( 21.0\% \)
and \( 0.9\% \) \cite{HCP}. The density variation of the atmosphere
with altitude is modeled by 5 layers. The pressure \( p \) as a function
of the altitude \( h \) is given by\begin{equation}
p(h)=a_{i}+b_{i}\exp (-h/c_{i}),\quad i=1,\ldots ,4
\end{equation}
in the lower four layers and by\begin{equation}
p(h)=a_{5}-h\frac{b_{5}}{c_{5}}
\end{equation}
in the fifth layer.

\begin{figure}
{\centering \resizebox*{10cm}{!}{\rotatebox{-90}
{\includegraphics{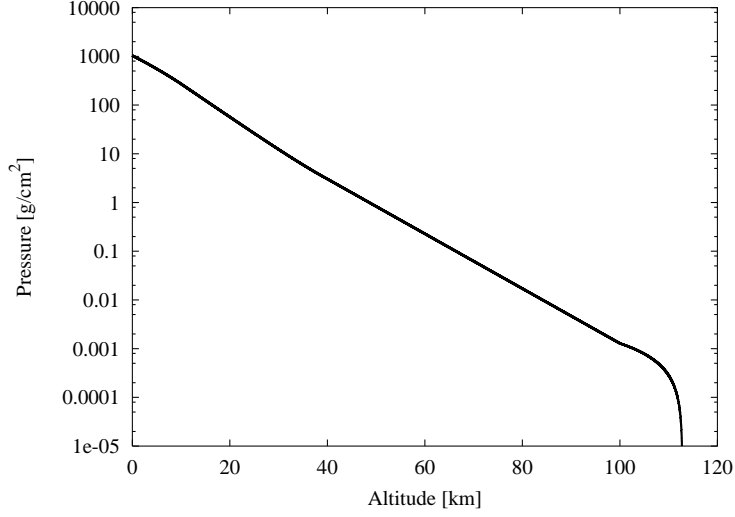}}} \par}
\caption{Pressure versus altitude for the U.S.~standard 
atmosphere model.\label{fig:USstandard}}
\label{pressure}
\end{figure}

The \( a_{i} \), \( b_{i} \), \( c_{i} \) parameters, that we report
in Table~\ref{tab:USstandard}, are those of the U.S.~standard atmospheric model
\cite{CORSIKA}. The boundary of the atmosphere in this model is
defined at the height \( 112.8\, \textrm{km} \), where the pressure vanishes.
In Figure~\ref{fig:USstandard} we show a plot of the pressure ($p=X_v$, also 
called vertical depth) as a function of the height. 
\begin{table}
\hspace{2cm}
\begin{tabular}{|c|c|c|c|c|}
\hline
Layer \( i \)&
Altitude \( h \) {[}km{]}&
\( a_{i}\, [\textrm{g}/\textrm{cm}^{2}] \)&
\( b_{i}\, [\textrm{g}/\textrm{cm}^{2}] \)&
\( c_{i}\, [\textrm{cm}] \)\\
\hline
\hline
1&
\( 0\ldots 4 \)&
\( -186.5562 \)&
\( 1222.6562 \)&
\( 994186.38 \)\\
2&
\( 4\ldots 10 \)&
\( -94.919 \)&
\( 1144.9069 \)&
\( 878153.55 \)\\
3&
1\( 0\ldots 40 \)&
\( 0.61289 \)&
\( 1305.5948 \)&
\( 636143.04 \)\\
4&
\( 40\ldots 100 \)&
\( 0.0 \)&
\( 540.1778 \)&
\( 772170.16 \)\\
\hline
5&
\( >100 \)&
\( 0.01128292 \)&
\( 1 \)&
\( 10^{9} \)\\
\hline
\end{tabular}
\caption{Parameters of the U.S.~standard atmosphere.}
\label{tab:USstandard}
\end{table}

\vspace{2cm}

This is defined via the integral  
\beq
X_v=\int_h^\infty \rho(h')d h'
\eeq
of the atmospheric density $\rho(h)$ for zero zenith angle, 
while the corresponding slant depth is given by 
\beq
X=\int_l^\infty \rho\left(l \cos\theta + \frac{1}{2} \frac{l^2}{R_T} 
\sin^2\theta\right) 
\eeq
for a zenith angle $\theta$ and $R_T$ is the radius of the earth. 

To put into perspective our Monte Carlo study it is convenient to briefly 
summarize the basic features of the theory of cascades on an analytical ground. 
The theory consists of the system of transport equations 
\cite{Gaisser,Farrar} for the numbers $N_n(E,X)$ of particles
of type $n$ with energy $E$ at height $X$
\begin{equation}
\frac{d N_{n}(E,X)}{d X}=-\frac{N_{n}(E,X)}{\lambda _{n}(E)}-\frac{1}{c\tau_n 
\gamma \rho _{\mathrm{Air}}}N_{n}(E,X)~,
\end{equation}
where \( \lambda _{n}(E) \) is their interaction length, $\tau_n$ is their 
lifetime and
\( \gamma  \) the Lorentz-factor corresponding to their given energy. 
In the simple case of an isothermal 
atmosphere \( \rho _{\mathrm{Air}}=\rho _{0}\exp (-h/h_{0})=X/h_{0} \), 
at a scale height $h_0$
\begin{equation}
\label{for:reacprob}
\frac{d N_{n}(E,X)}{d X}=-\frac{N_{n}(E,X)}{\lambda _{n}(E)}-
\frac{1}{d_{n}}N_n(E,X)
\end{equation}
where $d_n$ is their decay length, defined by
\beq
\frac{1}{d_n}=\frac{m c^2 h_0}{E c \tau_n X}.
\eeq
Particles produced at higher energies are also accounted for by an additional 
term in the cascade 
\begin{eqnarray}
\frac{\partial N_n(E,X)}{\partial X} & = & -N_{n}(E,X)\left[ \frac{1}
{\lambda _{n}(E)}+\frac{1}{d_n(E)} \right]\label{cascade} \\
&  & +\sum _{m}\int _{E}^{E^{max}} N_{m}(E',X)\left[ \frac{W_{mn}(E',E)}
{\lambda_{m}(E')}\right. \nonumber \\
&  & \, \, \, \, \, \, \, \, \, \, \, \, \, \, \, \left. +
\frac{1}{d_n(E')}D_{mn}(E',E)\right] dE'\nonumber ~~,
\end{eqnarray}
describing the change in the number of particles of type $n$ due to particles 
of type $m$ by interaction or decay, integrated over an allowed 
interval of energy. 
The functions \( W_{mn}(E',E) \) are the
energy-spectra of secondary particles of
type \( n \) in a collision of particle \( m \) with an air-molecule, while 
\( D_{mn}(E',E) \) are the corresponding decay-functions. The advantage of 
a transport equation compared to a Monte Carlo 
is, that it provides a rather simple analytical view of the development of the 
cascade across the atmosphere. 
Most common in the study of these equations is to use a factorized ansatz for 
the solution $N(E,X)=A(E)B(X)$, which 
assumes a scaling in energy of the transition functions \cite{Gaisser}. 
In our case an analytical treatment of the cascade corresponds to the 
boundary condition 
\beq
N_n(E,X_0)=\textrm{Pr}_n(E) \,\delta(E - f M_{BH}/\langle N\rangle)
\label{index_n}
\eeq
with $\textrm{Pr}_n(E)$ being the probability that the black hole decays 
into a specific state $n$. As we have already discussed 
above, these decays are uncorrelated and Eq.~(\ref{index_n}) is replicated 
for all the elementary states 
after hadronization. The emission probabilities $\textrm{Pr}_n(E)$ 
have been computed by us for a varying initial energy 
$E=f M_{BH}$ using renormalization group equations as described before, 
having corrected for energy loss in the bulk. The interactions 
in the injection spectrum of the original primaries at our $X_0$ 
($X_0=517$ g/cm$^2$) has been neglected since this is not implemented in CORSIKA. 
The showers have been performed independently and the results of the 
simulations have been statistically superimposed at the end 
with multiplicities computed at detector level ($X_1=553$ g/cm$^2$).
We have kept the gravity scale $M_*$ constant at 1 TeV and varied the mass 
of the black hole according to the available center of mass 
energy $E$. As we have already discussed in the previous sections, as a 
benchmark process we have selected a proton-air impact at the same 
$X_0$ with the boundary condition 
\beq
N_{p}(E,X_0)= \textrm{Pr}_p \delta(E - f M_{BH})
\eeq
which occurs with probability 1 ($Pr_p=1$).

We are interested both in the behaviour of the multiplicities and in the 
lateral distributions of the cascades developed at detector level. 
For this purpose we have
defined the opening of the conical shower after integration over the azimuthal 
angle, as in \cite{CCF}, 
and given the symmetry of the event, we plot only the distance from the center 
as a relevant parameter of the conical shower.

A varying number of extra dimensions \( n=0,\, 1,\ldots ,4 \) implies a 
different ratio for bulk-to-brane energy emission, 
a different average number of elementary decaying states and different energy 
distributions among these. 
We have varied the energy $E_1$ of the primary, thereby 
varying the mass of the black hole 
resonance, in the interval $10^{15}-10^{20}$\,
\textrm{eV}. The hadronization part of our code has been done by 
changing the $\textrm{Pr}_h(E)$ obtained from a numerical solution of the 
fragmentation functions separately for each value of the energy shared.

\begin{itemize}

\vspace{0.6cm}

\item{\em Preliminary studies}

We start with the numerical study of the partial and total multiplicities 
of the various sub-components and of their lateral distributions in the benchmark 
event. The results for photons and leptons are shown in 
Figures~\ref{multiphoton} to \ref{multiphoton3} as functions of the 
altitude of the impact and for two different observation points 
at 4,500 m and 5,000 m, approximately the altitudes 
of the detectors at Pamir and Chacaltaya, respectively.  
These preliminary plots, based on actual simulations of 
a proton-to-air nucleus impact at $10^{15}$ eV, 
show a steady growth of the 
multiplicities of the secondaries as we raise the point of first impact above the detector.
\begin{figure}
{\centering \resizebox*{10cm}{!}{\rotatebox{-90}
{\includegraphics{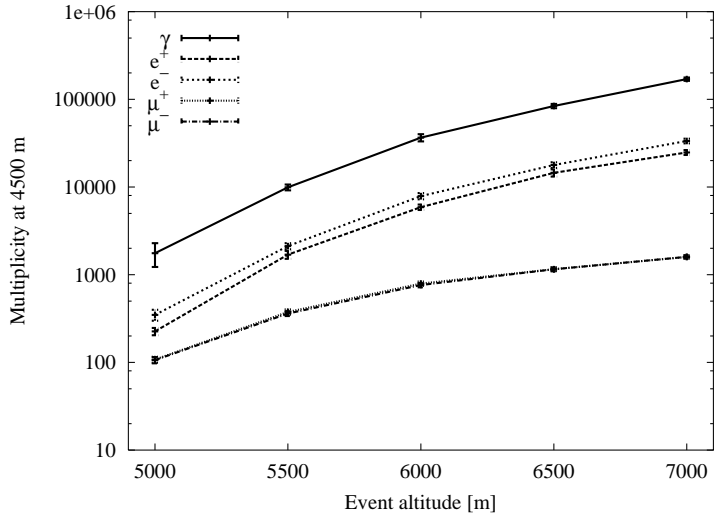}}} \par}
\caption{Benchmark event: Multiplicities of photons, 
\protect\( e^{\pm }\protect \), \protect\( \mu ^{\pm }\protect \)
at an observation level of 4500 m as a function of the altitude of the first impact.
\label{fig:moltVh_4500}}
\label{multiphoton}
\end{figure}
\begin{figure}
{\centering \resizebox*{10cm}{!}{\rotatebox{-90}
{\includegraphics{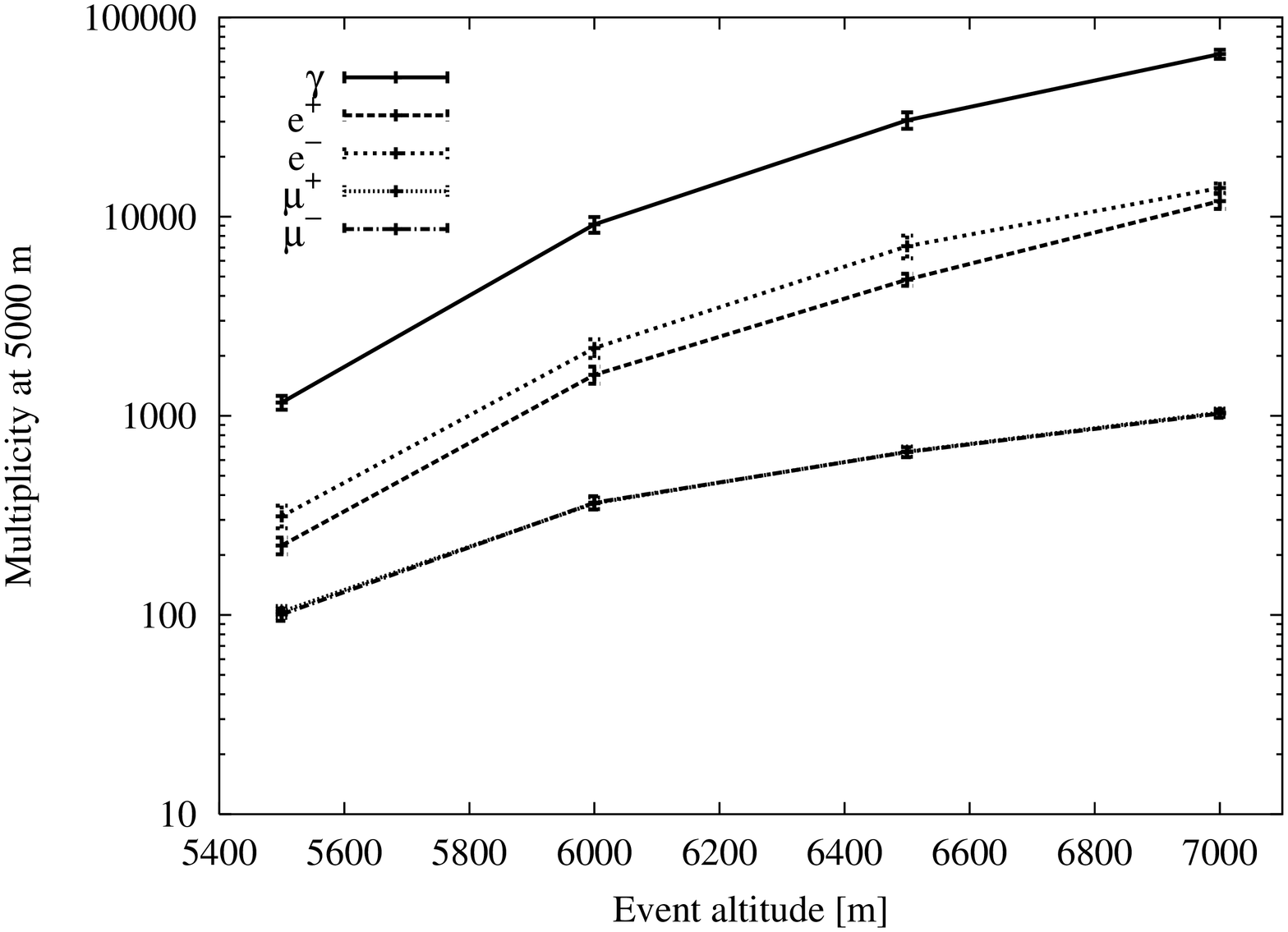}}} \par}
\caption{As in Figure \ref{fig:moltVh_4500}, but at an observation level of
5000 m.}
\label{multiphoton1}
\end{figure}
\begin{figure}
{\centering \resizebox*{10cm}{!}{\rotatebox{-90}
{\includegraphics{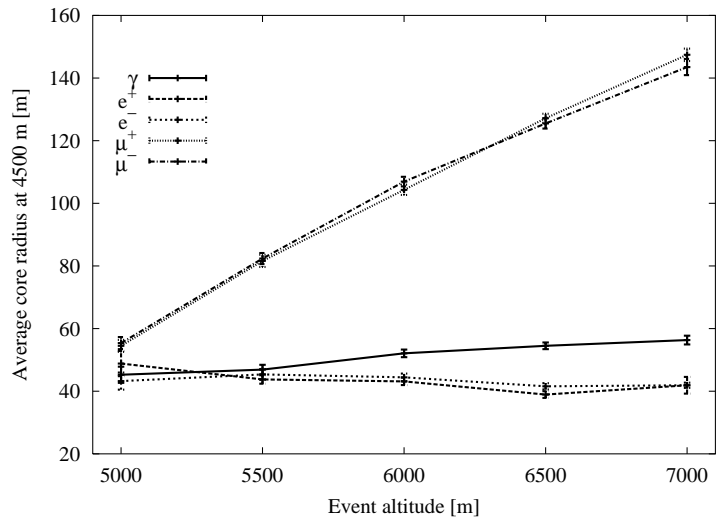}}} \par}
\caption{Average radial opening of the shower of photons, 
\protect\( e^{\pm }\protect \),
\protect\( \mu ^{\pm }\protect \) at an observation level of 4500
m as a function of the altitude of proton's first interaction.\label{distVh_4500}}
\label{multiphoton2}
\end{figure}
\begin{figure}
{\centering \resizebox*{10cm}{!}{\rotatebox{-90}
{\includegraphics{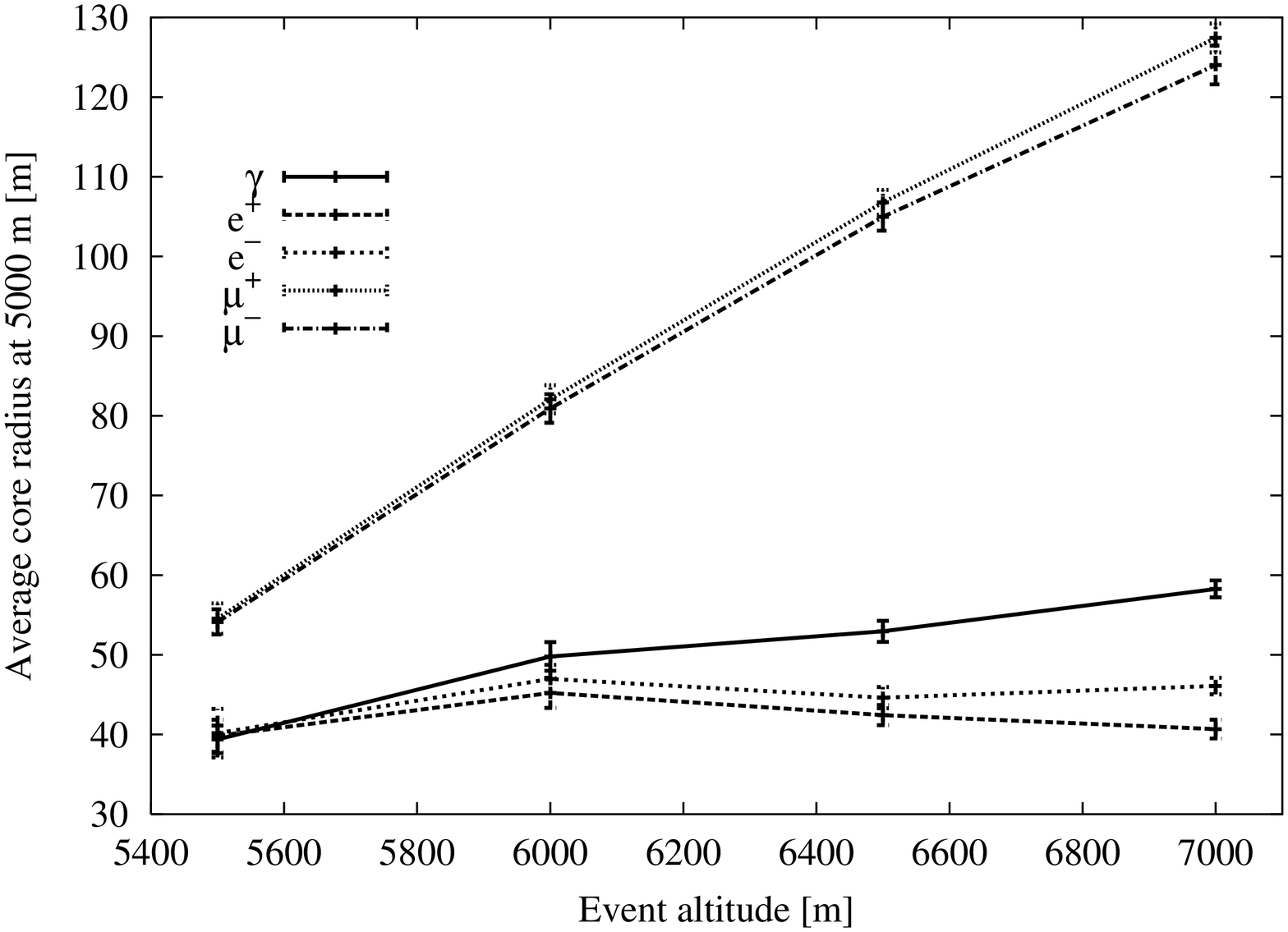}}} \par}
\caption{As in Fig. \ref{distVh_4500}, but at an observation level of 5000m.}
\label{multiphoton3}
\end{figure}
The statistical errors in the simulations 
(80 uncorrelated events have been collected per point) are 
rather small, quite uniformly over all the altitudes of the impact, 
and indicate a satisfactory stability of the result. 
The positions of the detectors 
do not seem to have an appreciable impact on the characteristics of the 
secondaries. As for the lateral distributions we observe an increase in 
the opening of the showers with the event altitude, 
which is more enhanced for the muonic component and for the photons and 
less for the electrons and positrons. Also in this case 
the statistical fluctuations are rather small. On the basis of these 
results we have selected for the remaining simulations 
a first impact at 5,500 \textrm{m} and the observation (detector) level at 
5,000 m. However, for comparison, we will also show later the results
of a second set of simulations that have been 
performed with the first impact at 15,000 m.

\vspace{0.6cm}

\item{\em Choice of scales and corrections}

To compare standard and black holes events, 
we have selected a gravity scale of $M_*=1$TeV and varied 
the black hole mass, here taken to be equal to the available 
center of mass energy during the collision. Therefore, a varying $E_1$ 
is directly related to a varying $M_{\rm{BH}}$ 
and we have corrected, as explained above, for the energy 
loss into gravitational emission. Unfortunately, this can be 
estimated only heuristically, with bounds largely 
dependent on the impact parameter of the primary collision. 
A reasonable estimate may be of the order of $10-15 \%$ \cite{kanti1}.   
Corrections related to emission in the bulk have also been included, 
in the way discussed in previous sections.

\vspace{0.6cm}

\item{\em Energy ratios: electromagnetic versus hadronic} 

Not all observables are statistically insensitive to the natural 
fluctuations of the air showers. In the study of black hole versus 
standard (benchmark) events, the study of the ratios 
$N_{\rm{em}}/N_{\rm{hadron}}$ and 
$E_{\rm{em}}/E_{\rm{hadron}}$ have been proposed 
as a way to distinguish between ordinary showers and other extra-ordinary 
ones. {\em Centauro} events, for instance, have been claimed to be 
characterized by a rather small ratio of electromagnetic over hadronic 
energy deposited in the detectors, contrary to normal showers, in which
this ratio is believed to be $E_{\rm{em}}/E_{\rm{hadron}}\sim 2$. 
Instead, as one can easily recognize from the results presented in 
Figures \ref{ratio1} and \ref{ratio2}, the multiplicity 
ratio takes values in two different 
regimes. In the ``band'' of values $1-5$ for the case of 
the lower first impact and $100-160$ for the higher impact. 
The larger values of the band in this latter case are justified by the 
fact that the shower is far more developed, given the altitude of the 
impact, and therefore is characterized by an even more dominant
electromagnetic component. The energy ratio, on the other hand
can take small values, in agrement with the values observed in Centauros.
However, notice that both the black hole and standard simulations
show a complex pattern for these ratios and in addition they are 
characterized by large 
fluctuations for varying energy and number of extra dimensions. 
\begin{figure}
{\centering \subfigure[ ]{\resizebox*{10cm}{!}{\rotatebox{-90}
{\includegraphics{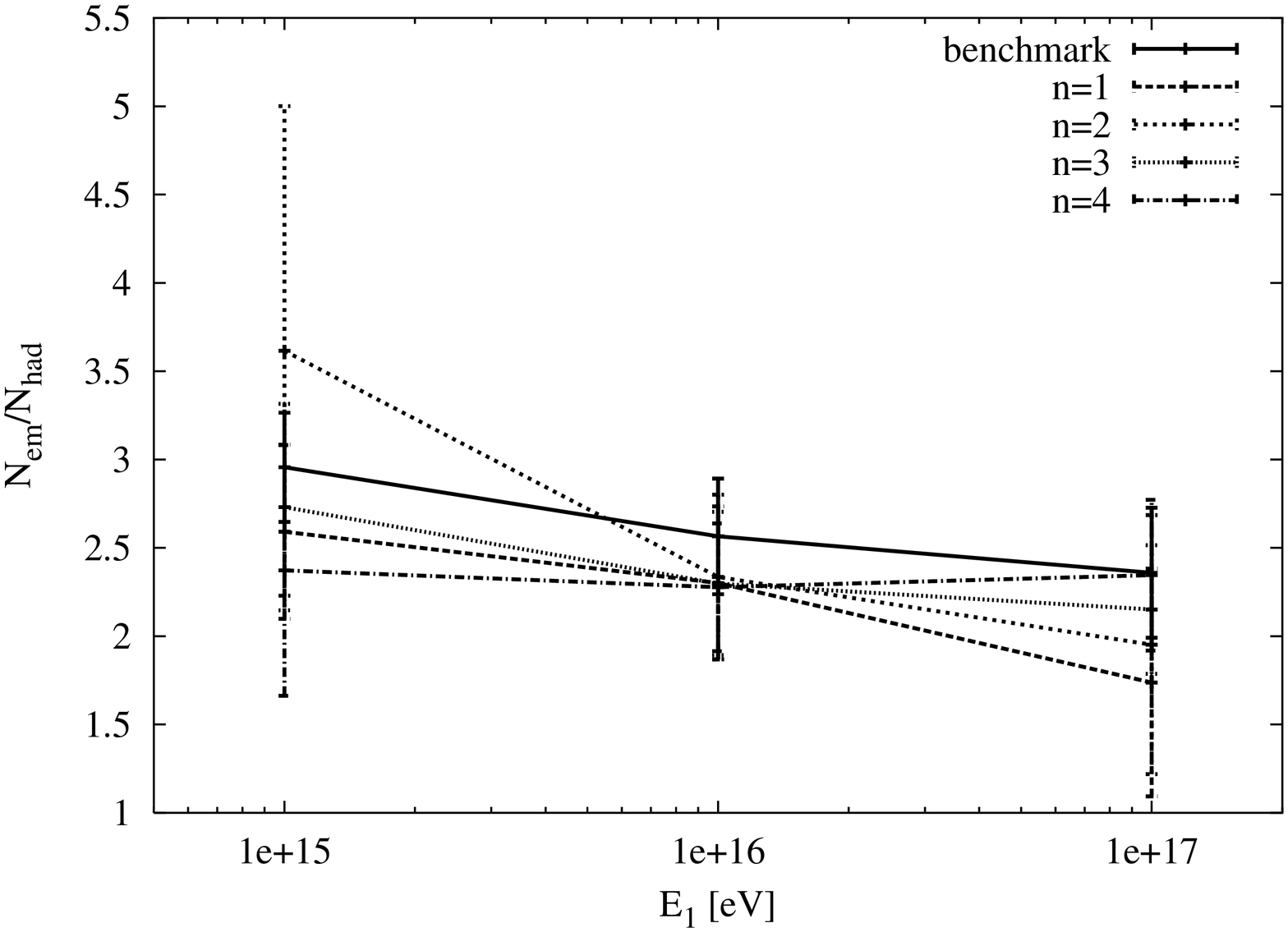}}}} \par}
{\centering \subfigure[ ]{\resizebox*{10cm}{!}{\rotatebox{-90}
{\includegraphics{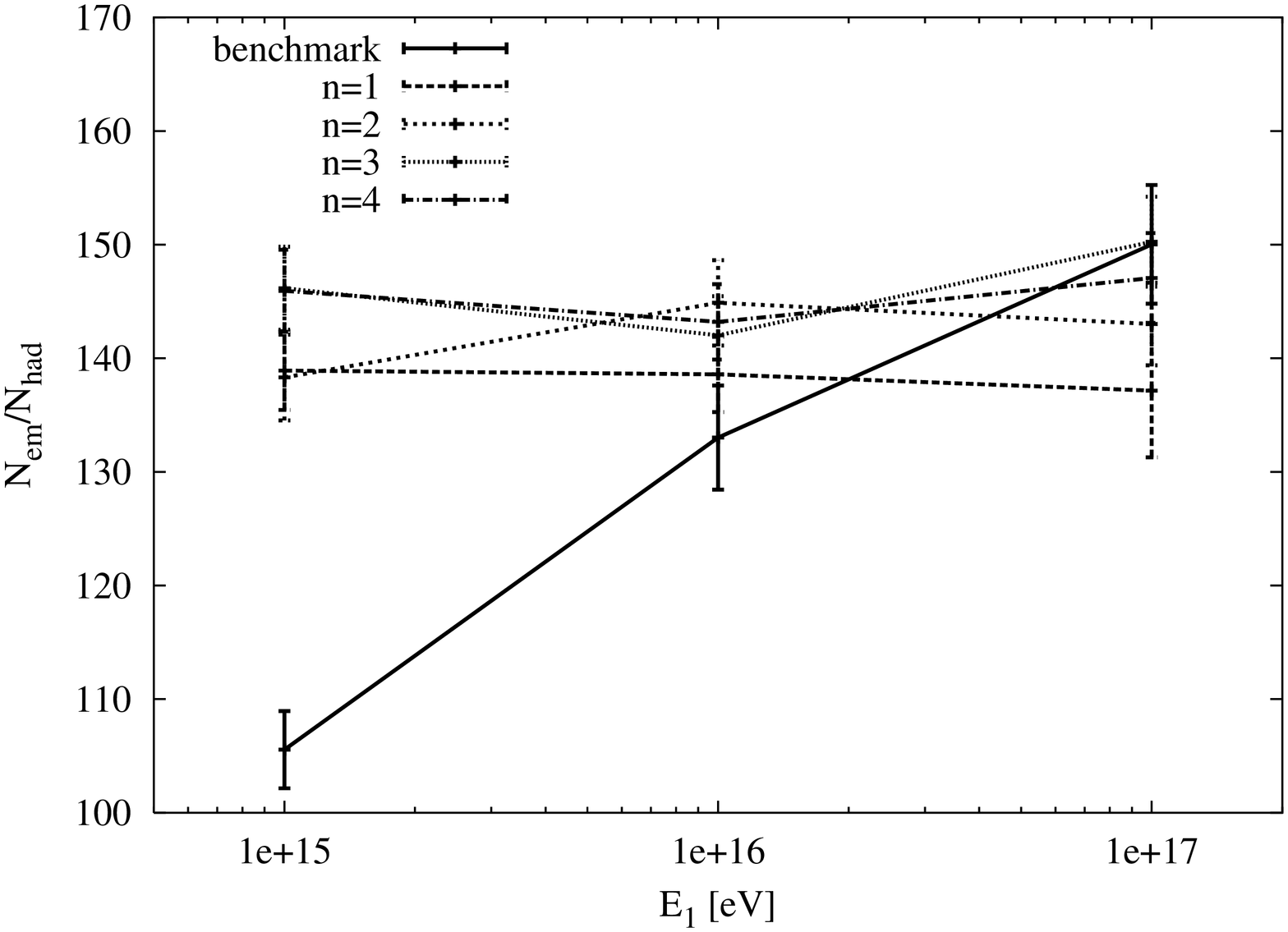}}}} \par}
\caption{Ratio between \protect\( N_{em}\protect \) (total multiplicity of
photons and \protect\( e^{\pm }\protect \)) and \protect\( N_{had}\protect \)
(total multiplicity of everything else) as a function of \protect\( E_{1}\protect \).
The first interaction is kept fixed at \protect\( 5500\, \textrm{m}\protect \)
(517 $g/cm^2$) (a), or at \protect\( 15000\, \textrm{m}\protect \)
(124 $g/cm^2$) (b),
and the observation level is \protect\( 5000\, \textrm{m}\protect \)
(553 $g/cm^2$).
We show in the same plot the benchmark (where the primary is a proton)
and mini black holes with different numbers of extra-dimensions 
\protect\( n\protect \).\label{fig:bhwl_NrappVE}}
\label{ratio1}
\end{figure}
\begin{figure}
{\centering \subfigure[ ]{\resizebox*{10cm}{!}{\rotatebox{-90}
{\includegraphics{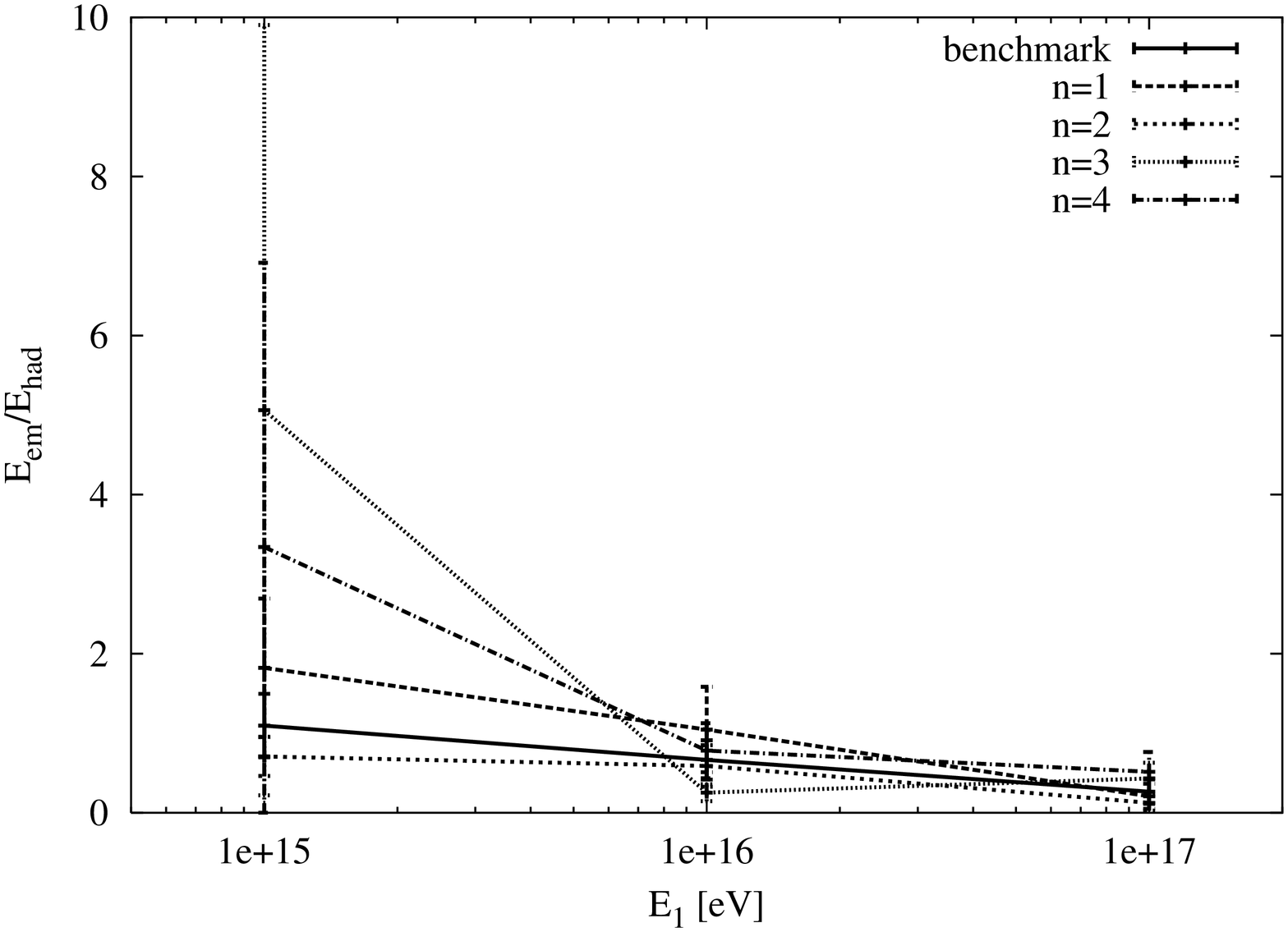}}}} \par}
{\centering \subfigure[ ]{\resizebox*{10cm}{!}{\rotatebox{-90}
{\includegraphics{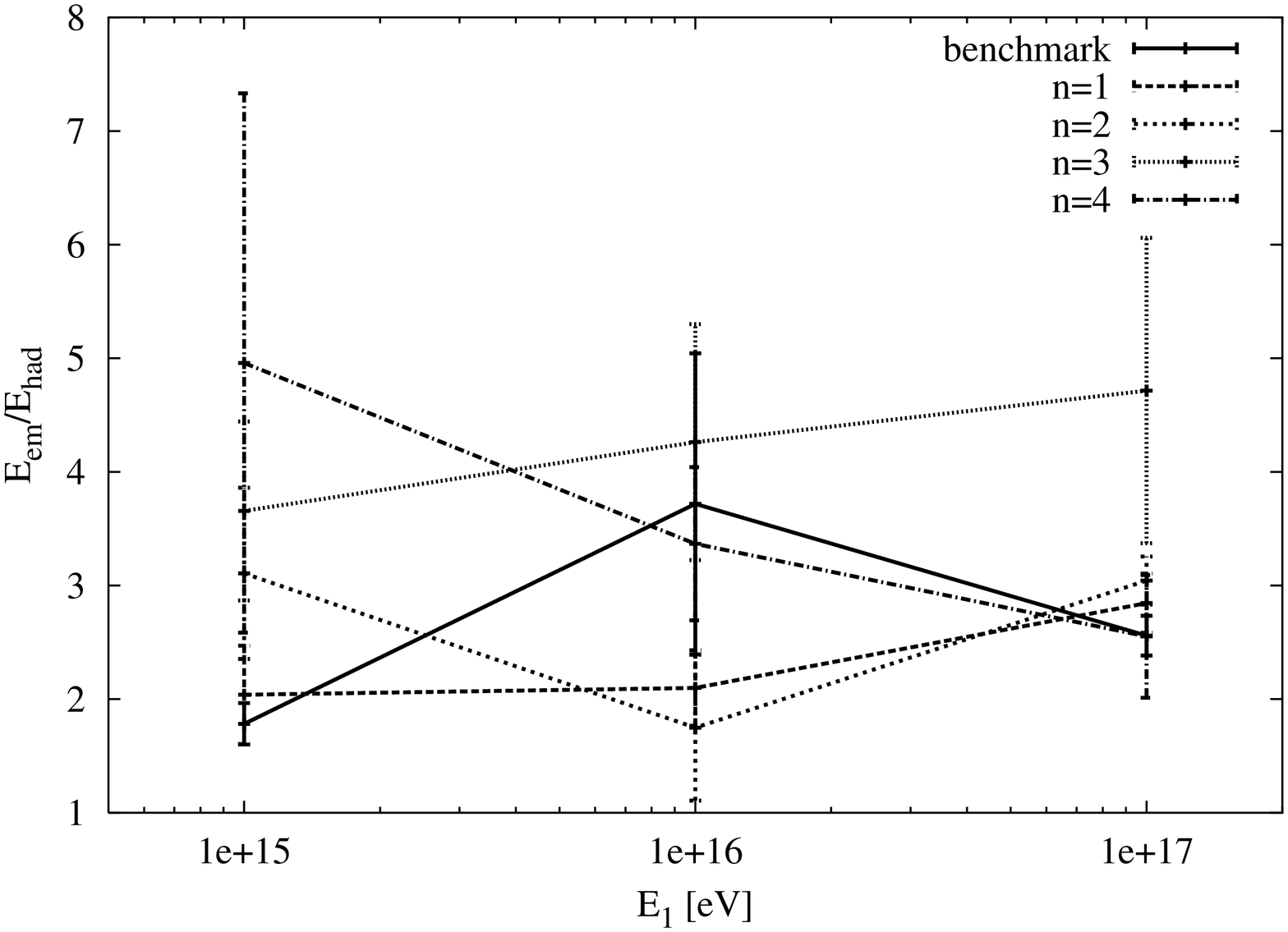}}}} \par}
\caption{As in Figure~\ref{fig:bhwl_NrappVE}, but this time we show the ratio
between \protect\( E_{em}\protect \) (total energy of photons and
\protect\( e^{\pm }\protect \)) and \protect\( E_{had}\protect \)
(total energy of everything else) as a function of \protect\( E_{1}\protect \).}
\label{ratio2}
\end{figure}
Furthermore, there 
does not seem to be a statistically significant difference in these observables 
between the benchmark event and the black hole mediated ones, at least
for event heights greater than 500 m above the detector. 
We conclude that (a) either these observables may not be suitable, 
especially given the limited 
statistics of the existing and future experiments, 
to discriminate between black hole mediated events versus standard ones, or (b) that
the differences in these ratios appear in events with initial impact
in the range $0-500$ m from the detector. This latter issue, of relevance
to the Centauro events, is currently under study \cite{CCT2}.

%\vfill

%\eject

\vspace{0.6cm}

\item{\em Multiplicities}

For the multiplicities themselves the situation is much cleaner.
In Figures~\ref{prima}-\ref{multipro} we show the behaviour of the 
total as well as of some partial multiplicities in black hole mediated 
events and in standard 
events as a function of the energy and for a varying number of extra 
dimensions. 

\begin{figure}
{\centering \subfigure[ ]{\resizebox*{10cm}{!}{\rotatebox{-90}
{\includegraphics{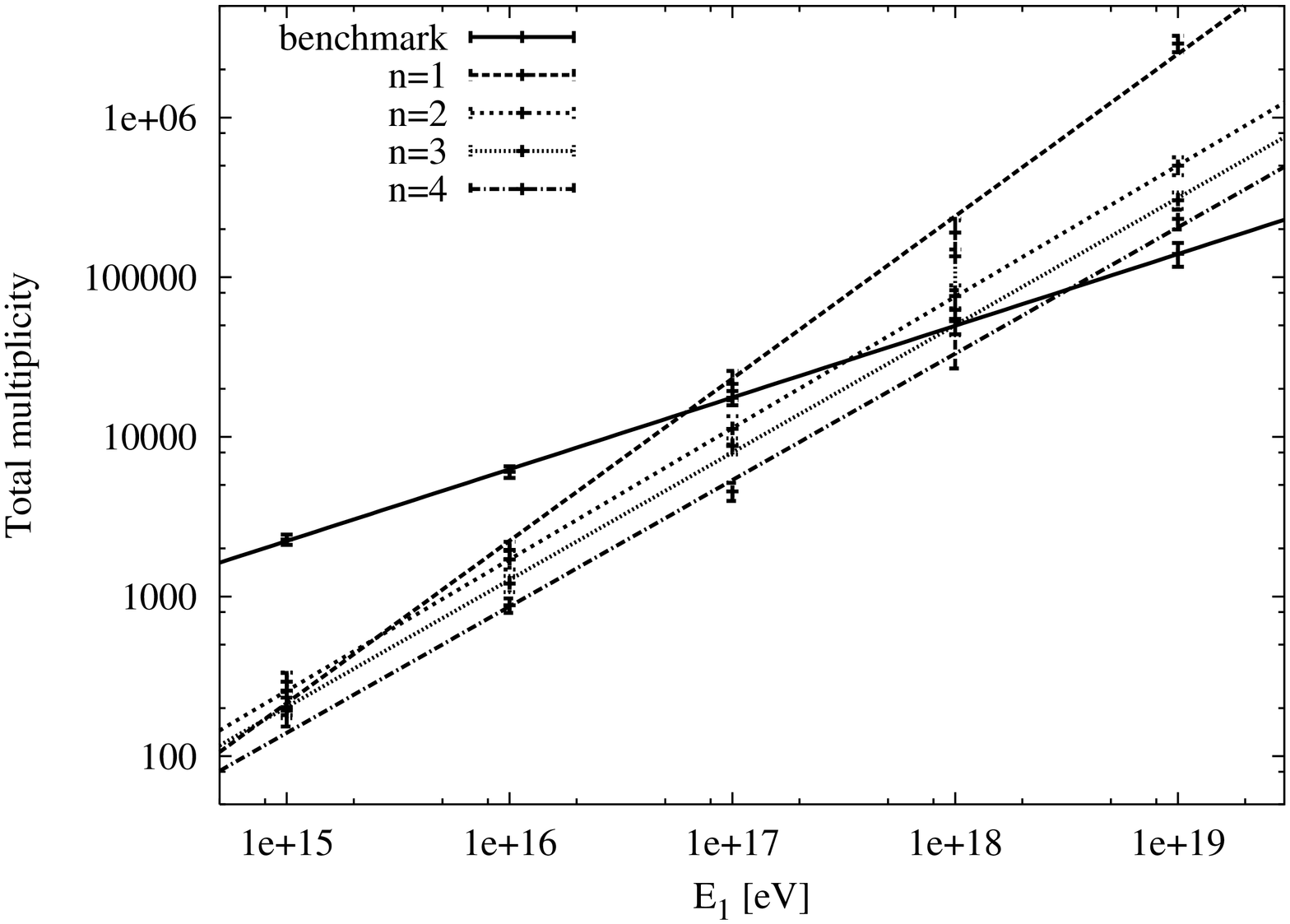}}}} \par}
{\centering \subfigure[ ]{\resizebox*{10cm}{!}{\rotatebox{-90}
{\includegraphics{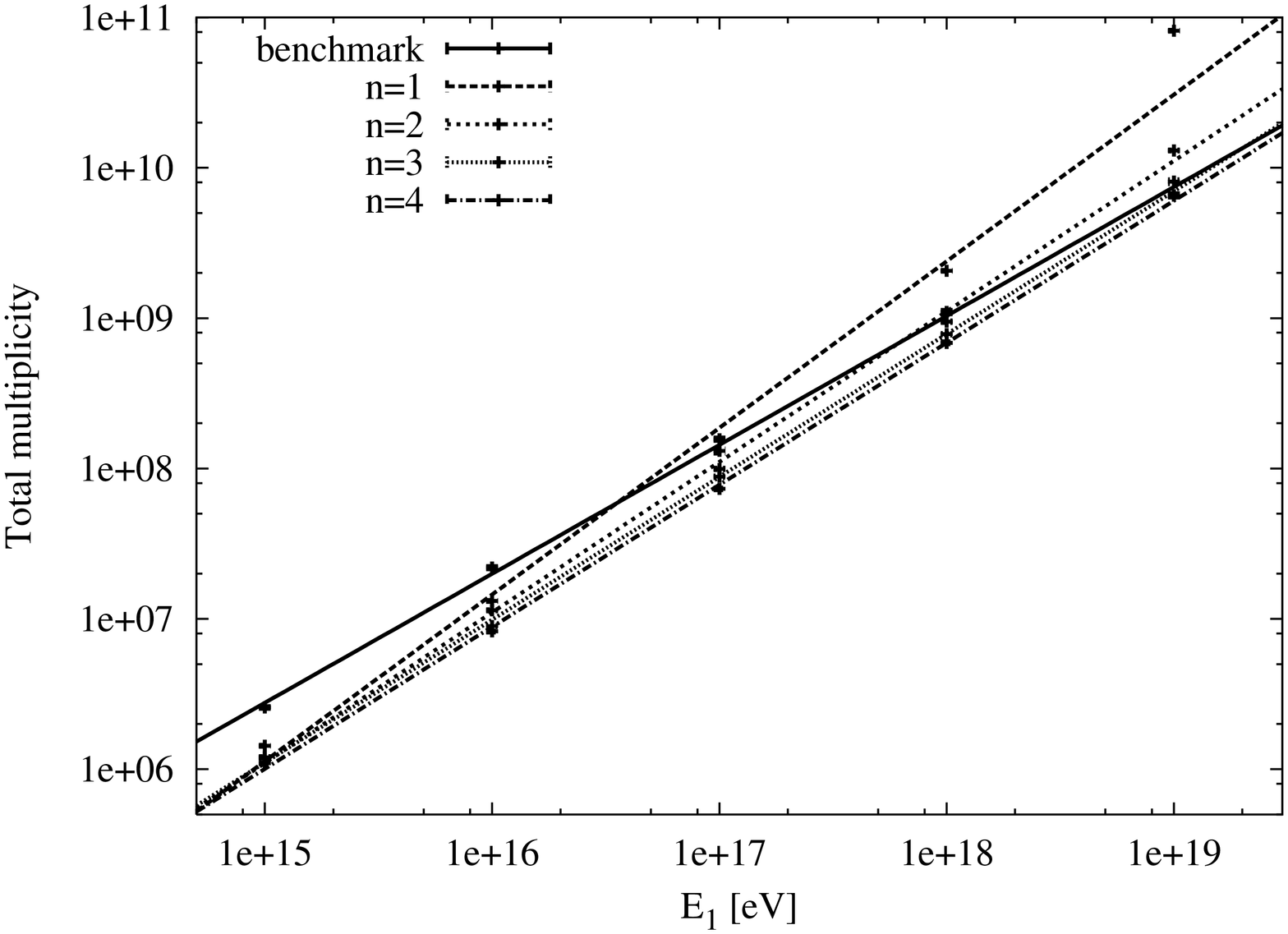}}}} \par}
\caption{Plot of the total particle multiplicity
as a function of \protect\( E_{1}\protect \). Case (a) is for an impact point 
of 5,500 m and case (b) for 15,000 m.}
\label{prima}
\end{figure}

\begin{figure}
{\centering \subfigure[ ]{\resizebox*{10cm}{!}{\rotatebox{-90}
{\includegraphics{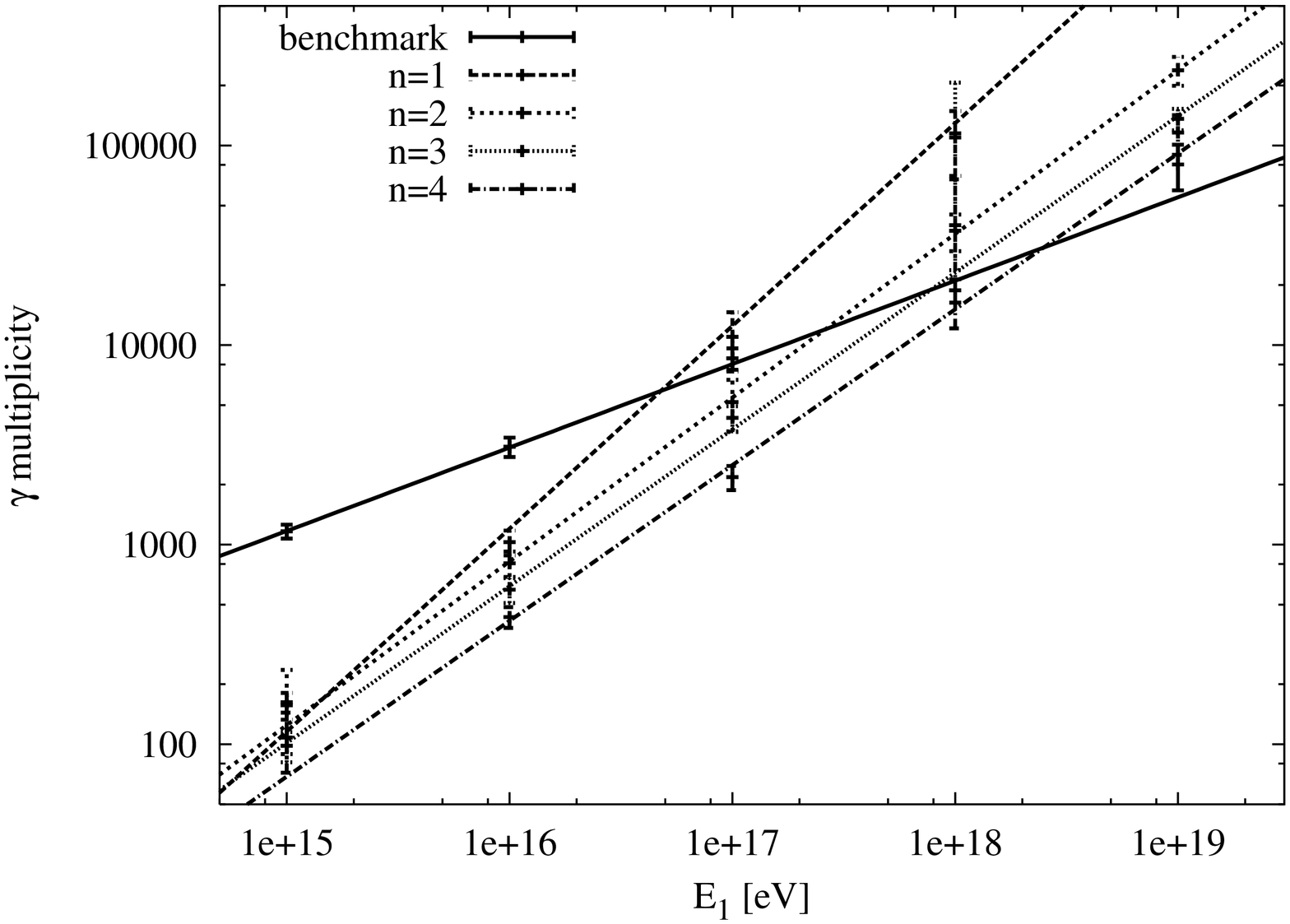}}}} \par}
{\centering \subfigure[ ]{\resizebox*{10cm}{!}{\rotatebox{-90}
{\includegraphics{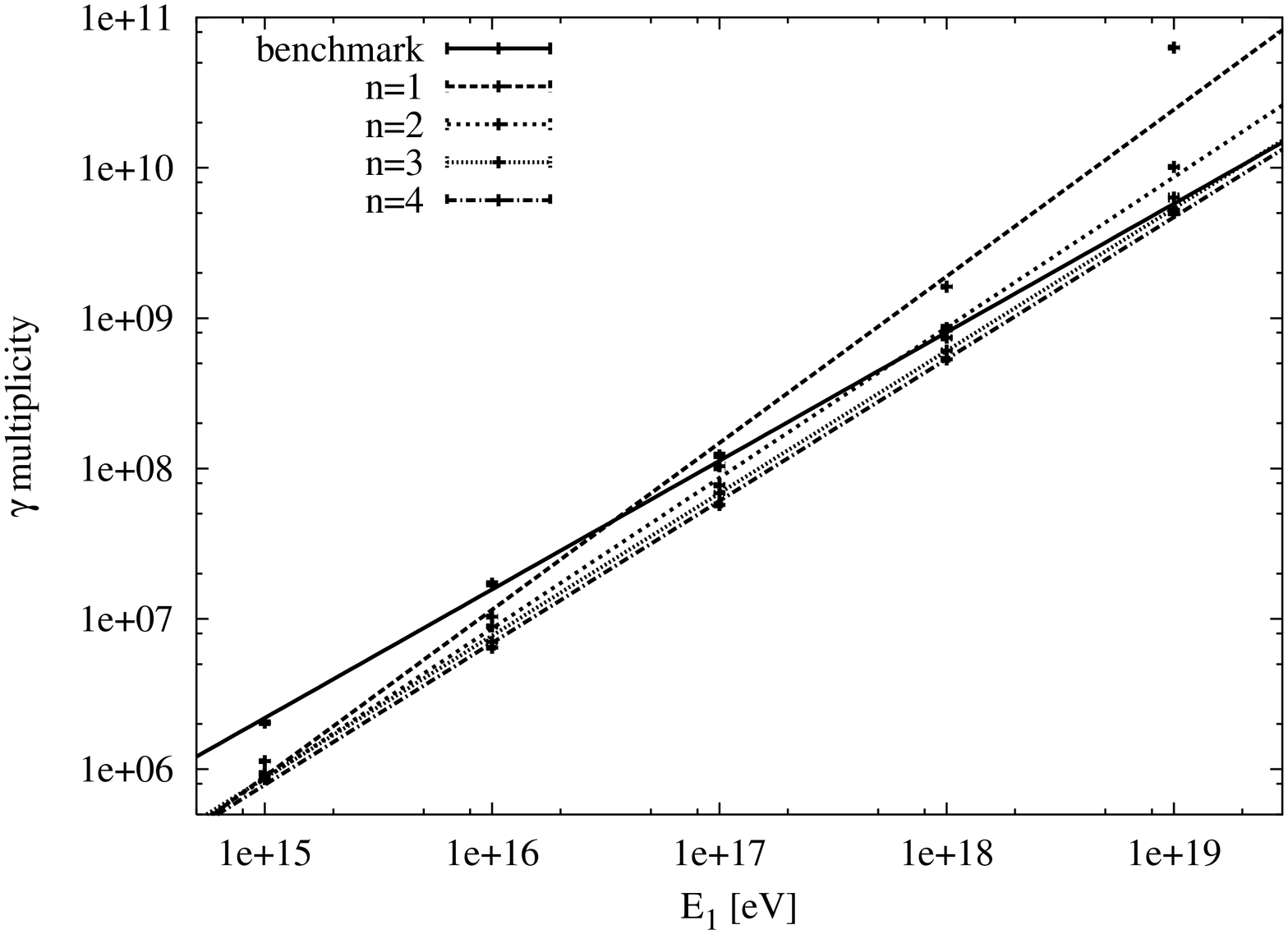}}}} \par}
\caption{Plot of the multiplicity of photons as a function of 
\protect\( E_{1}\protect \), (a) 5,500 m, (b) 15,000 m }
\label{seconda}
\end{figure}

\begin{figure}
{\centering \subfigure[ ]{\resizebox*{10cm}{!}{\rotatebox{-90}
{\includegraphics{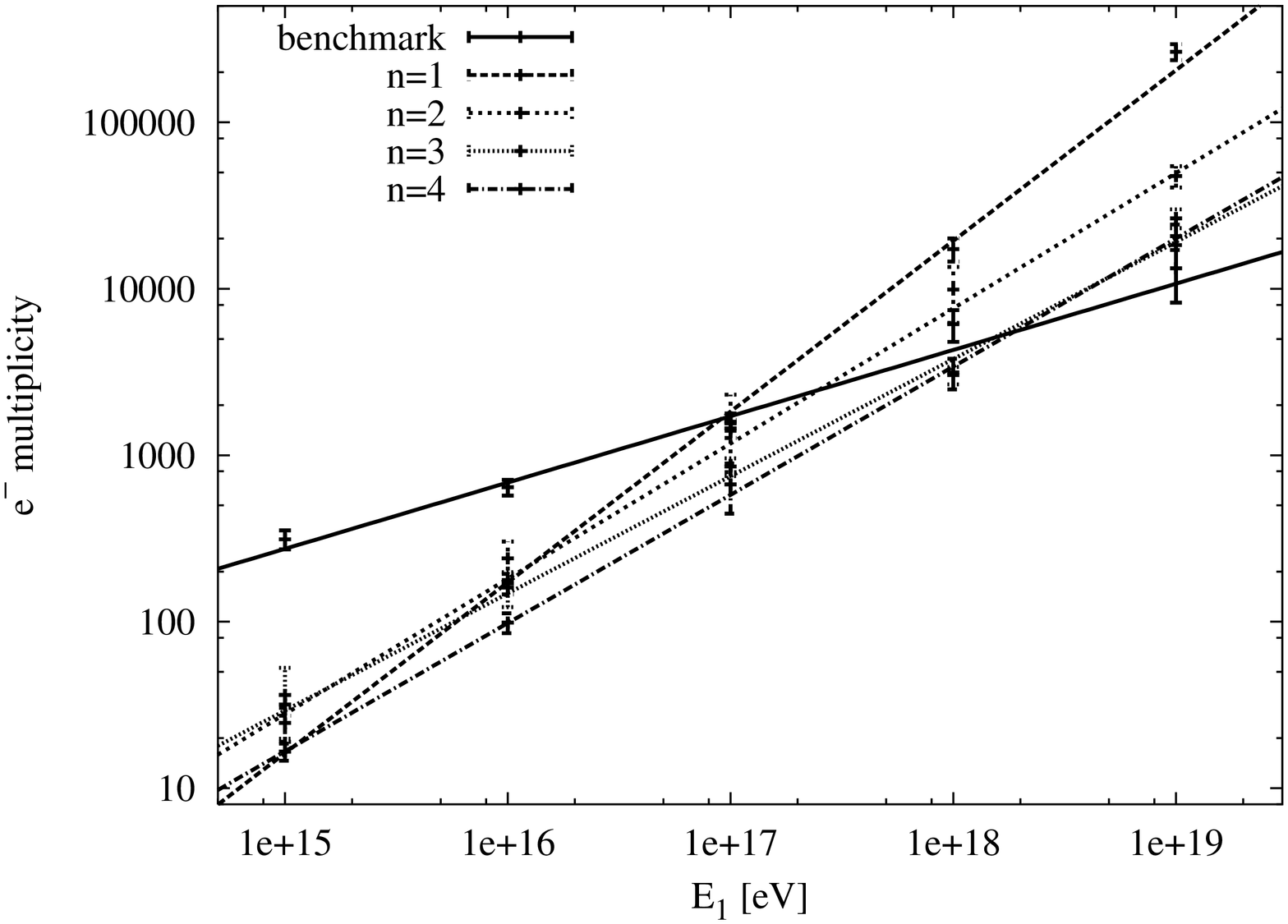}}}} \par}
{\centering \subfigure[ ]{\resizebox*{10cm}{!}{\rotatebox{-90}
{\includegraphics{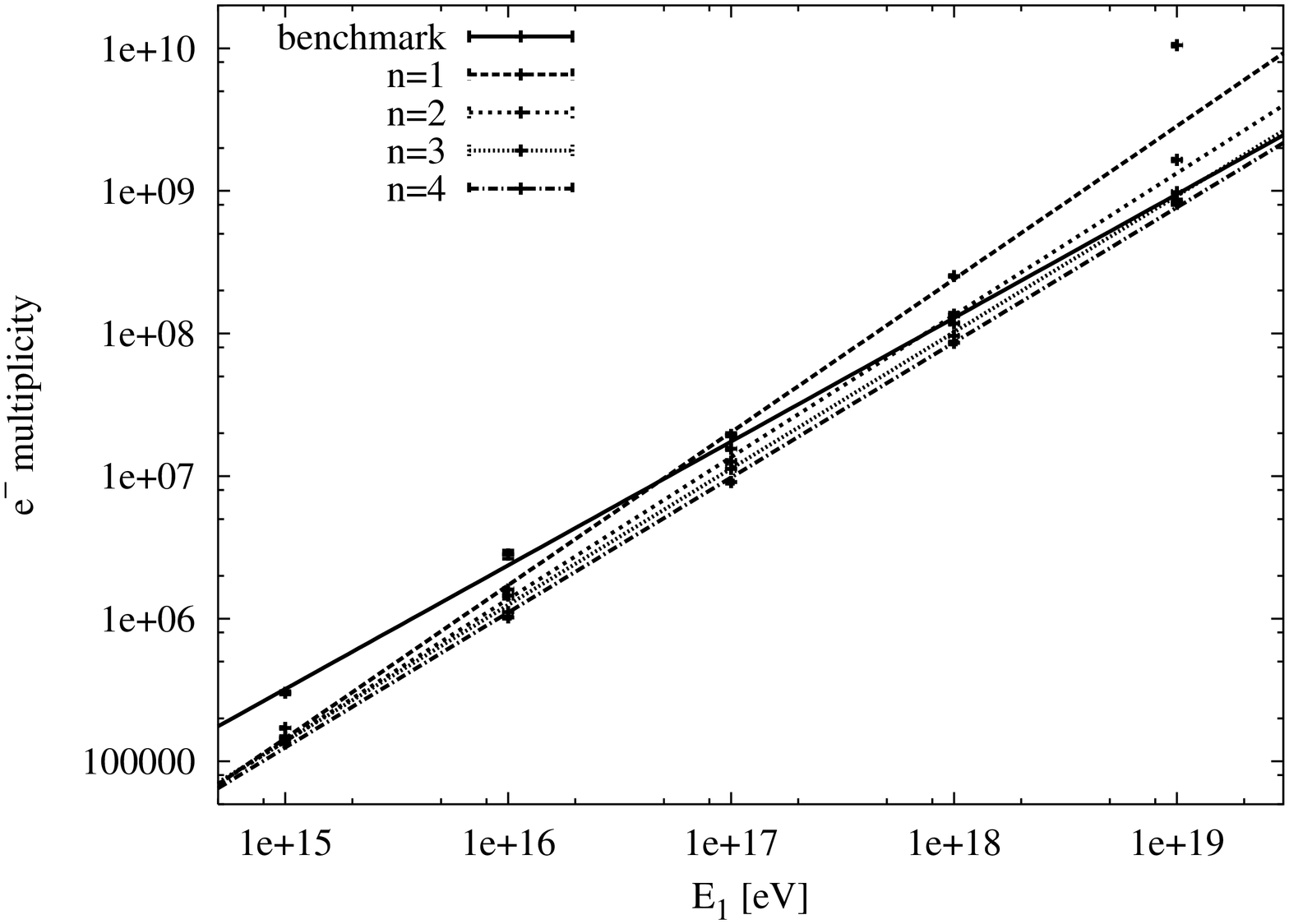}}}} \par}
\caption{Plot of the multiplicity of \protect\( e^{-}\protect \) 
as a function of \protect\( E_{1}\protect \), 
(a) 5,500 m, (b) 15,000 m}
\label{terza}
\end{figure}

\begin{figure}
{\centering \subfigure[ ]{\resizebox*{10cm}{!}{\rotatebox{-90}
{\includegraphics{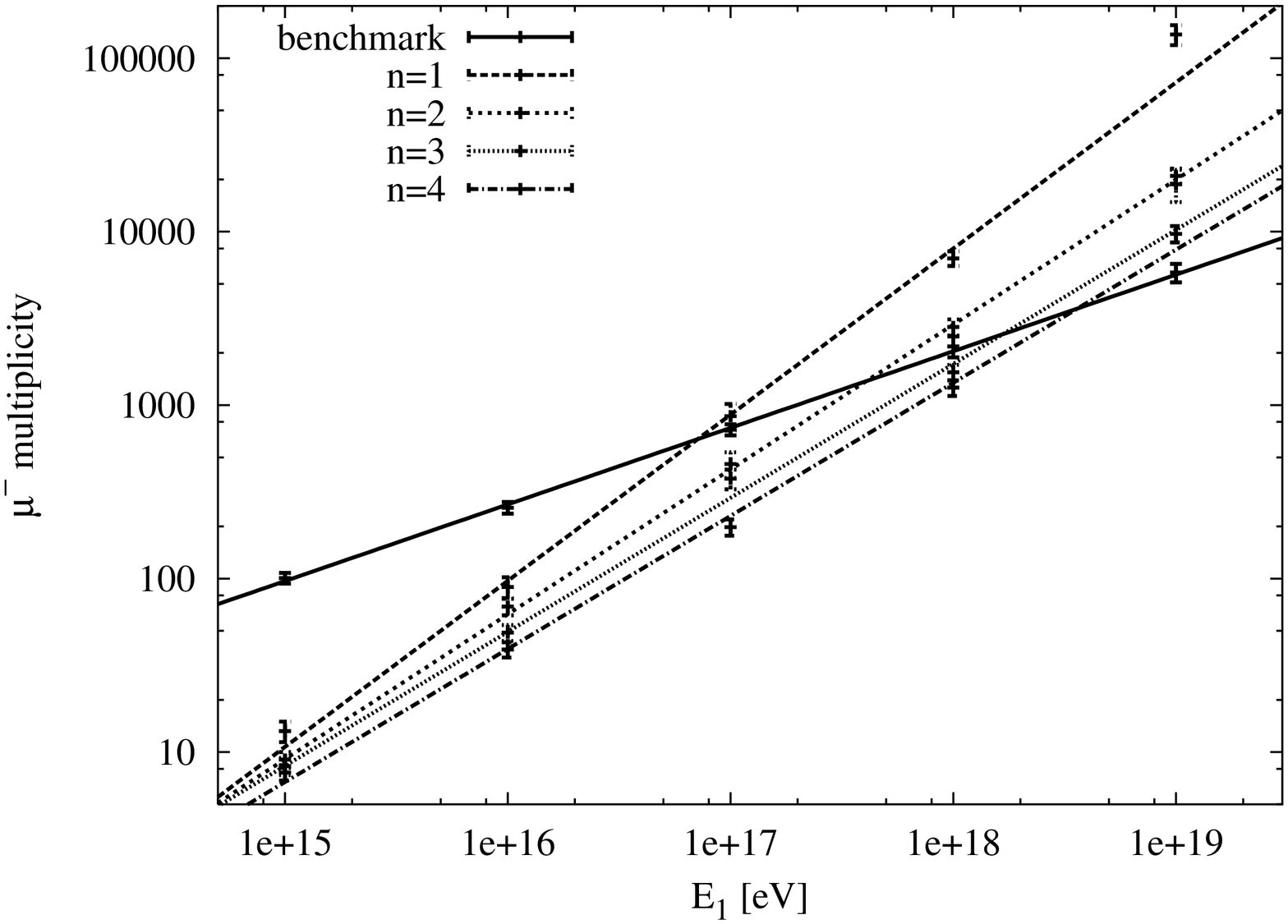}}}} \par}
{\centering \subfigure[ ]{\resizebox*{10cm}{!}{\rotatebox{-90}
{\includegraphics{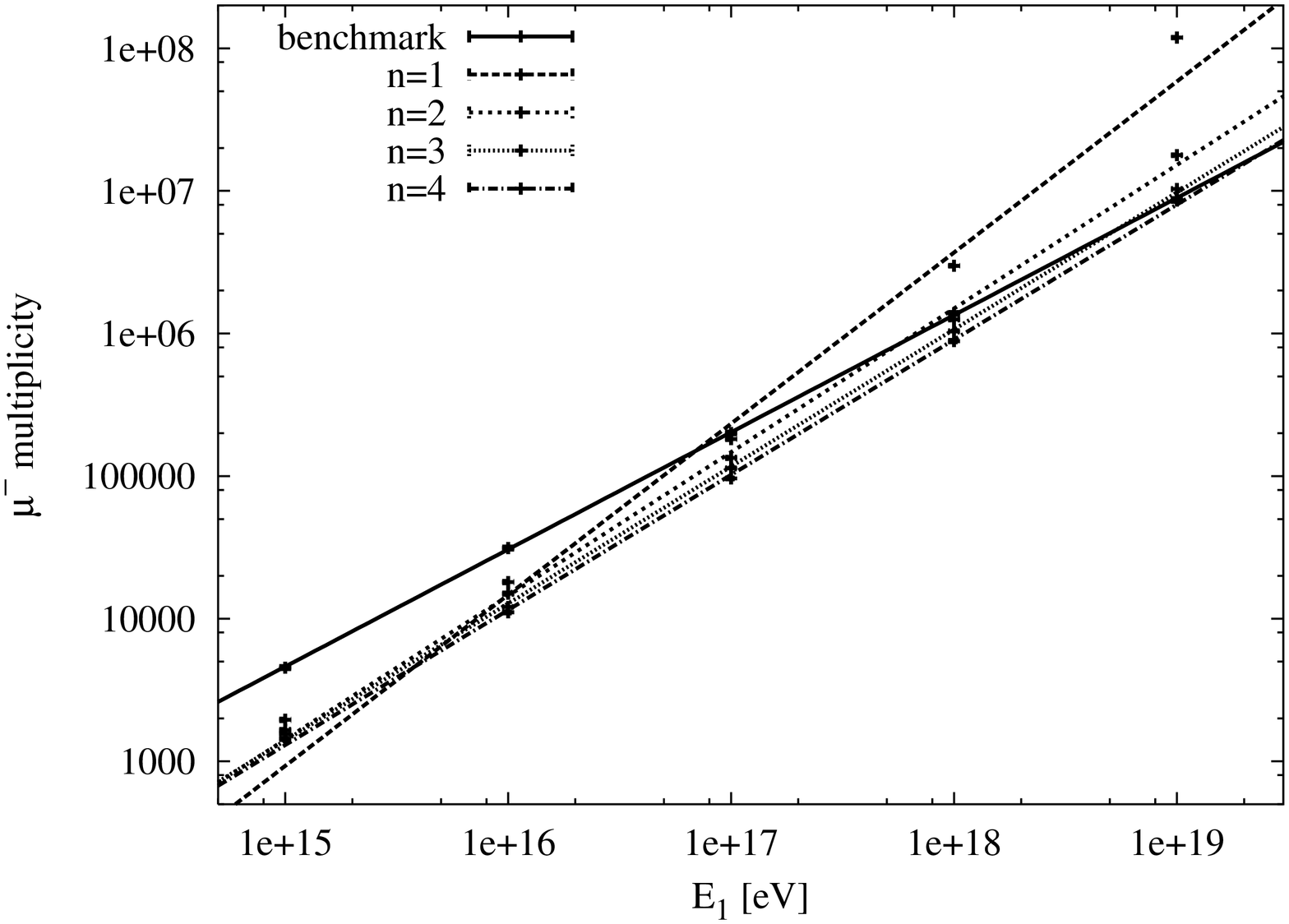}}}} \par}
\caption{Plot of the multiplicity
of \protect\( \mu ^{-}\protect \) as a function of \protect\( E_{1}\protect \), 
(a) 5,500 m, (b) 15,000 m}
\label{quarta}
\end{figure}

\begin{figure}
{\centering \subfigure[ ]{\resizebox*{10cm}{!}{\rotatebox{-90}
{\includegraphics{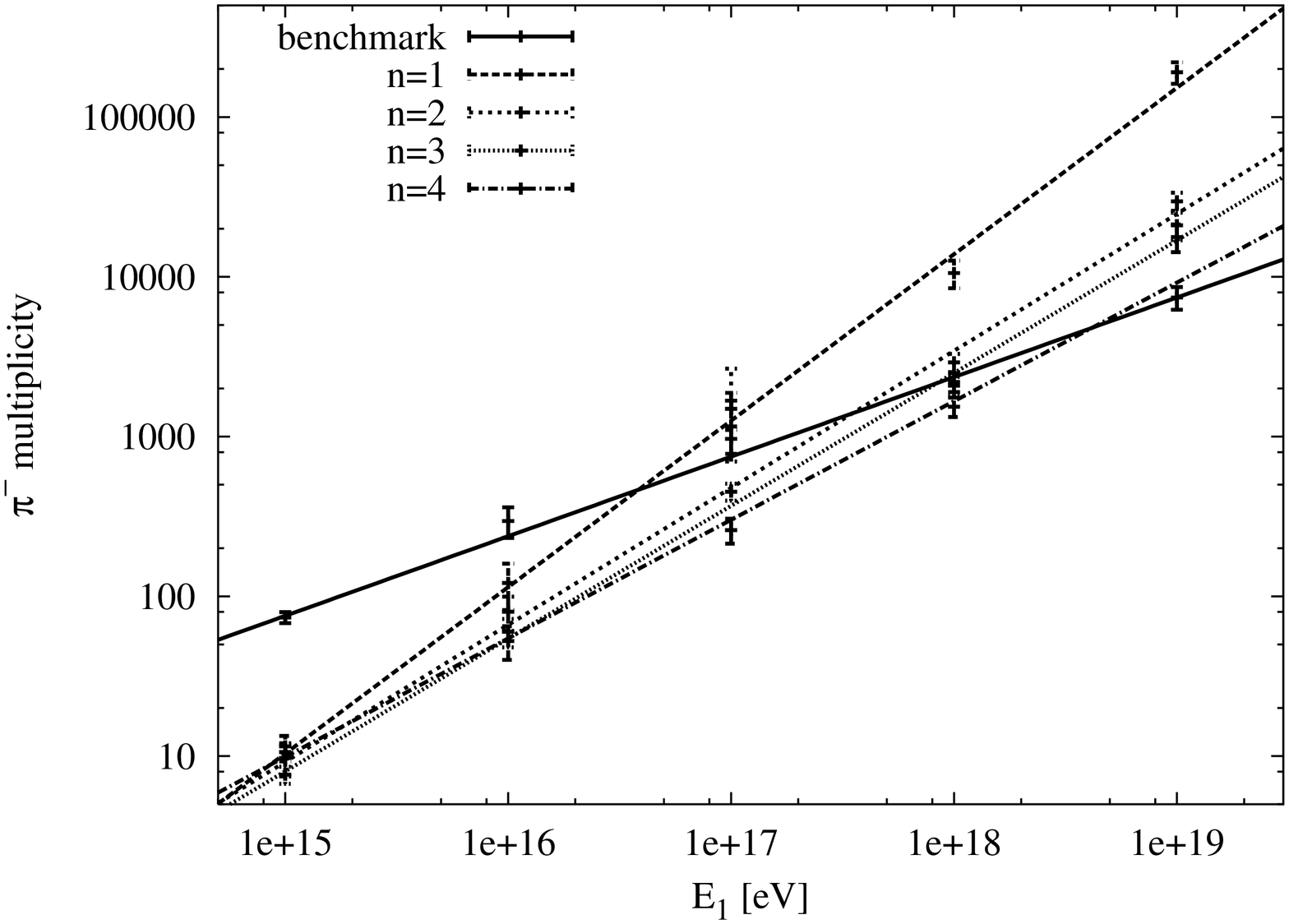}}}} \par}
{\centering \subfigure[ ]{\resizebox*{10cm}{!}{\rotatebox{-90}
{\includegraphics{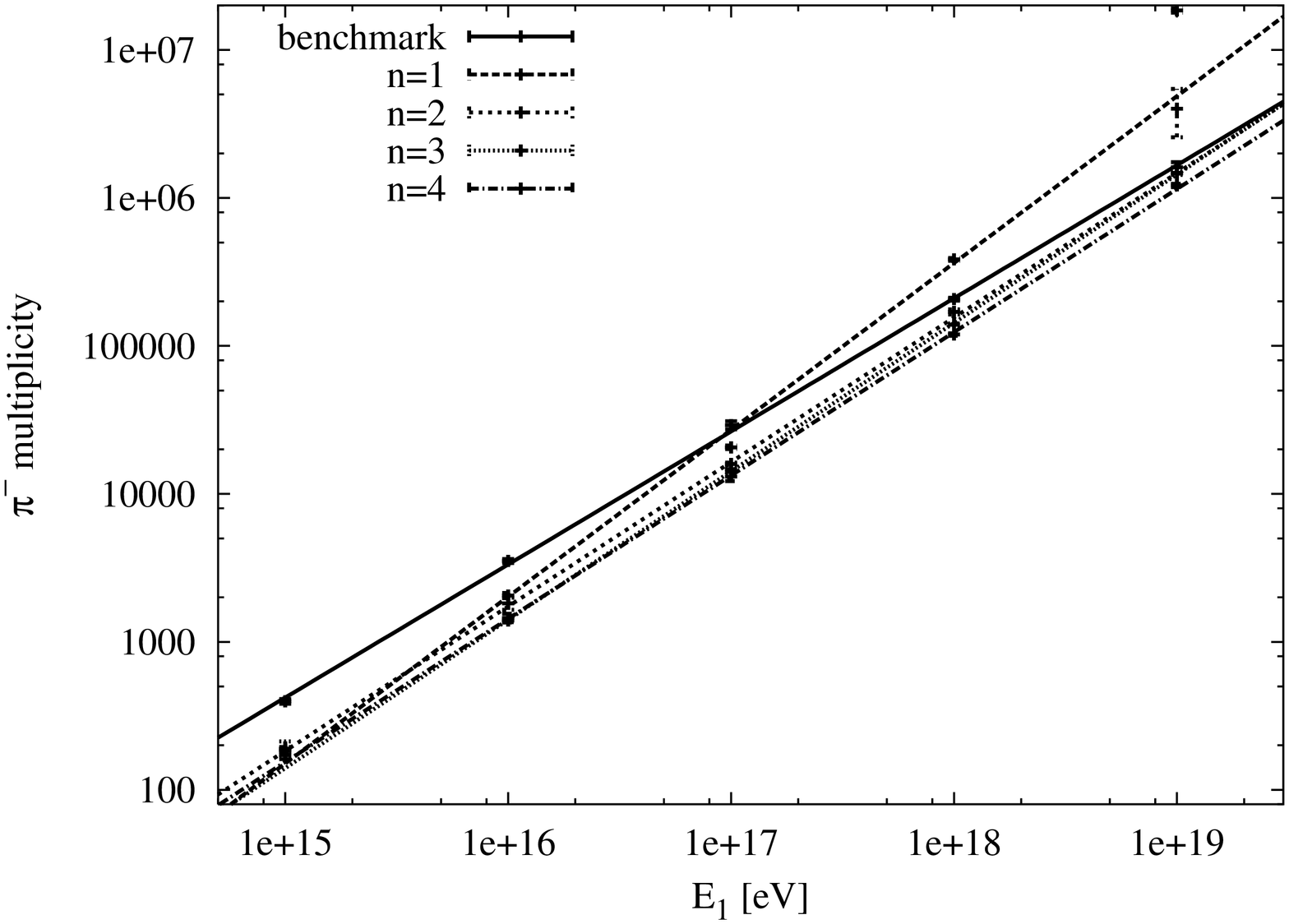}}}} \par}
\caption{Plot of the multiplicity
of \protect\( \pi ^{-}\protect \) as a function of \protect\( E_{1}\protect \), 
(a) 5,500 m, (b) 15,000 m.}
\label{multipi}
\end{figure}

\begin{figure}
{\centering \subfigure[ ]{\resizebox*{10cm}{!}{\rotatebox{-90}
{\includegraphics{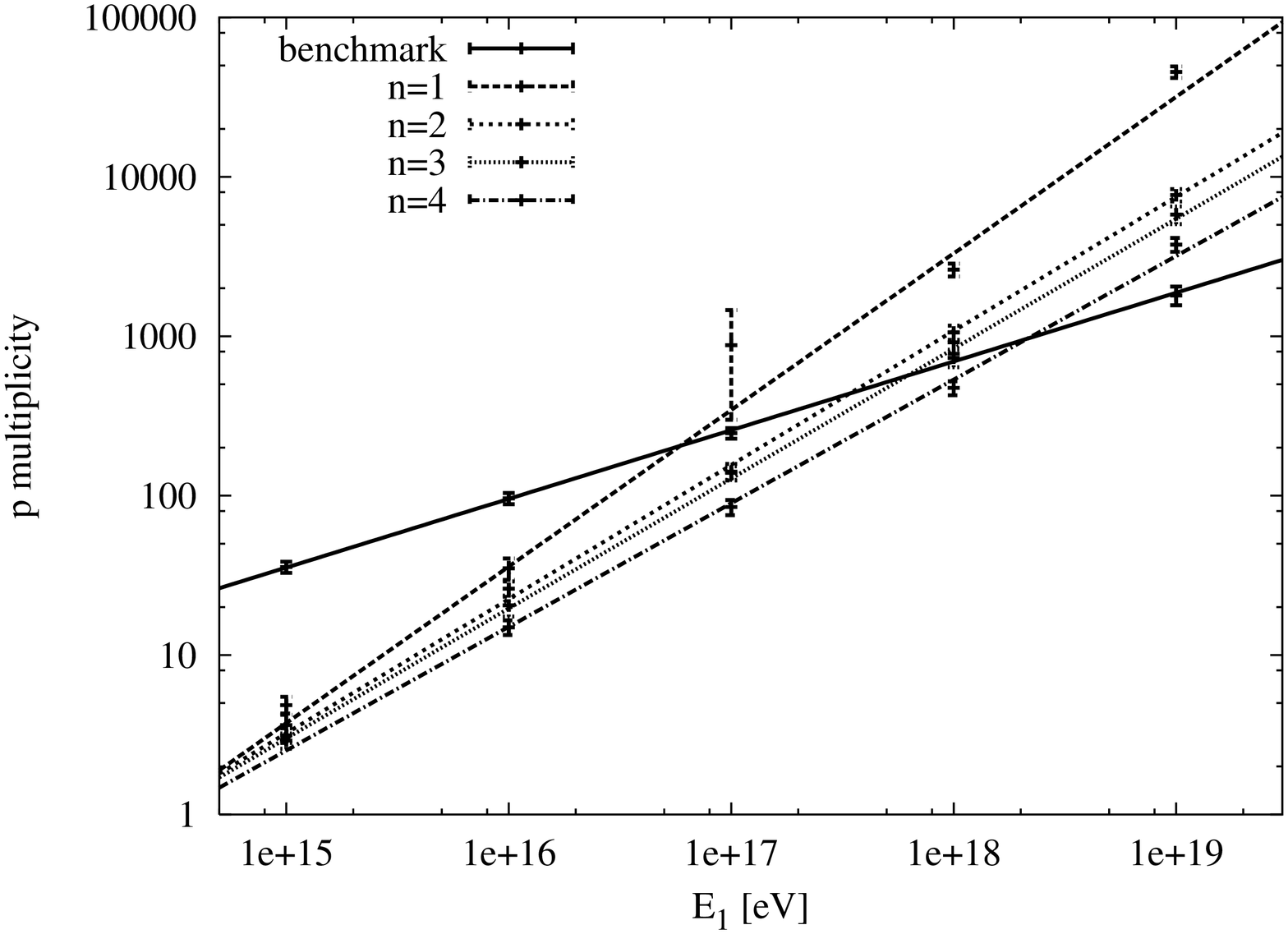}}}} \par}
{\centering \subfigure[ ]{\resizebox*{10cm}{!}{\rotatebox{-90}
{\includegraphics{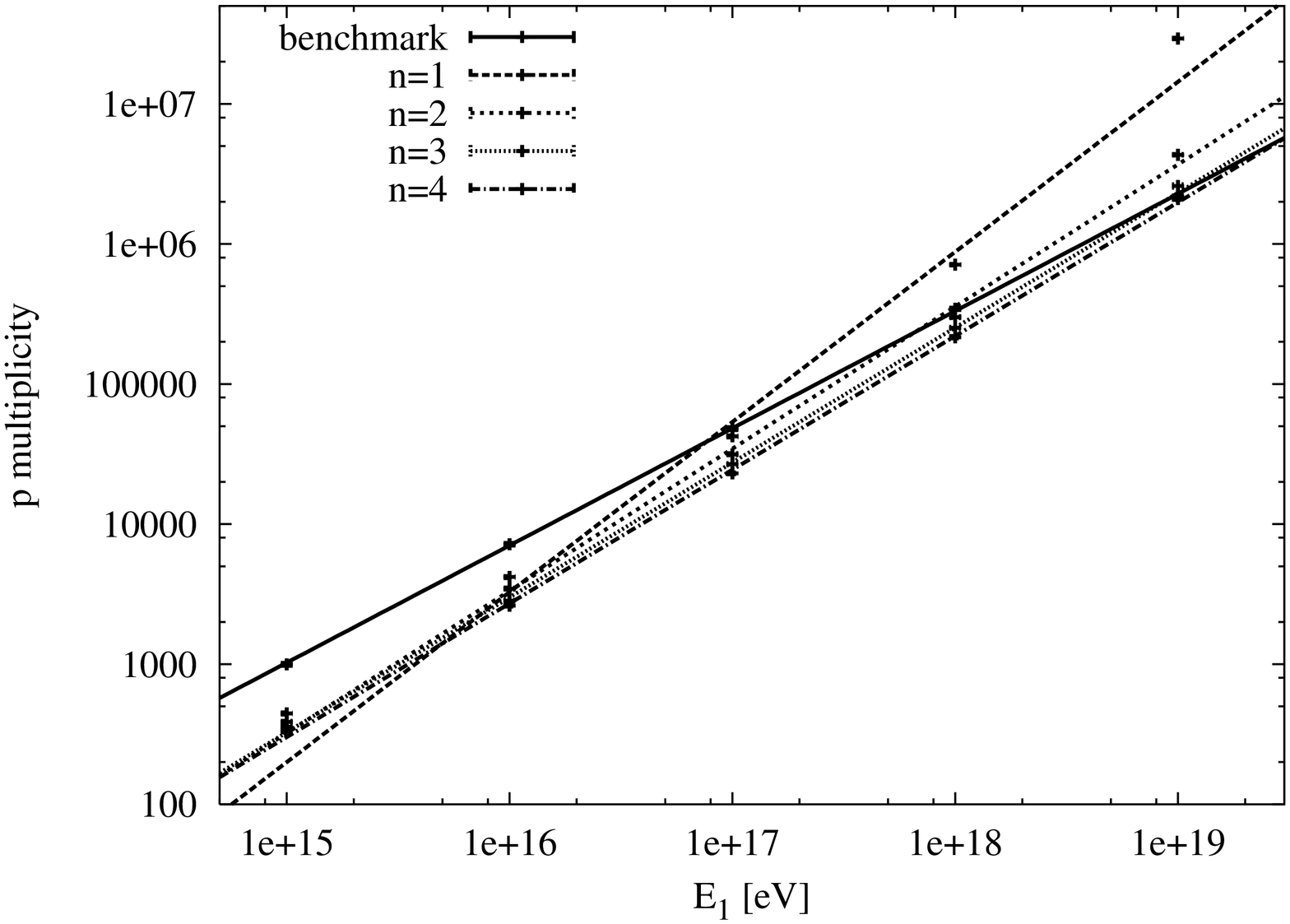}}}} \par}
\caption{Plot of the multiplicity
of protons as a function of \protect\( E_{1}\protect \), (a) 5,500 m, (b) 15,000 m.}
\label{multipro}
\end{figure}
The curves are very well fitted in a log-log plot by 
a linear relation of the form 
\beq
N=10^{q(n)} E_1^{\sigma(n)}  
\label{ntotal}
\eeq
with intercepts $q(n)$ and slopes $\sigma(n)$, 
that increase with the number of extra dimensions $n$. 
We present in Figures (a) results of the simulations performed with a first 
impact taken at 5,500 m, while Figures (b) refer to a first collision at 
15,000 m. In Figures of type (a) the slopes of the benchmark events are 
smaller than those of the black hole events and show a larger intercept. 
This feature is common to all the  
sub-components of the air showers. A simple explanation of this fact is 
that at lower value of the impact energy, the 
number of states available for the decay of the black hole is smaller 
than the number of partonic degrees of freedom available in 
a proton-proton collision. We recall, as we have already discussed in 
the previous sections, that our benchmark results define in this case 
an upper bound for the total multiplicities expected in a 
neutrino-proton collision. Therefore, in a more realistic comparison, 
we would discover that the black hole and the standard results should 
differ more noticeably. 
The large multiplicity of the states available for the decay of the black 
hole dominates over that of a standard hadronic interaction, and this 
justifies the larger multiplicities produced at detector level.  
As we increase the altitude of the impact, in plots of type (b) we find 
a similar trend but the differences in the total and partial multiplicities 
are much harder to discern for black holes and benchmark events. In fact, 
for collisions starting at higher altitudes the showers are all 
fully developed and the differences between the two underlying events are 
less pronounced. 

Another feature of the black hole events is that the slopes and the 
intercepts of the various plots, for a given choice of altitude of 
the impact, are linearly correlated. To illustrate this 
point we refer to Figures~\ref{fitcurve1}-\ref{fitcurve2} from which 
this behaviour is 
immediately evident. To generate each of these figures we have plotted 
the parameters 
$(\sigma,q)$ of a corresponding plot - for the total or for the partial 
multiplicities - 
independently of the specific number of extra dimensions. The results 
shown in these 
figures clearly indicate that the relation between the intercept q and 
the slope $\sigma$ 
appearing in Eqn.~(\ref{ntotal}) is linear and independent of $n$ 
\beq
q=\alpha\,\sigma+\beta
\eeq
with $\alpha$ and $\beta$ typical of a given setup (photons, total multiplicities,
etc.) 
but insensitive to the parameter $n$. Therefore, black hole events are characterized 
by particle multiplicities on the ground of the form  
\beq
N_{ground}=10^{\alpha\,\sigma + \beta} E_1^{\sigma}.  
\label{ntotalfinal}
\eeq

\begin{figure}
{\centering \subfigure[ ]{\resizebox*{10cm}{!}{\rotatebox{-90}
{\includegraphics{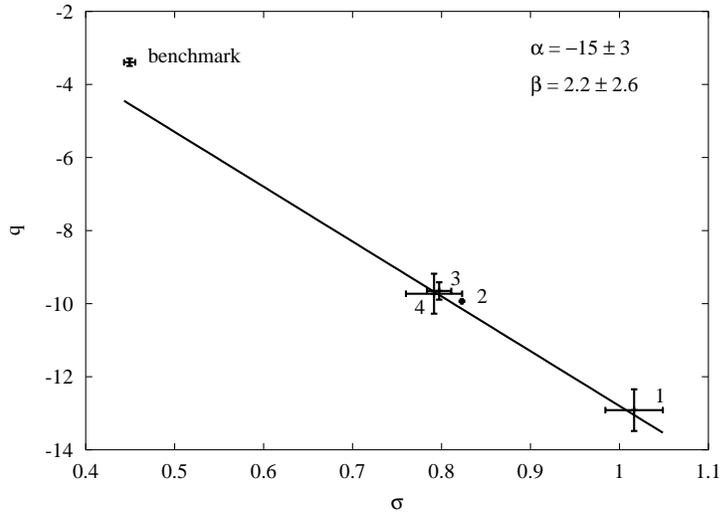}}}} \par}
{\centering \subfigure[ ]{\resizebox*{10cm}{!}{\rotatebox{-90}
{\includegraphics{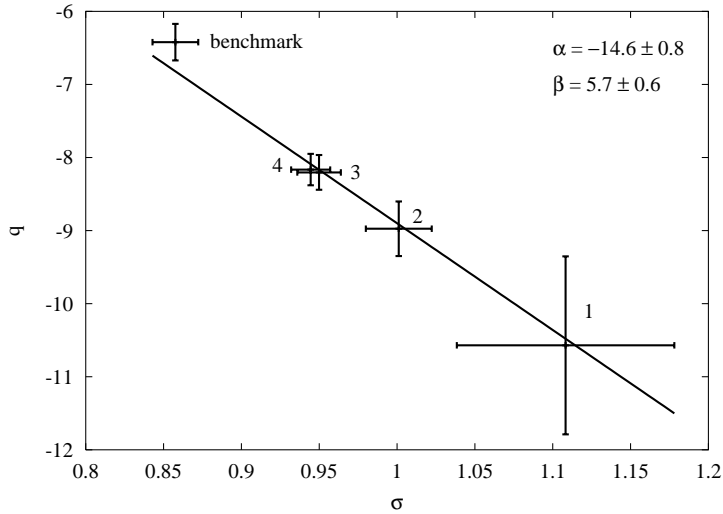}}}} \par}
\caption{Parameter fit for the intercepts and the
slopes of the curves in Fig.~\ref{prima} for the total
multiplicities. The numbers over each point in this plot indicate the
value of the extra-dimensions. The \protect\( (\sigma,q)\protect \) parameters
are fitted to a straight line \protect\( q= \alpha\,\sigma+\beta \protect \)
independently of the
numbers of extra dimensions. (a) is the fit for 5,500 m, (b) for 15,000 m.
The benchmark is also shown in the plot, but has not been used in the fit.}
\label{fitcurve1}
\end{figure}
\begin{figure}
{\centering \subfigure[ ]{\resizebox*{10cm}{!}{\rotatebox{-90}
{\includegraphics{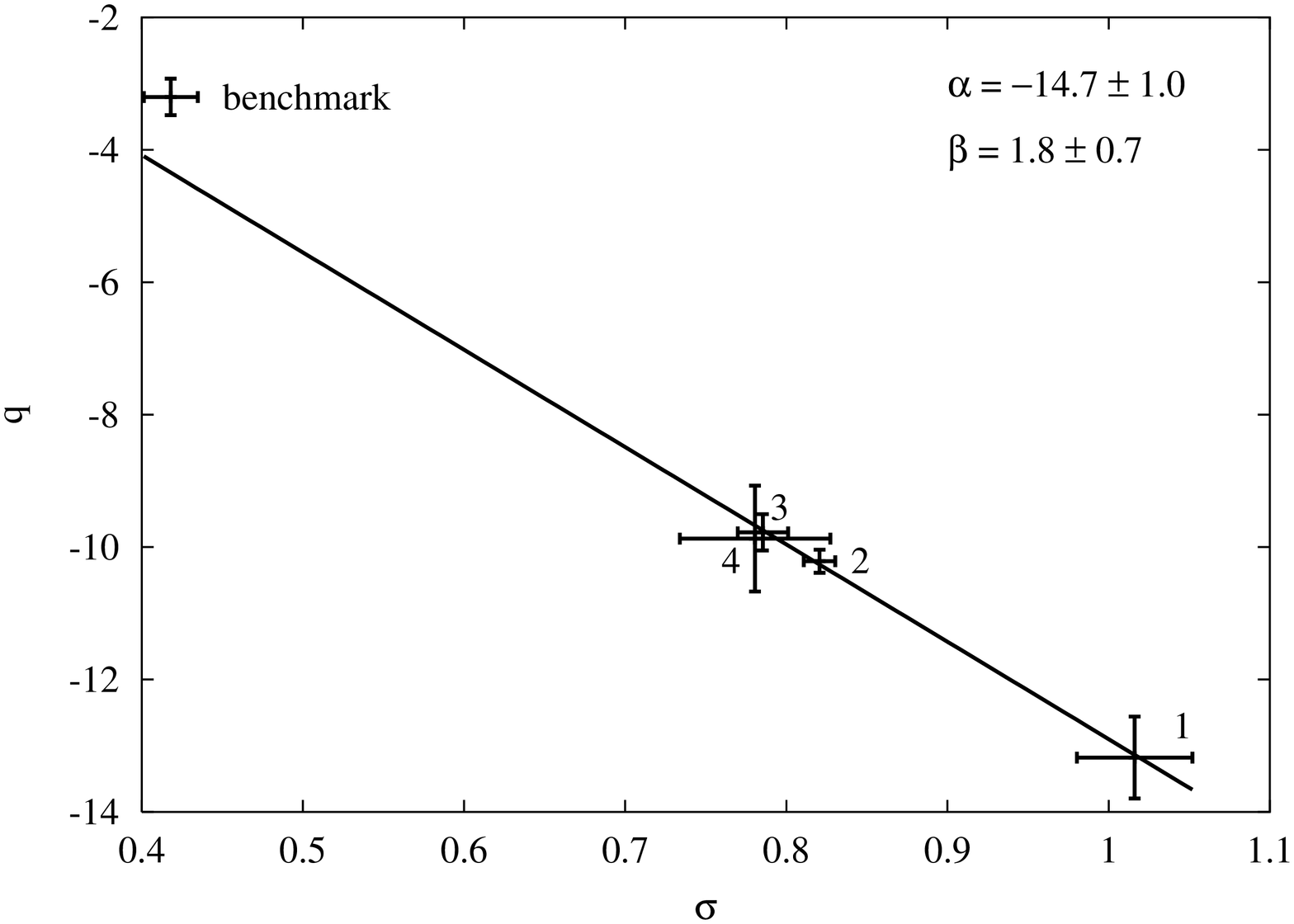}}}} \par}
{\centering \subfigure[ ]{\resizebox*{10cm}{!}{\rotatebox{-90}
{\includegraphics{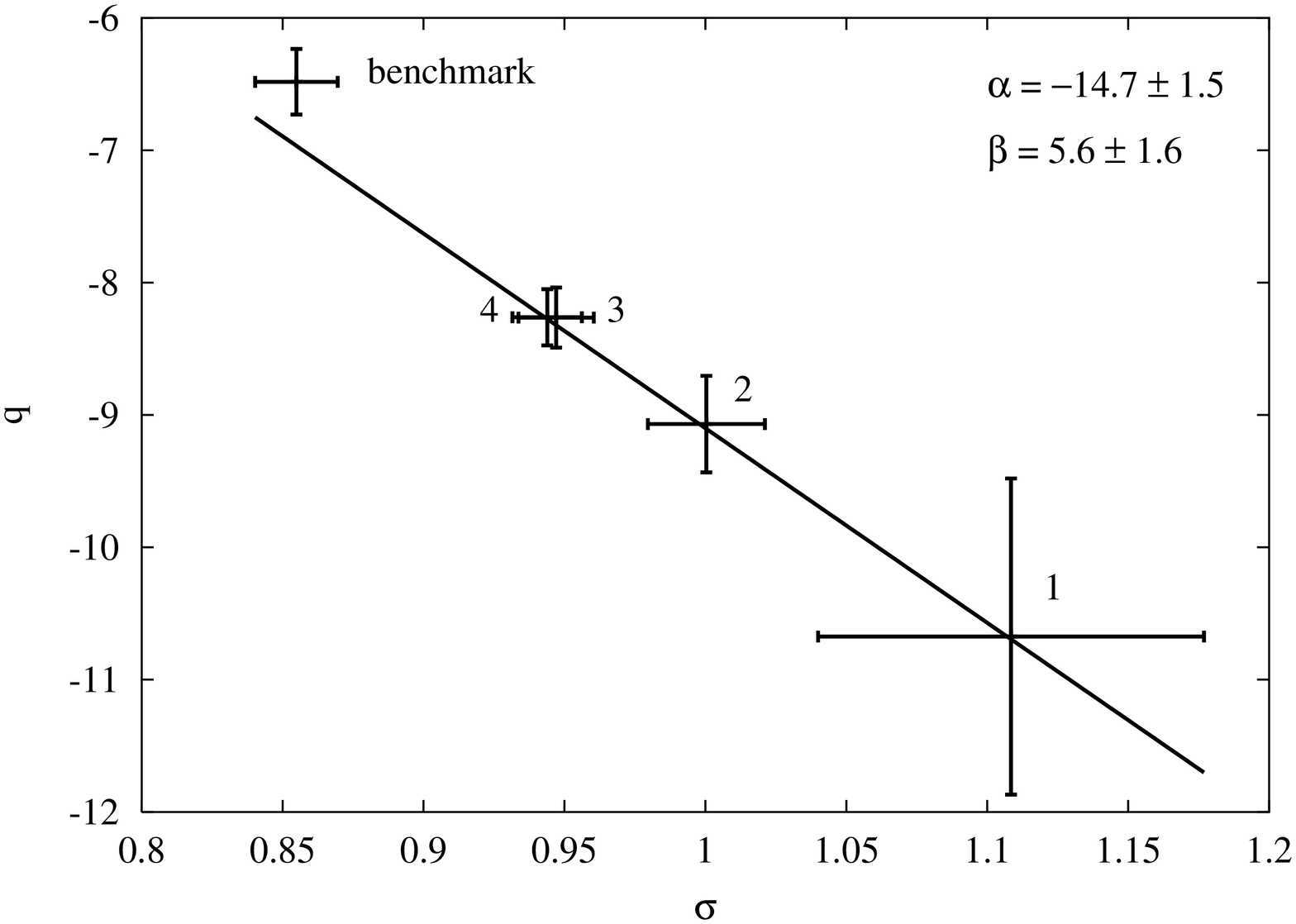}}}} \par}
\caption{Paramater fit for the intercepts and the slopes for the
curves in Fig.~\ref{seconda}, now for the multiplicity of photons. 
(a) is the fit for 5,500 m, (b) for 15,000 m.}
\label{fitcurve2}
\end{figure}

%\vfill

%\eject
\vspace{0.6cm}

\item{\em Lateral distributions}

In Figures~\ref{core1}-\ref{core5} we illustrate 
the results of our study of the lateral 
distributions for the total inclusive shower and the various sub-components 
as a function of the incoming energy $E_1$. The average opening 
of the shower as measured at detector level is plotted versus energy in a log-log 
scale. Notice a growing opening of the shower as 
we raise the energy of the black hole resonance, which is more 
remarked for a lower number of extra dimensions. In contrast, 
the benchmark simulation 
shows a small decrease (negative slope) with energy. The larger opening 
of the shower in black hole mediated events - compared to standard air 
showers - is due to the s-wave emission typical of a black hole decay, 
which is very different from an ordinary collision. Contrary to the 
case of multiplicities, here simulations of type (a) and 
(b) show a similar trend, with very distinct features between standard and 
black hole events. 
Notice that in this case the difference in the partonic content of the 
two different events (benchmark versus black hole mediated) is less relevant, 
since it is the geometrical fireball emission in the black hole case 
which is responsible for the generation of larger lateral distributions.

\begin{figure}
{\centering \subfigure[ ]{\resizebox*{10cm}{!}{\rotatebox{-90}
{\includegraphics{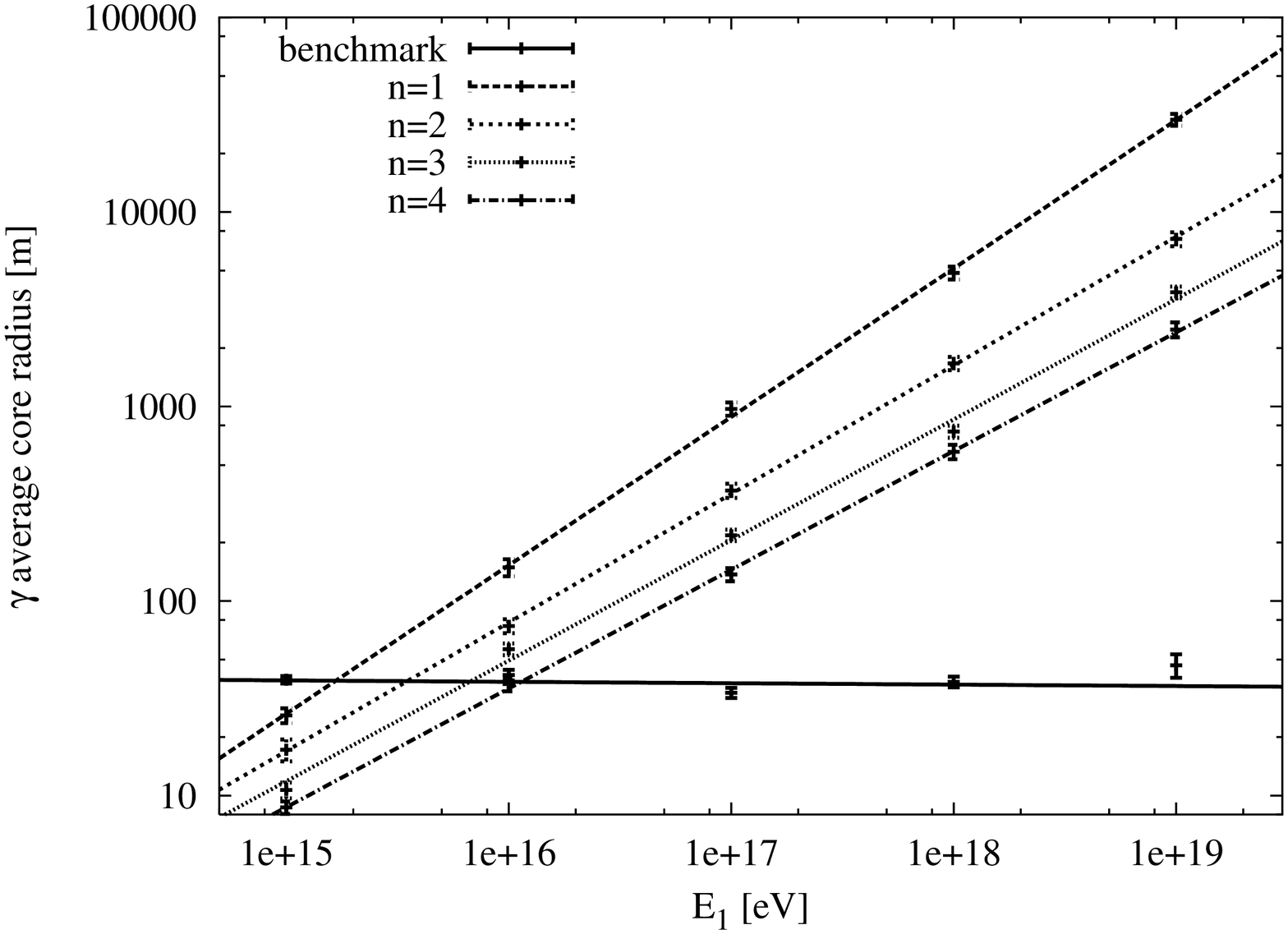}}}} \par}
{\centering \subfigure[ ]{\resizebox*{10cm}{!}{\rotatebox{-90}
{\includegraphics{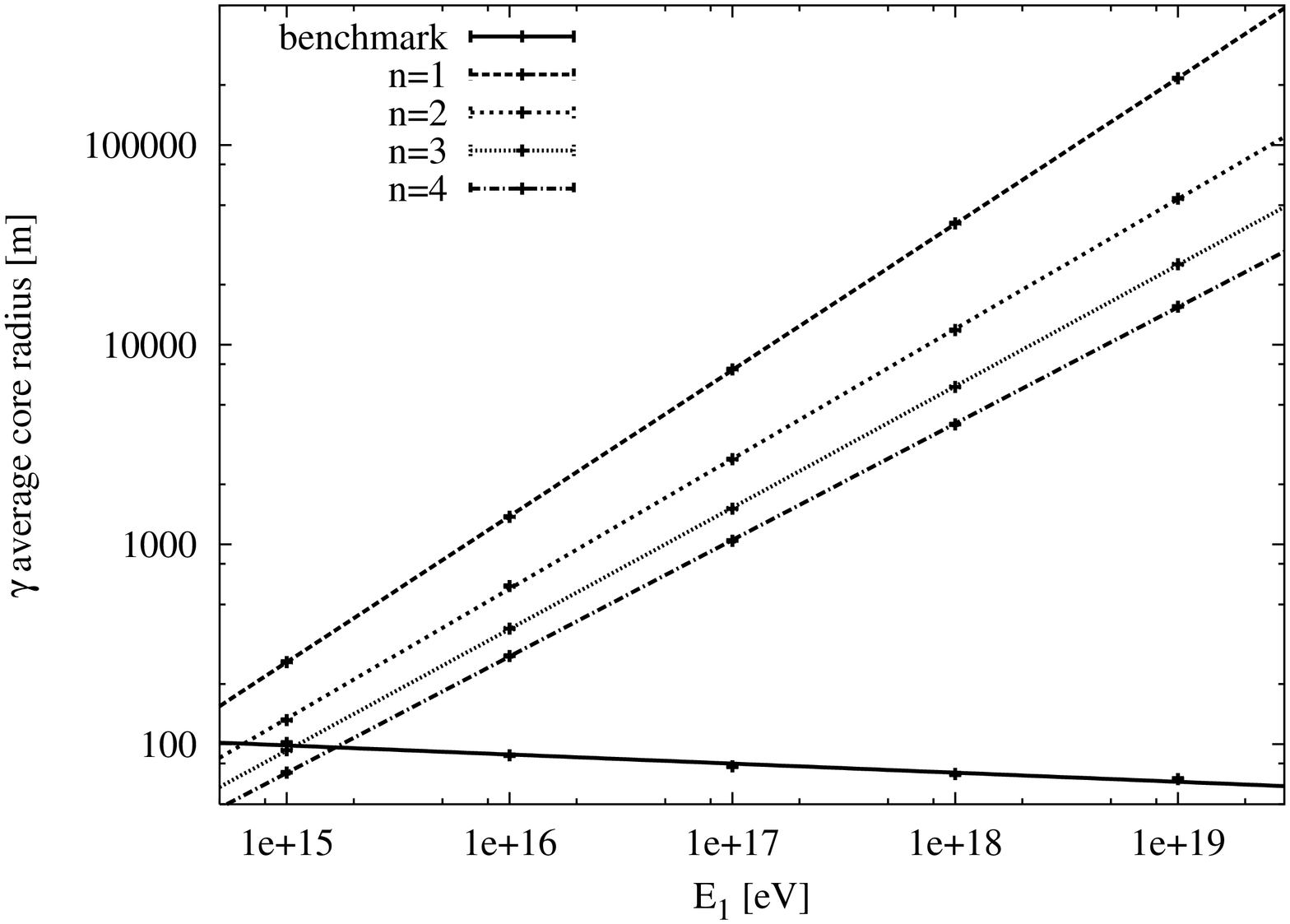}}}} \par}
\caption{Plot of the average
radius $R$ of the core of the shower of photons as a function 
of \protect\( E_{1}\protect \) for a black hole with a varying number 
of extra dimensions.
The benchmark result is also shown for comparison. (a) 5,500 m, (b) 15,000 m.}
\label{core1}
\end{figure}

\begin{figure}
{\centering \subfigure[ ]{\resizebox*{10cm}{!}{\rotatebox{-90}
{\includegraphics{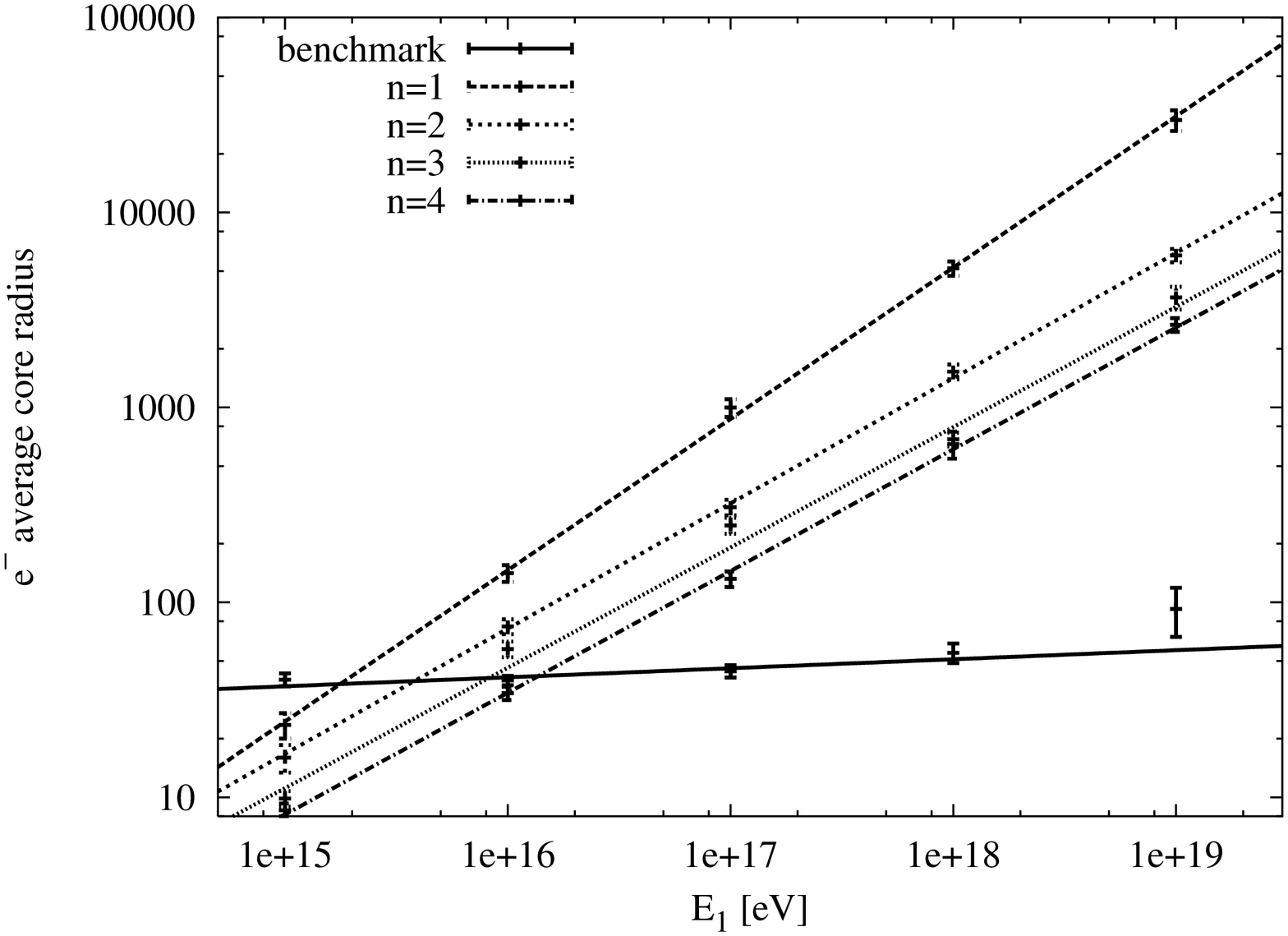}}}} \par}
{\centering \subfigure[ ]{\resizebox*{10cm}{!}{\rotatebox{-90}
{\includegraphics{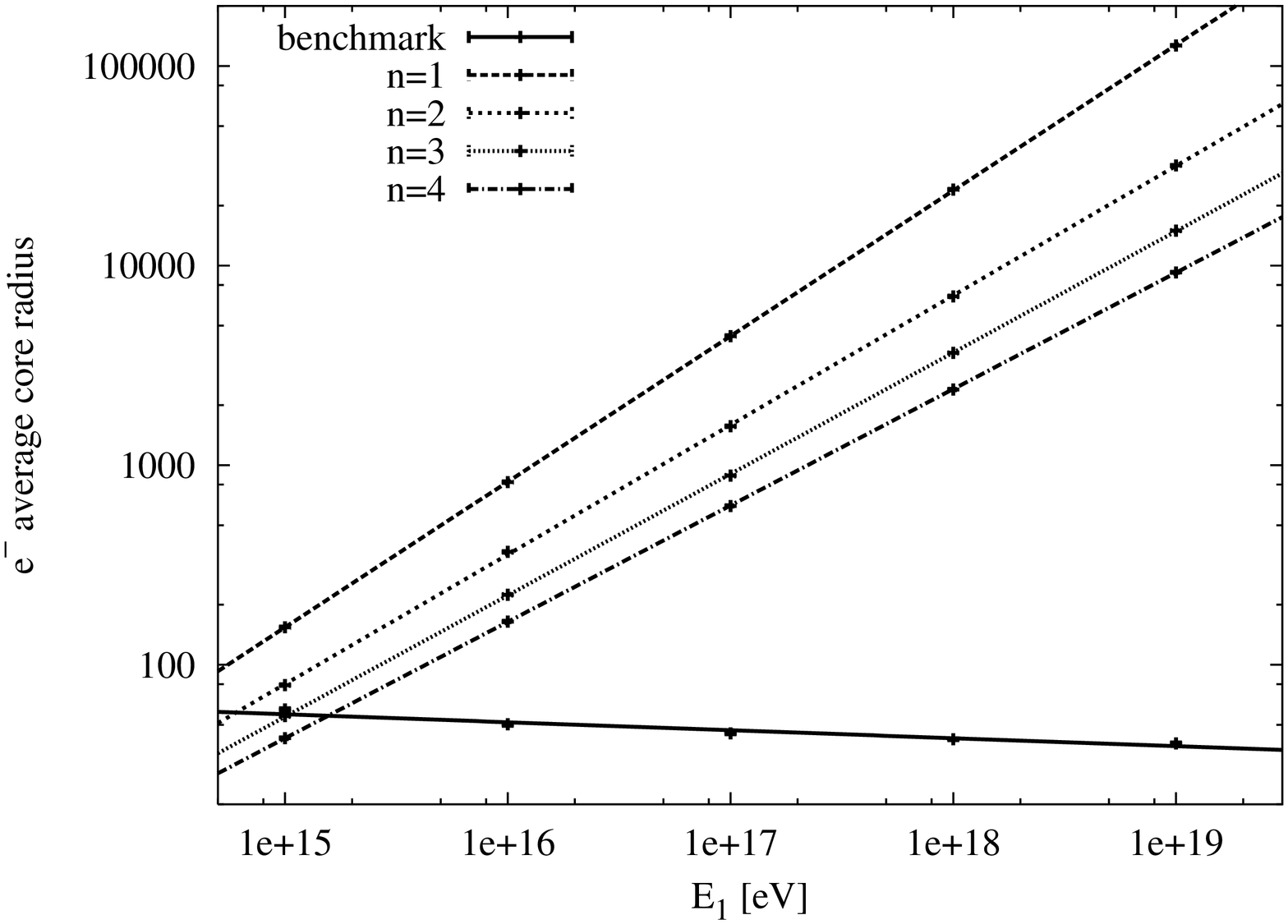}}}} \par}
\caption{Plot of the average
size of the core $R$ of the shower of \protect\( e^{-}\protect \)
as a function of \protect\( E_{1}\protect \). (a) 5,500 m, (b) 15,000 m.}
\label{core2}
\end{figure}

\begin{figure}
{\centering \subfigure[ ]{\resizebox*{10cm}{!}{\rotatebox{-90}
{\includegraphics{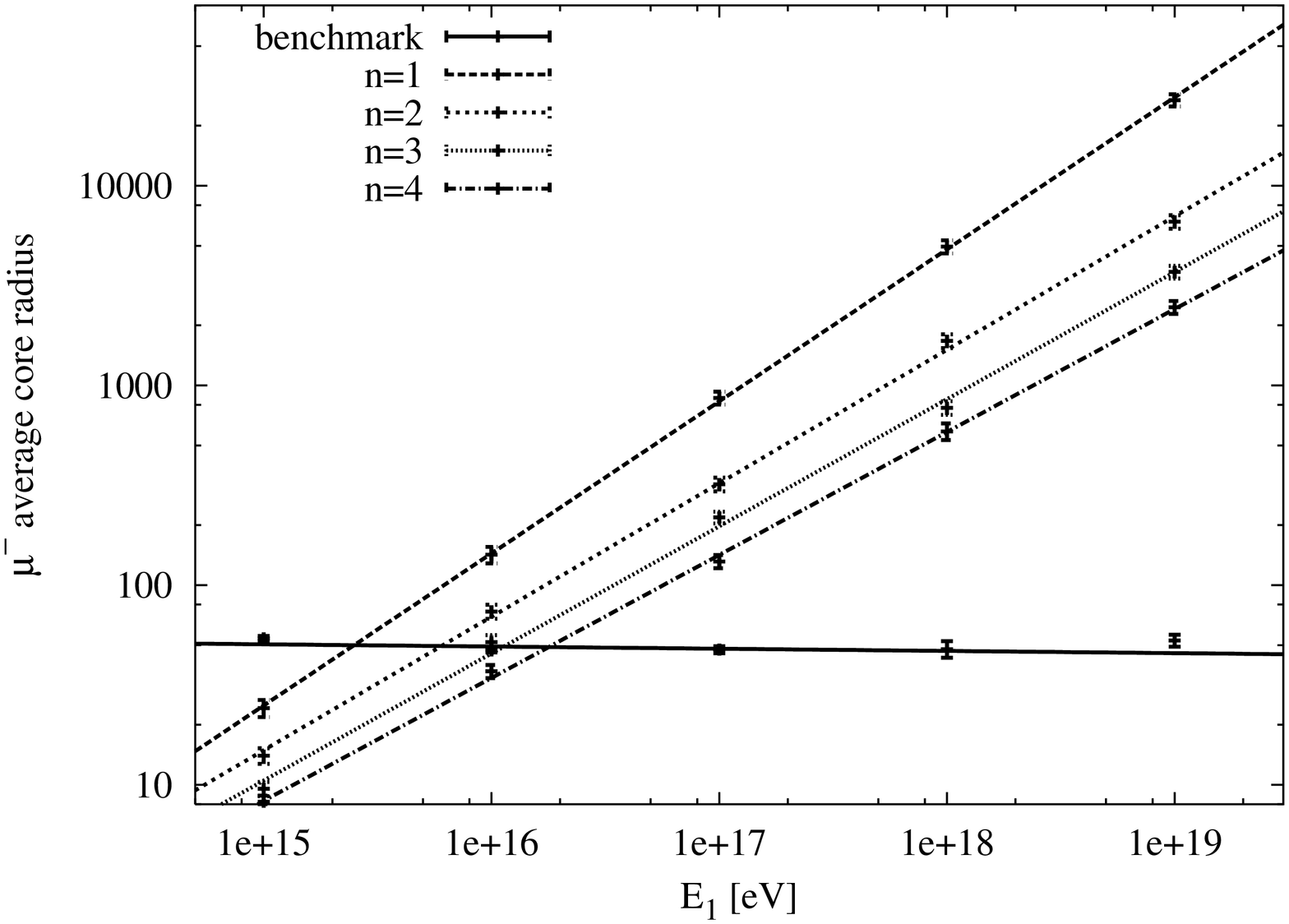}}}} \par}
{\centering \subfigure[ ]{\resizebox*{10cm}{!}{\rotatebox{-90}
{\includegraphics{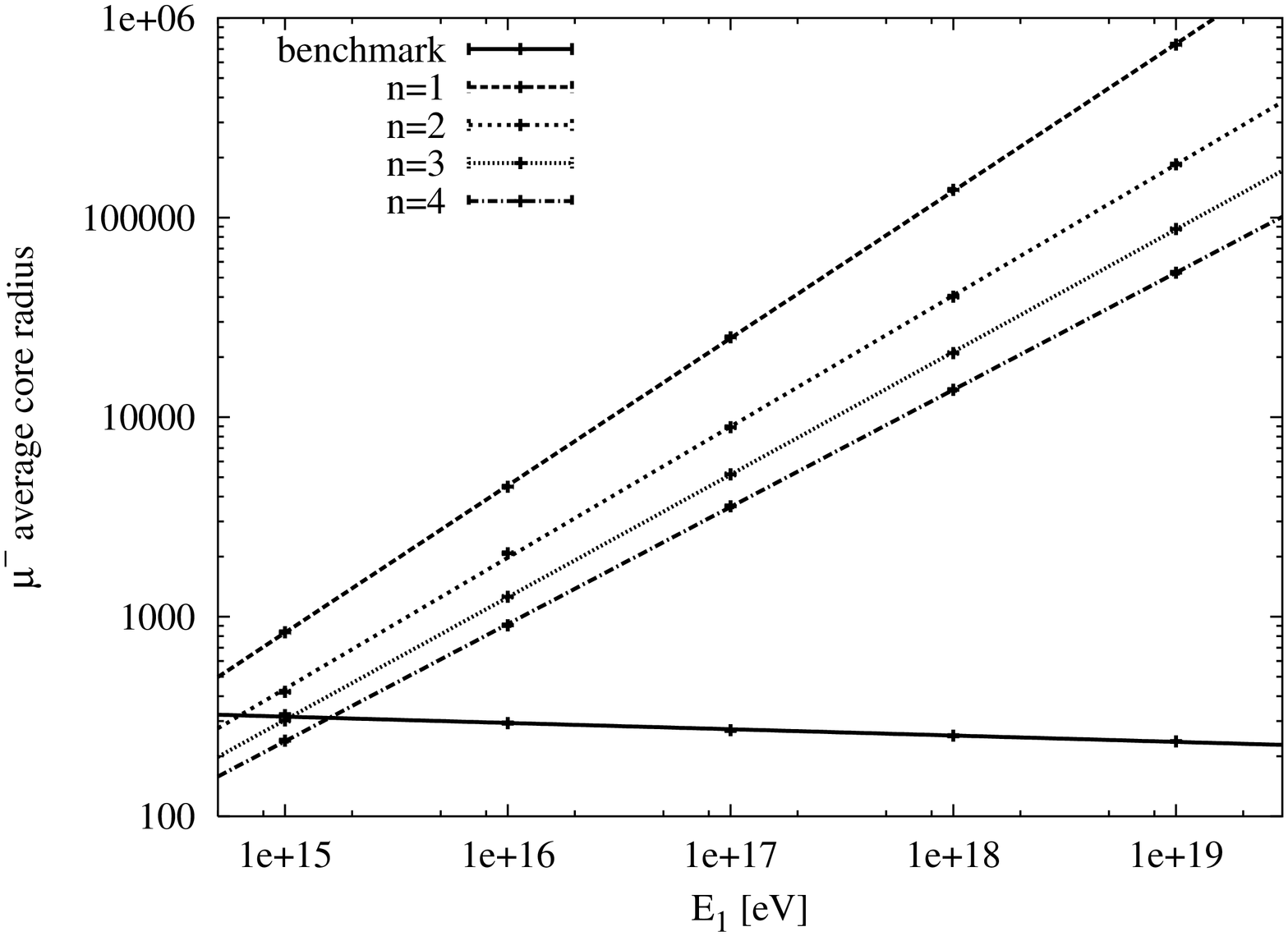}}}} \par}
\caption{Plot of the average
size $R$ of the core of the shower of \protect\( \mu ^{-}\protect \)
as a function of \protect\( E_{1}\protect \), (a) 5,500 m, (b) 15,000 m.}
\label{core3}
\end{figure}

\begin{figure}
{\centering \subfigure[ ]{\resizebox*{10cm}{!}{\rotatebox{-90}
{\includegraphics{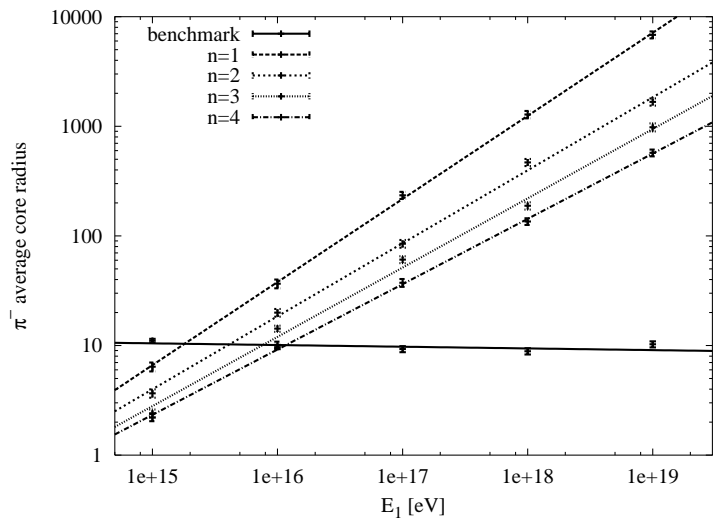}}}} \par}
{\centering \subfigure[ ]{\resizebox*{10cm}{!}{\rotatebox{-90}
{\includegraphics{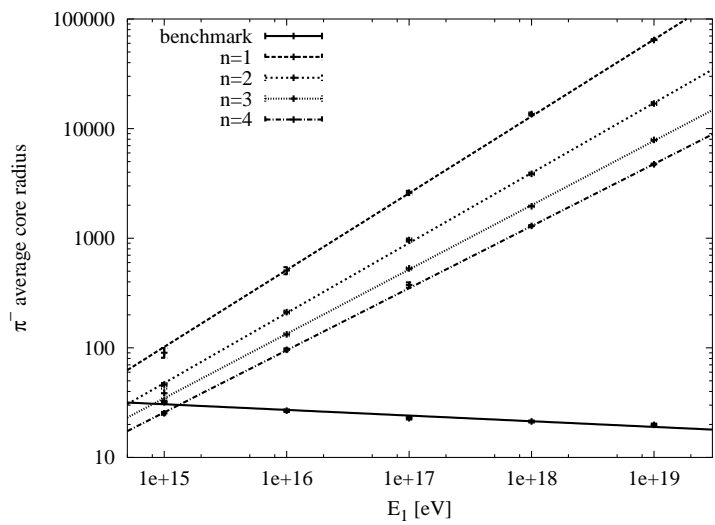}}}} \par}
\caption{As above for \protect\( \pi ^{-}\protect \)}
\label{core4}
\end{figure}

\begin{figure}
{\centering \subfigure[ ]{\resizebox*{10cm}{!}{\rotatebox{-90}
{\includegraphics{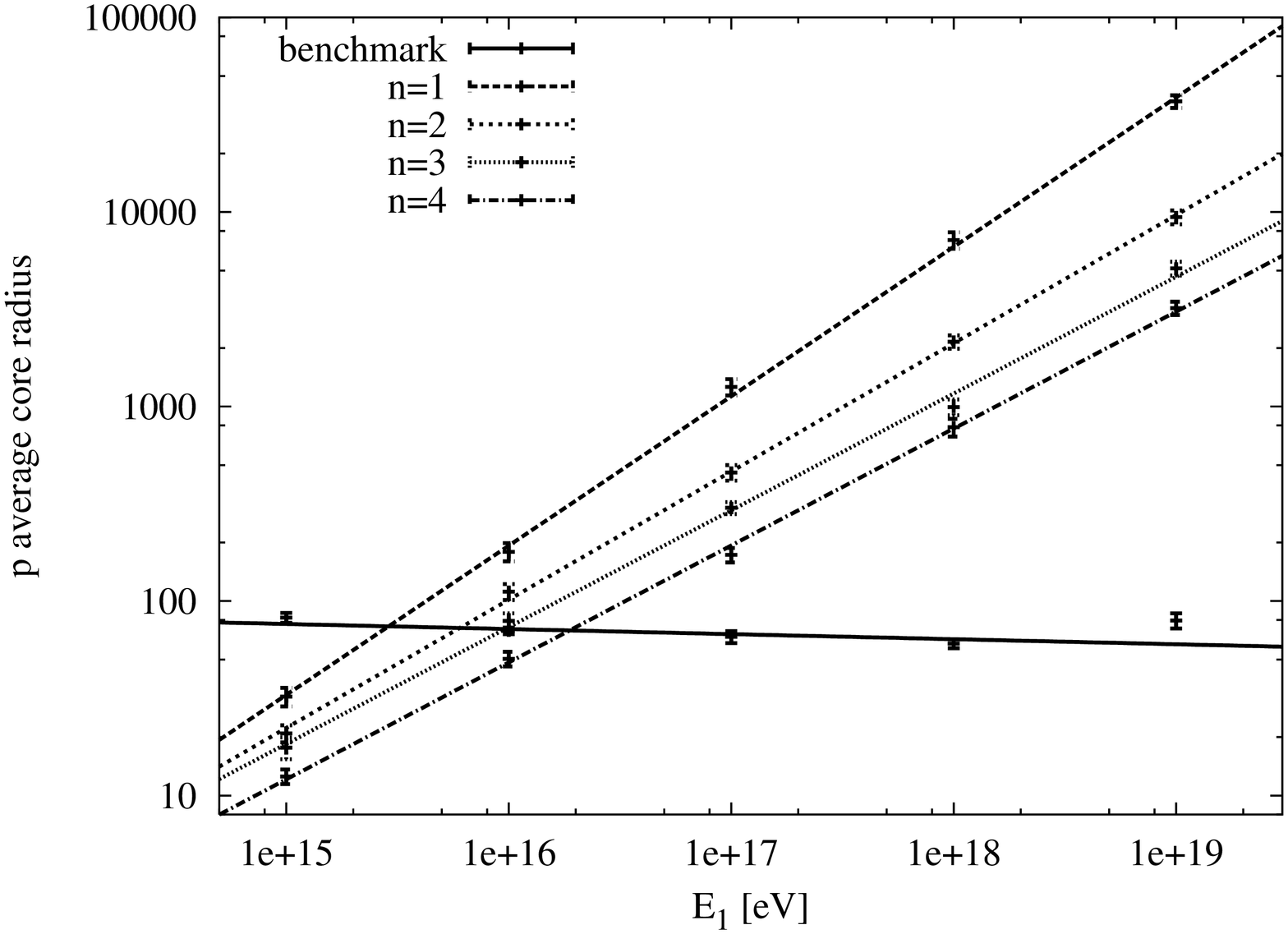}}}} \par}
{\centering \subfigure[ ]{\resizebox*{10cm}{!}{\rotatebox{-90}
{\includegraphics{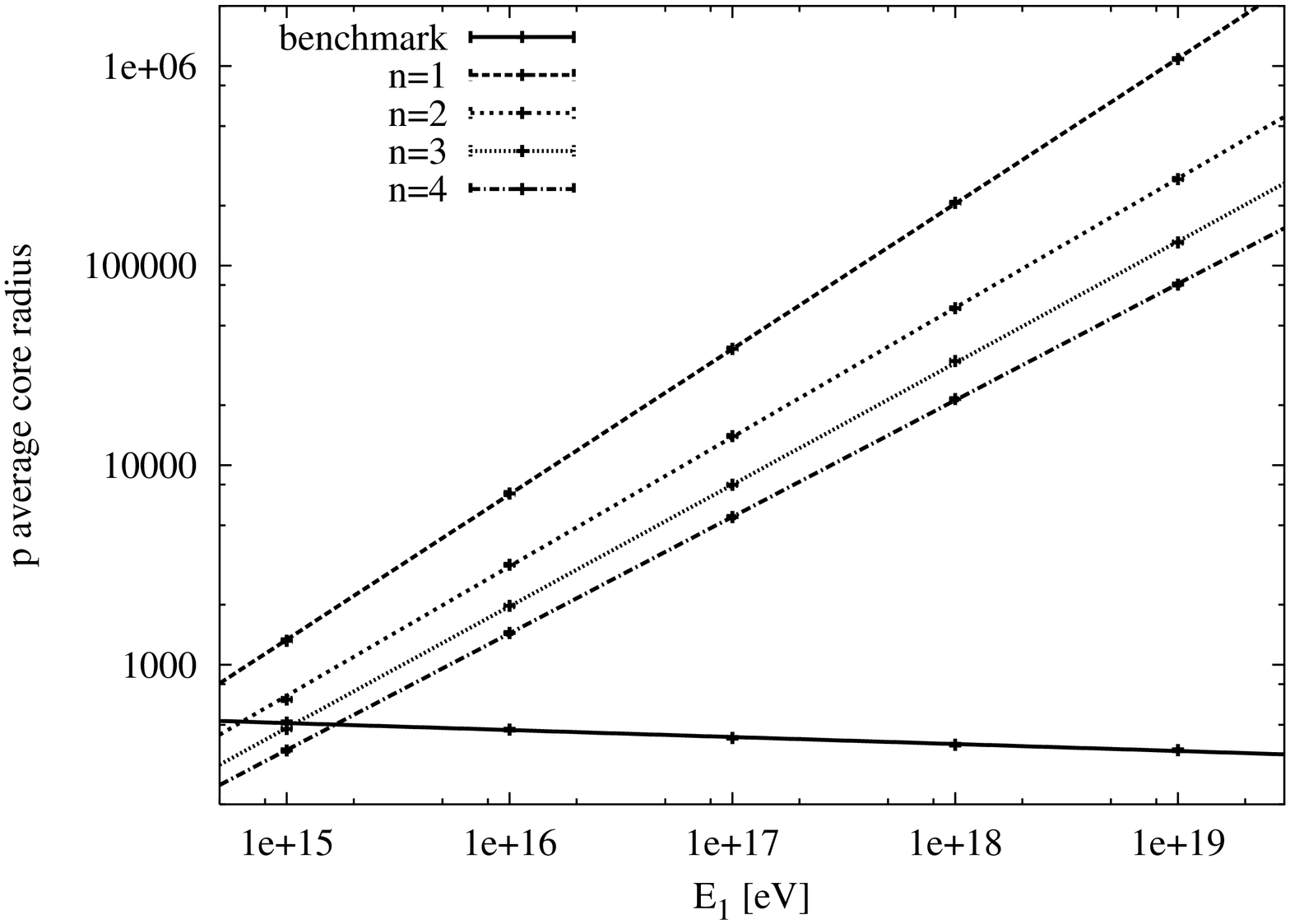}}}} \par}
\caption{The proton core size as a function of \protect\( E_{1}\protect \), 
(a) 5,500 m, (b) 15,000 m.}
\label{core5}
\end{figure}

Also in this case we discover a linear relation between 
average radius $R$ of the conical openings and energy, relation that can be fitted 
to a simple power law
\beq
R=10^{q'(n)} E_1^{\sigma'(n)}.  
\label{ntotal1}
\eeq
In analogy to Figures~\ref{fitcurve1} and \ref{fitcurve2}, we show in 
Figure~\ref{fitcurve3} that for a given setup there is a linear relation between 
slopes and intercepts of Eq.~(\ref{ntotal1})
\beq
q'=\alpha'\,\sigma' +\beta'
\eeq
with $\alpha'$ and $\beta'$ typical for a given setup, 
but again independent of $n$.

\end{itemize}

\begin{figure}
{\centering \subfigure[ ]{\resizebox*{10cm}{!}{\rotatebox{-90}
{\includegraphics{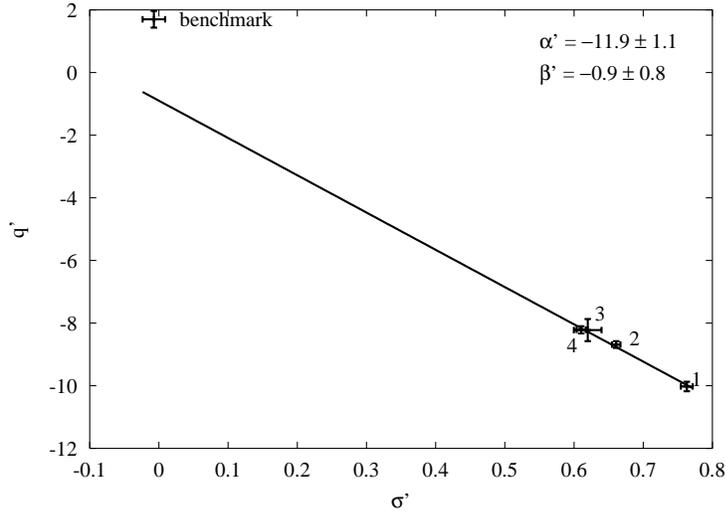}}}} \par}
{\centering \subfigure[ ]{\resizebox*{10cm}{!}{\rotatebox{-90}
{\includegraphics{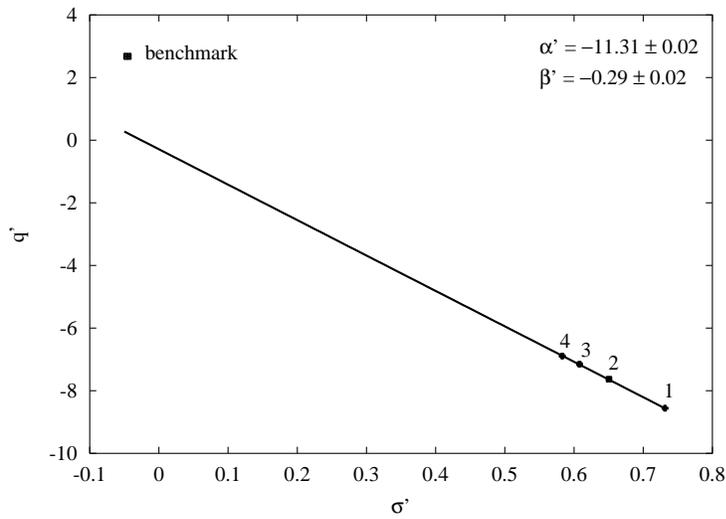}}}} \par}
\caption{Parameter fit for the curves
in Fig.~\ref{core1}, describing the openings of the showers of photons. 
(a) is the fit for 5,500 m, (b) for 15,000 m.}
\label{fitcurve3}
\end{figure}

%\vfill

%\eject

%\clearpage
\section{Discussion}

A rather detailed analysis was presented of 
some of the main observables which characterize the air showers
formed, when a high energy collision in the atmosphere
leads to the formation of a mini black hole. We have decided
to focus our attention on the particle
multiplicities of these events, on the geometrical opening
of the showers produced, and on the ratio of their electromagnetic to
hadronic components, as functions of the entire ultra high energy
spectrum of the incoming primary source.
We have shown that in a double logarithmic scale
the energy vs multiplicity as well as the energy vs shower-size 
plots are linear, characterized
by slopes which depend on the number of extra dimensions.
We have compared these predictions with standard (benchmark) simulations 
and corrected for the energy which escaped in the bulk, or emitted 
by the black holes at stages prior to the Schwarzschild phase.
Black hole events are characterized by faster growing 
multiplicities for impacts taking place close to the detector; impacts 
at higher altitudes share a similar trend, but less pronounced. The
multiplicities from the black hole are larger in the lower part of the energy
range, while they become bigger for higher energies.
We should also mention that, given the choice made for our benchmark 
simulations, here we have been considering the worst scenario: in a 
simulation with an impacting neutrino 
it should be possible to discern between the two underlying events, 
whether they are standard or black hole mediated.   
The lateral distributions appear to be the most striking signature 
of a black hole event. Due to the higher $p_T$s involved, they are 
much larger than in the benchmark standard simulations. 

Our analysis can be easily generalized 
to more complex geometrical situations, where 
a slanted entry of the original primary is considered and, in particular, 
to horizontal air showers, which are relevant for the detection of 
neutrino induced showers, on which we hope to return in a future work. 

To address issues related to the Centauros, one has to concentrate on
black holes produced closer to the detectors, since the 
main interaction in these events is believed to have taken place somewhere 
between 0 and 500 m above the detectors \cite{CCT2}. 
The multiplicities in the showers will of course decrease and are expected,
based on the above graphs, to get very close to 
the values observed, while the ratio of $N_{{\rm em}}/N_{{\rm hadron}}$ 
will also approach the observed values, 
since the kaons produced by the black hole's "democratic" evaporation
will not have enough time to decay and will be counted as hadrons.
Another crucial question in connection with the mini black hole
interpretation of the Centauro events 
is whether the partonic decay products 
of the black hole passed through a high temperature 
quark-gluon plasma phase and whether a Disoriented Chiral
Condensate (DCC) was formed, before it finally turned into hadrons. In the present 
analysis it was implicitly assumed that no such phase was developed.
However, if a DCC forms, the prediction for the ratio of the 
electromagnetic to hadronic component will be drastically different.

Needless to say, the subject of black hole production and evaporation
touches upon several different subfields of theoretical and experimental 
high energy physics and 
astrophysics. The fields of Cosmic ray physics and of 
the air shower formation in the atmosphere; 
of quantum gravity/string theory, of non-perturbative low energy QCD and 
of the quark-gluon plasma phase, just to mention a few, 
and this may not even be a complete list. 
Given our incomplete understanding of all of these, it is clear that
a lot more has to be done, before one can safely compare the theory to the observational data. The analysis of diffractive interactions at those energies, 
in particular, will require considerable attention.
Nevertheless, in our view, the issues involved are very important and
deserve every effort.

\vspace{2cm}

\centerline{\bf Acknowledgments}

\vspace{0.5cm}

We thank Raf Guedens, Dominic Clancy, Andrei Mironov, Alexey Morozov and 
Greg Landsberg for helpful discussions. 
The work of T.N.T. is partially supported by the grant HRPN-CT-2000-00122 
from the EU, as well by the Hellenic Ministry of Education grants 
``O.P.Education - Pythagoras'' and ``O.P.Education - Heraklitos''.
The work of C.C. and A.C. is supported by INFN of Italy (BA21).
The simulations have been performed at the INFN-Lecce computer cluster.
C.C. thanks the Theory Group at the University of Crete and in particular
E. Kiritsis for hospitality, and the Theory Division at the 
Max Planck Institute in Munich for hospitality while completing this work.

\vspace{1.5cm}
\centerline{\bf POSTSCRIPT}
After completing these studies we have been informed by D. Heck 
that a new version of CORSIKA has been released, which 
is able to deal with neutrino primaries. As we have discussed in our work, 
the use of a more realistic benchmark based on neutrino primaries does not 
invalidate the results of our simulations, but should actually render 
the differences between mini black hole mediated and standard events 
even more pronounced.

%%%%%%%%%%%%%%%%%%%%%%%%%%%%%%%%%%%%%%%%%%%%%%%%%%%%%%%%%%%%%%%%%%%%%%%%%%%%%%%%%%%%%%%%%%%%%%%%

\end{document}